\documentclass[iop,numberedappendix]{emulateapj}
\usepackage{natbib}
\usepackage{amsmath}
\usepackage[dvipsnames]{xcolor}
\newcommand{\Msun}{\ensuremath{\mathrm{M_{\odot}}}}

\newcommand{\Hy}{\ensuremath{\textrm{H}}}
\newcommand{\HI}{\ensuremath{\textrm{H} \, \textsc{i}}}
\newcommand{\HII}{\ensuremath{\textrm{H} \, \textsc{ii}}}
\newcommand{\He}{\ensuremath{\textrm{He}}}
\newcommand{\HeI}{\ensuremath{\textrm{He} \, \textsc{i}}}
\newcommand{\HeII}{\ensuremath{\textrm{He} \, \textsc{ii}}}
\newcommand{\HeIII}{\ensuremath{\textrm{He} \, \textsc{iii}}}
\newcommand{\nH}{n_{\Hy}}
\newcommand{\nHI}{n_{\HI}}
\newcommand{\nHII}{n_{\HII}}
\newcommand{\nHe}{n_{\He}}
\newcommand{\nHeI}{n_{\HeI}}
\newcommand{\nHeII}{n_{\HeII}}
\newcommand{\nHeIII}{n_{\HeIII}}
\newcommand{\nel}{n_{\rm e}}

\newcommand{\mean}[1]{\langle #1 \rangle}
\newcommand{\lyalpha}{Lyman-$\alpha$}
\newcommand{\Deltaopt}{\Delta_{\bigstar}}
\newcommand{\kurv}{\langle|\kappa|\rangle}
\newcommand{\lj}{\lambda_{\rm P}}
\newcommand{\taue}{\tau_{\rm e}}

\shorttitle{UVB in Hydrodynamical Simulations}
\shortauthors{O\~norbe et al.}

\begin{document}

\title{Self-Consistent Modeling of Reionization in Cosmological Hydrodynamical Simulations}

\author{Jose~O\~norbe\altaffilmark{1}, Joseph~F.~Hennawi}
\affil{Max-Planck-Institut f\"ur Astronomie, K\"onigstuhl 17, 69117 Heidelberg, Germany}
\and
\author{Zarija~Luki\'c}
\affil{Lawrence Berkeley National Laboratory, Berkeley, CA 94720-8139, USA}
\altaffiltext{1}{onorbe@mpia.de}

\begin{abstract}
  The ultraviolet background (UVB) emitted by quasars and galaxies
  governs the ionization and thermal state of the intergalactic medium
  (IGM), regulates the formation of high-redshift galaxies, and is
  thus a key quantity for modeling cosmic reionization. The vast
  majority of cosmological hydrodynamical simulations implement the
  UVB via a set of spatially uniform photoionization and photoheating
  rates derived from UVB synthesis models. We show that simulations
  using canonical UVB rates reionize and, perhaps more importantly,
  spuriously heat the IGM, much earlier $z \sim 15$ than they should. 
  This
  problem arises because at $z > 6$, where observational constraints
  are nonexistent, the UVB amplitude is far too high.  We introduce a
  new methodology to remedy this issue, and we generate self-consistent
  photoionization and photoheating rates to model any chosen
  reionization history.  Following this approach, we run a suite of
  hydrodynamical simulations of different reionization scenarios and
  explore the impact of the timing of reionization and its concomitant
  heat injection on the the thermal state of the IGM. We present a comprehensive
  study of the pressure smoothing scale of IGM gas, illustrating its
  dependence on the details of both hydrogen and helium reionization,
  and argue that it plays a fundamental role in interpreting
  \lyalpha{} forest statistics and the thermal evolution
  of the IGM.  The premature IGM heating we have uncovered implies that
  previous work has likely dramatically overestimated the impact of
  photoionization feedback on galaxy formation, which sets the minimum
  halo mass able to form stars at high redshifts. We make our new UVB
  photoionization and photoheating
  rates publicly available for use in future simulations.
\end{abstract}

\keywords{intergalactic medium --- cosmology: early universe --- cosmology: large-scale structure of universe --- 
galaxies: formation --- galaxies: evolution --- methods: numerical }

\maketitle

\section{Introduction} \label{sec:intro}

In our current standard model of the universe, hydrogen and helium
account for 99\% of the baryonic mass density \citep{Planck:2015}.
After the recombination epoch,
these elements remain neutral until ultraviolet radiation from star-forming galaxies
and active galactic nuclei reionizes them. 
Therefore, this ultraviolet background (UVB) governs
the ionization state of intergalactic gas and plays a
key role in its thermal evolution through photoheating.
During the reionization of $\HI$ and later $\HeII$,
ionization fronts propagate supersonically
through the  intergalactic
medium (IGM), impulsively heating gas to $\sim10^{4}$ K
\cite[see, e.g.,][]{Abel:1999,McQuinn:2012,Davies:2016}.
As the universe evolves,
it is well known that the balance between cooling due
to  Hubble  expansion and inverse-Compton scattering of
 cosmic microwave background (CMB) photons
and  heating  due  to  the  gravitational
collapse and photoionization heating give rise to a
well-defined temperature-density relationship in the IGM
\citep[][]{Hui:1997,McQuinn:2012}:
 \begin{equation}
 T=T_{0}\times \Delta^{\gamma-1}
 \label{eq:T0gamma}
 \end{equation}
where $\Delta=\rho/\bar{\rho}$ is the overdensity with respect to the mean and
$T_{0}$ is the temperature at the mean density.
Immediately after the reionization of $\HI$
($z \lesssim 6$) or $\HeII$ ($z \lesssim 3$), $T_{0}$ is likely to be around 
$\sim2\times 10^{4}$ K and $\gamma\sim1$ \citep{Bolton:2009,McQuinn:2009}; at lower redshifts,
$T_{0}$ decreases as the universe expands,
while $\gamma$ is expected to increase and 
asymptotically approach a value of $1.62$ \citep{Hui:1997}.

Another important physical ingredient to describe the thermal
state of the IGM is the gas pressure support.
At small scales and high densities, baryons experience pressure 
forces that prevent them  from  tracing  the  collisionless  
dark  matter.  This pressure results in an effective 
3D smoothing of the baryon distribution
relative to the dark matter, at a characteristic scale.
known as the Jeans pressure smoothing scale, $\lj$.
In an  expanding  universe  with  an  evolving  thermal  state,
this scale at  a  given  epoch  is  expected  to  depend
on the entire thermal history, because fluctuations
at earlier times expand or fail to collapse depending on
the IGM temperature at that epoch \citep{Gnedin:1998,Kulkarni:2015}. 
Recently, \citet{Rorai:2013} and \citet{Rorai:2015} have shown
that an independent measurement of the pressure smoothing scale
can be obtained using the coherence of \lyalpha{} 
forest absorption in close quasar pairs \citep{Hennawi:2006,Hennawi:2010}.

\lyalpha{} forest observations between $2<z<6$ probe
the moderate overdensities characteristic of the IGM and 
therefore are a crucial tool to understand the properties of the UVB.
In the last decade, the precision of these measurements 
has continued to grow both in terms of their numbers
(BOSS\footnote{Baryon Oscillation Spectroscopic Survey (BOSS): https://www.sdss3.org/surveys/boss.php}
survey) and
in quality \citep[high signal-to-noise ratio spectrum from, e.g.][]{OMeara:2015}.
However, while it seems that we keep learning
more and more about the ionization history of the universe,
for both $\HI$ and $\HeII$ reionizations
\citep[e.g.][]{Becker:2013,Syphers:2014,Worseck:2014,Becker:2015}
the thermal history of the universe is still far from certain.
The statistical properties of the \lyalpha{} forest are sensitive to
the thermal state of the gas, trough both thermal broadening of lines
and pressure support.
When constraints on the thermal history are reviewed,
they yield very puzzling results.
Measurements of $T_{0}$ from different groups utilizing
different methodology are in poor agreement
\citep{Schaye:2000,Bolton:2008,Lidz:2010,Becker:2011,
Rudie:2012,Garzilli:2012,Boera:2014,Bolton:2014}.
A similar problem appears when measurements of the slope of 
the temperature–density relation, $\gamma$, are compared. At $z\simeq3$ some authors
have even found that $\gamma$ is either close to isothermal ($\gamma=1$)
or even
inverted \citep[$\gamma<1$;][but see \citealt{Lee:2015}]{Bolton:2008,Viel:2009}.
Most studies of the thermal state of the IGM ignore uncertainties
resulting from the unknown pressure smoothing scale \citep[but see][]{Becker:2011,Puchwein:2015}, which
produces a 3D smoothing that is difficult to disentangle from the the similar but 1D
smoothing resulting from thermal broadening
\citep{Peeples:2010a,Peeples:2010b,Rorai:2013}.
Therefore, ignoring this effect has probably
contributed to the confusing and sometimes contradictory
published constraints on $T_{0}$ and $\gamma$ \citep{Puchwein:2015}.

With the help of accurate models of the IGM, the statistics of the \lyalpha{} forest can
be used to constrain its thermal parameters and ultimately cosmic reionization.
Ideally one will run coupled radiative transfer hydrodynamical simulations that include
extra physics governing the sources of ionizing photons (stars, quasars, etc.). 
Despite significant progress on this front \citep{Wise:2014,So:2014,Gnedin:2014,Pawlik:2015,Norman:2015,Ocvirk:2015}
these simulations are still too costly for sensible exploration of the parameter space.
For this reason, the dominant approach, implemented in the vast majority of hydrodynamical codes, 
is to assume that all gas elements are optically thin to ionizing photons, such that their ionization
state can be fully described by a uniform and isotropic UV+X-ray background radiation field.
Thus, the radiation field is encapsulated by a set of photoionization
and photoheating rates that evolve with redshift for each relevant
ion.  The minimal set of ions are $\HI$, $\HeI$ and $\HeII$ in order
to track the most relevant ionization events, as well as the thermal
heating associated with them.  Of course, although this optically thin
approximation is a valid assumption once the mean free path of
ionization photons, $\lambda_{\rm mfp,\nu}$, is large enough, it is
certainly not true during cosmic reionization events.  As such, this
optically thin approach is not meant to provide an accurate
description of reionization itself, but it should at least provide a reasonable description
of the heat injection associated with reionization.

This is important since galaxies forming during the
reionization epoch are sensitive to the thermal state of the gas, and even well after reionization
gas elements can retain thermal memory of reionization heating \citep{Gnedin:1998, Kulkarni:2015}.
It is important to remark here that these UVB models 
have relevant consequences for galaxy formation and evolution models and
hydrodynamical simulations. Several
groups have already shown how important the UVB model is to
determine the star formation of the first galaxies and their evolution
by not only setting the minimum halo mass able to form stars \citep[i.e., halos
massive enough to overcome gas pressure forces;][]{Rees:1986,Sobacchi:2013}
but also regulating the gas accretion from the IGM into the more massive halos
\citep{Quinn:1996,Simpson:2013,BenitezLlambay:2015,Wheeler:2015a}.

\begin{figure*}
\begin{center}
\includegraphics[angle=0,width=\textwidth]{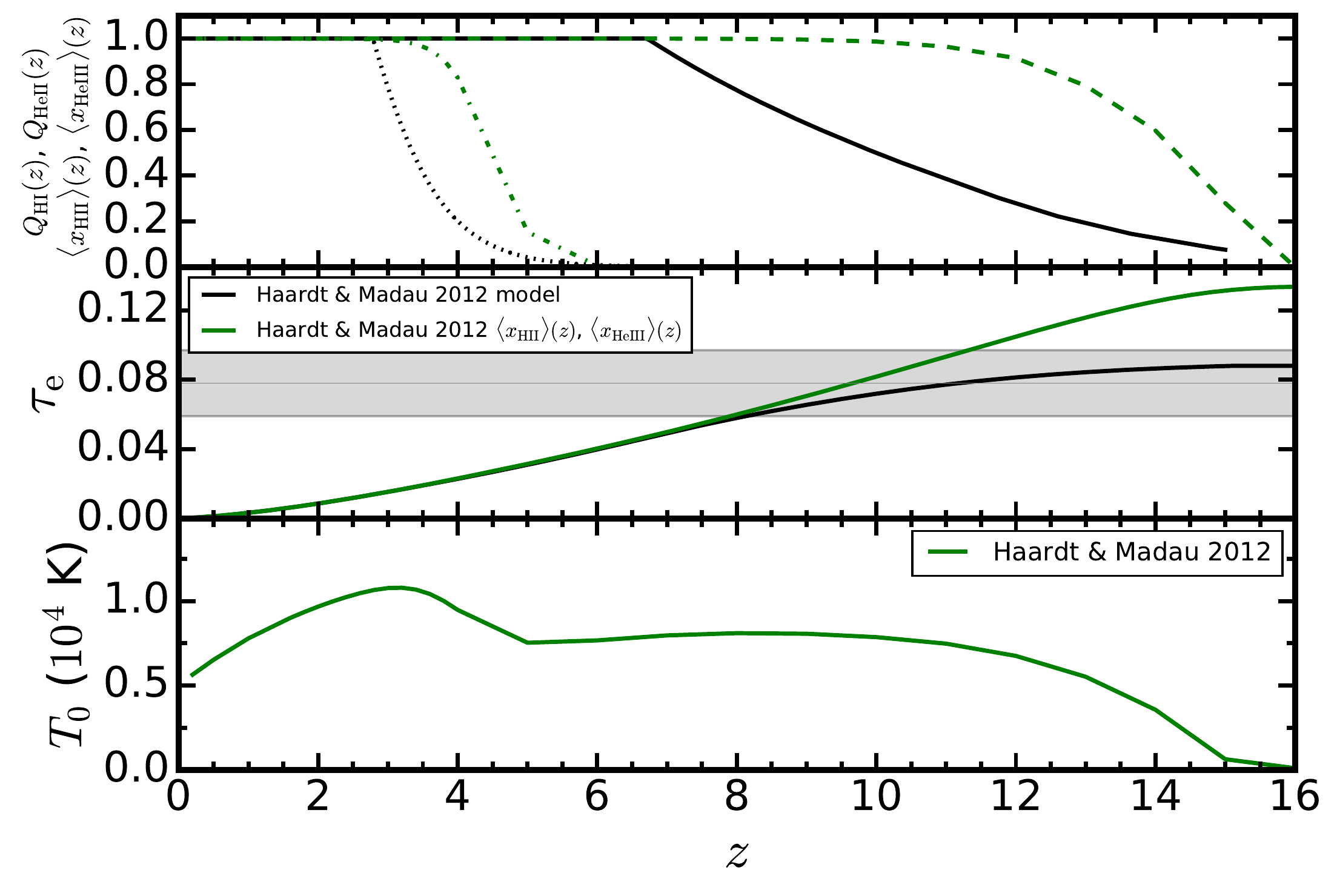}
\caption{Ionization and thermal history of the universe obtained in hydrodynamical 
simulations using standard tables from \citet{Haardt:2012}.
The upper panel shows the evolution of the 
$\HII$, $\HeIII$ volume-averaged ionized fractions in the simulation.
Dashed lines stand for the $\mean{x_{\HII}}$ and dot-dashed lines for the $\mean{x_{\HeIII}}$)
The full and the dot-dashed black lines stand for the
$Q_{\HII}$ (solid lines), $Q_{\HeIII}$ (dot-dashed lines) volume filling factors 
calculated by \citet{Haardt:2012} for their model.
The middle panel shows the integrated electron scattering optical depth, $\taue$.
The gray band stands for \citet{Planck:2015} constraints on $\taue$ coming from the CMB.
The lower panel shows the evolution of the temperature at mean density.
Notice that reionization finishes much earlier in the simulation.
All parameters presented in this 
figure are converged within $<5\%$ accuracy (see Section~\ref{ssec:convergence}).
\label{fig:Qhistgas0}}
\end{center}
\end{figure*}

The standard approach is to adopt photoionization and
photoheating rates from semianalytical synthesis models
of the UVB \citep{Haardt:1996,Haardt:2001,FaucherGiguere:2009,Haardt:2012}. 
However, these UVB synthesis models surely break down during
reionization events, and the validity of using them in optically
thin simulations (during reionization) is questionable.
Moreover, as we will show, these models are 
fundamentally inconsistent during reionization, 
leading to different reionization histories in the simulations
than the ones given by the authors. Specifically, they
reionize the universe too early, and as a result they
produce spurious heating of the IGM at early times
(see Section~\ref{sec:typmodels} and Figure~\ref{fig:Qhistgas0}).
In this paper, we improve on the limitations of current UVB models 
to provide reliable ionization and thermal histories during reionization
by developing a new method to model 
ionization and heating during reionization in hydrodynamical simulations.
In the context of this method, we demonstrate how to run simulations
with self-consistent ionization and thermal histories that agree with 
constraints from the CMB
and IGM measurements. 
Moreover, we make these new tables publicly available in the default
format used by most cosmological codes.

The outline of the paper is as follows. 
In Section~\ref{sec:typmodels} we discuss in detail current standard methods 
that include the effect of the UVB in optically thin hydrodynamical simulations.
We show that these models have problems reproducing the desired
ionization and thermal histories.
In Section~\ref{sec:newmodel} we present a new method to improve the 
current models of the UVB during reionization events. 
The different
reionization models considered in this work, based on current observational constraints,
are motivated in Section~\ref{sec:reionmodels}.
We describe the basic details of the hydrodynamical cosmological 
code that we have used in this work, the analysis pipeline, and the properties
of the simulations in Section~\ref{sec:code}.
The ionization and thermal histories of the simulations using the 
new UVB models are shown and examined in Section~\ref{sec:results}. 
We explore the possibility of reproducing observational
constraints on the $\HI$ and $\HeII$ transmission in
Section~\ref{sec:meanflux}.
In Section~\ref{sec:discuss} we discuss the limitations of our new approach,
provide a comparison to previous work, and discuss previous work
using incorrect UVB models in galaxy formation simulations that likely overestimate the impact
of photoionization feedback.
We conclude in Section~\ref{sec:conc}. In Appendix~\ref{app:volave} we provide
details on 
how the new photoionization and photoheating rates are derived in our method.
In Appendix~\ref{app:HMold} we present the ionization and thermal histories 
of several widely used UVB models.
The effects of cosmology on the new models are discussed in
Appendix~\ref{app:cosmo}. Finally, in Appendix~\ref{app:tables}
we present the photoionization and photoheating rates of the new
models.

\section{Ionization and Thermal Histories of Common UVB Models} 
\label{sec:typmodels}

Current UVB models used in optically thin hydrodynamical simulations
give the evolution of both photoionization, $\Gamma_{\gamma}$ ($\HI,\HeI,\HeII$), and
photoheating $\dot{q}$ ($\HI,\HeI,\HeII$) rates with redshift.
Therefore, during reionizations the simulated IGM is photoheated everywhere by
the same spectrum using a homogeneous value of $\dot{q}$.
It is important to remark that the photoheating rates are given
as energy per second per ion. Then, for example, in the case of $\HI$ reionization,
the photoheating rate per volume for each resolution element is
$\nHI\dot{q}_{\HI}$.

\citet{Haardt:1996} were the first to try to develop self-consistent UVB
models in a cosmological context using radiative transfer methods
and taking into account observations of the ionizing sources (namely,
quasars and galaxy luminosity functions) and the absorption of the ionizing photons 
(column density distribution of neutral hydrogen, $N_{\HI}$, absorbers, and $\HI$ mean flux).
An ionization front heats up the gas behind it \citep[see, e.g.,][]{Abel:1999,McQuinn:2012,Davies:2016}
and therefore it is crucial to include radiative transfer effects in the UVB modeling. 
Self-consistent methods use photoionization modeling codes that implement
1D radiative transfer (e.g. \textsc{cloudy}) to try to take this effect into account.
Subsequent efforts have developed these
models further \citep{FaucherGiguere:2009,Haardt:2012}.
These UVB models are adopted in essentially all nonadiabatic cosmological hydrodynamic
simulations to compute the ionization state and photoheating rates of intergalactic gas
\citep{Somerville:2015}.
These include simulations focusing on the properties of 
the IGM \citep[e.g.][]{Katz:1996,MiraldaEscude:1996,Lukic:2015},
but also simulations modeling galaxy formation and evolution
\citep[e.g.][]{Vogelsberger:2013,Hopkins:2014,Shen:2014,Governato:2015,Dave:2016}.

We first want to present the ionization and thermal histories obtained
when one of the most widely used UVB models (e.g. tabulated photoionization and photoheating rates) 
is used \citep[][hereafter HM12]{Haardt:2012}.
The upper panel of Figure~\ref{fig:Qhistgas0} shows the $\HII$ (solid black line)
and $\HeIII$
(dot-dashed black line) ionization history calculated by 
the HM12 
model (black lines), 
which indicates that $\HI$ reionization should
finish at $z_{\rm reion,\HI}=6.7$ and $\HeII$ at $z_{\rm reion,\HeII}=2.75$. 
These are given by the volume filling fraction evolution, $Q_{\HII}(z)$
which can be thought of the probability that the hydrogen in a given region is ionized
\citep{Madau:1999a}, and $Q_{\HeIII}(z)$ the analogous quantity for a
doubly ionized helium region. 
In this panel we also show the ionization history of an
hydrodynamical cosmological simulation using the
HM12 UVB model
(green lines). 
This simulation uses
the standard methodology employed in other optically thin simulations
that we describe it in detail in Section~\ref{sec:code}.
To obtain the ionization history from the simulations we computed the ionization fraction
of each volume element in the simulation, $x_{\HII}=\nHII/\nH$ and $x_{\HeIII}=\nHeIII/\nHe$, 
and then calculate the average weighting in volume (i.e. averaging all
of the cells)\footnote{Notice that
calculating the volume filling factor, $Q_{\HII}$, is not the optimal way to describe
reionization in optically thin simulations. One needs to set up an ionization threshold
(standard values are 0.999-0.9) and compute how many cells in the simulation
have an ionization fraction above this level. It is easy to see that in optically
thin simulations the filling factor evolution will be just a step function
that jumps to 1 as soon as the volume-averaged ionization fraction, $\mean{x_{\HII}}$, reaches
the chosen threshold. The temperature evolution shown in Figure~\ref{fig:Qhistgas0} illustrates
why this approach will not be the correct description of how reionization took place in
these simulations.}.
The first striking thing that we learn from this comparison is
that using the HM12 photoionization rates
effectively reionizes the universe much earlier than the
reionization redshift reported by HM12. It appears that some aspect
of the HM12 calculation is not internally consistent.

The middle panel of Figure~\ref{fig:Qhistgas0} shows the 
integrated electron scattering optical depth, defined as
\begin{equation}
\begin{aligned}
 & \taue(z)=c\sigma_T \int_0^{z}\nel{(1+z')^2dz'\over H(z')}\\
 \end{aligned}
 \label{eq:tauCMB}
\end{equation}
where $c$ is the velocity of light, 
$\sigma_T$ is the Thomson cross section, $H(z)$ is the Hubble parameter,
and $\nel$ is the proper electron density.
We have computed the electron
density in our simulation as $\nel=\mean{\nH}(1+\chi)\mean{x_{\HII}} +\chi \mean{x_{\HeIII}}$
where $\chi=Y_{\rm p}/(4X_{\rm p})$ and $X_{\rm p}$ and $Y_{\rm p}$ are the hydrogen and helium mass abundances,
respectively 
(see Appendix~\ref{ssec:volave} for a detailed derivation of this equation). 
We also make the standard assumption that the reionization of $\HeI$ is perfectly coupled with
that of $\HI$.
The observational constraints on $\taue$ 
coming from the CMB 
\citep[$\taue=0.078\pm 0.019$][]{Planck:2015} are indicated
by a gray band\footnote{During the making of this paper, new constraints
on reionization from Planck were published \citep{Planck:2016a}, moving
these constraints to a lower value and reducing the errors: $\taue=0.058\pm 0.012$.
These results do no change any of the conclusions of this paper.}.
The results of the simulation (green) not only differ from the expected results
of the model (black) but also they are in strong disagreement with the observational constraints.

Moreover, the lower panel of Figure~\ref{fig:Qhistgas0} shows the thermal history
of this simulation (green line) via the evolution of the temperature at mean
density, $T_{0}$ defined in eqn.~(\ref{eq:T0gamma}).
See  Section~\ref{ssec:samplesum} for a discussion of the procedure used to fit
the $\rho$-$T$ relation.
We can see that, not only do reionization events occur too
early, but also that the heating associated with them starts at much earlier
times, $z\sim15$.
We have confirmed that this result is not due to resolution,
assumed cosmological model, atomic rates considered,
or some particular code characteristic \citep[see Figures 3 and A1 in][for 
the same effect but using a SPH Lagrangian code, GADGET-3, and using both an equilibrium
and non-equilibrium ionization solver]{Puchwein:2015}.
Thus, why do the reionization histories in the simulation and the one calculated
by these authors differ so much?

UVB semianalytic synthesis models --- such as the one used by HM12 --- 
rely on two main assumptions: (1) the photoionizing background
is everywhere uniform, and (2) radiative transfer is optically thin.
Clearly, both assumptions break down during reionization. While a full solution
requires a radiative transfer simulation, 
a large number of IGM studies are insensitive to the reionization details.
Thus, the goal is to nevertheless have an approximate
UVB background representing the mean UVB to adopt in an optically thin simulation.
Therefore, it is important to state upfront that the rates obtained under
these sets of assumptions 
and the ones obtained in a patchy reionization model (e.g., a
radiation transfer simulation) can differ significantly during
reionization. \citet[][]{Lidz:2007} illustrate this point
in a simple way by considering two toy models.
In the first case, representing patchy reionization, we
imagine equal-sized ionized bubbles each with an interior neutral fraction
$x_{\HI,\rm  IN}<<1$, filling a fraction $Q_{\HII}$ of the volume of the IGM,
which is otherwise completely neutral.
For simplicity, we neglect helium and consider an IGM with a
uniform density and temperature. In the second model, representing uniform ionization
the neutral fraction is identical at each location within the IGM with
$\mean{x_{\HI}}=1-Q_{\HII}$. In each case, photoionization equilibrium tells us that
$\mean{\Gamma_{\HI}}\sim\mean{\alpha\nH}\mean{(1-x_{\HI})^{2}/x_{\HI}}$, where
$\alpha$ here is the recombination factor. 
Now, in a uniformly ionized IGM we get 
$\mean{\Gamma_{\HI}}_{\rm uniform}\sim \mean{\alpha\nH} Q_{\HII}^{2}/(1-Q_{\HII})$.
In the toy patchy model, we have $(1-x_{\HI})^{2}/x_{\HI}\sim 1/x_{\HI, \rm IN}$
inside each ionized bubble and $\sim 0$ outside of ionized regions. Hence,
$\mean{\Gamma_{\HI}}_{\rm patchy}\sim Q_{\HII} \mean{\alpha\nH}/x_{\HI, \rm IN}$.
The ratio of the volume-averaged photoionization rates is just
$\mean{\Gamma_{\HI}}_{\rm patchy}/\mean{\Gamma_{\HI}}_{\rm uniform}\sim (1-Q_{\HII})/Q_{\HII}x_{\HI, \rm IN}$.
This will typically be a very large number: for example, if $50$\% of the volume
is filled by ionized bubbles $Q_{\HII}=0.5$ each with an interior neutral
fraction of $x_{\HI, \rm IN}=10^{-4}$, the volume-averaged photoionization 
rate is a factor of $10^{4}$ times larger in the patchy reionization
model than in the uniform model.

This is also relevant to understanding how the ionization history 
given by the volume filling factor can be compared with the one
given by the volume-averaged ionization fractions and the 
intrinsic differences between the two.
The $Q$ formalism only
knows about sources and sinks of ionizing photons and
does not tell us anything about the value of
the neutral fraction in highly ionized regions.
In what follows we focus on hydrogen reionization,
but analogous considerations also apply to helium.
The volume filling factor evolution in UVB synthesis models
is computed as
\begin{equation}
\frac{dQ_{\HII}}{dt}=\frac{\dot{n}_{\rm photon,\HII}}{\mean{\nH}}-\frac{Q_{\HII}}{\mean{t_{rec,H}}}
\label{eq:Q}
\end{equation}
where 
$t_{\rm rec,\Hy}$ is the hydrogen recombination time
and $\dot{n}_{\rm photon,\HI}$ is the mean number of ionizing photons emitted 
by all radiation sources available per second\footnote{The
recombination time for $\HI$ reionization 
 is generally defined as $t_{\rm rec}=[(1+\chi)\alpha_{B} C_{IGM} \mean{\nH}]^{-1}$ where $\alpha_{B}$
 is the recombination coefficient to the excited states of hydrogen,
 $\chi$ accounts also for the presence of photoelectrons
 from singly ionized helium,
 and $C_{IGM}\equiv\mean{\nHII^2}/\mean{\nHII}^2$ 
 is the clumping factor of ionized hydrogen.
   A practical issue is how $t_{\rm rec}$ should
   be evaluated when $Q < 1$, and in particular
   when $Q \ll 1$.
  We refer to the nice and detailed discussion on this
 issue done by \citep[][]{So:2014}.
 In any case, recombination rates are relatively unimportant at high redshifts,
 and this possibility cannot explain the big discrepancy between
 the model and the simulation.}.
In the context of these models,
$\dot{n}_{\rm photon,i}=\int_{\nu_{L,i}}^{\infty} \frac{\epsilon_{\nu}}{h\nu}d\nu$, is considered 
to be the number of ionizing photons
emitted into the IGM by all radiation sources \citep[HM12;][]{So:2014} 
where $\epsilon_{\nu}$ is the total emissivity as a function of frequency
obtained by the assumption of the sources (i.e. galaxies and quasars
luminosity functions)\footnote{These models also need to assume some galaxy escape
fraction at each redshift that is generally chosen by first iteratively solving the 
integrated cosmological radiative transfer equation}
This model uses observational constraints of 
the distribution of absorbers along the line of sight, and it is able to reproduce
available measurements of the mean free path at $1$ Ryd and the 
\lyalpha{} effective opacity. 

\begin{figure}
\begin{center}
\includegraphics[angle=0,width=0.48\textwidth]{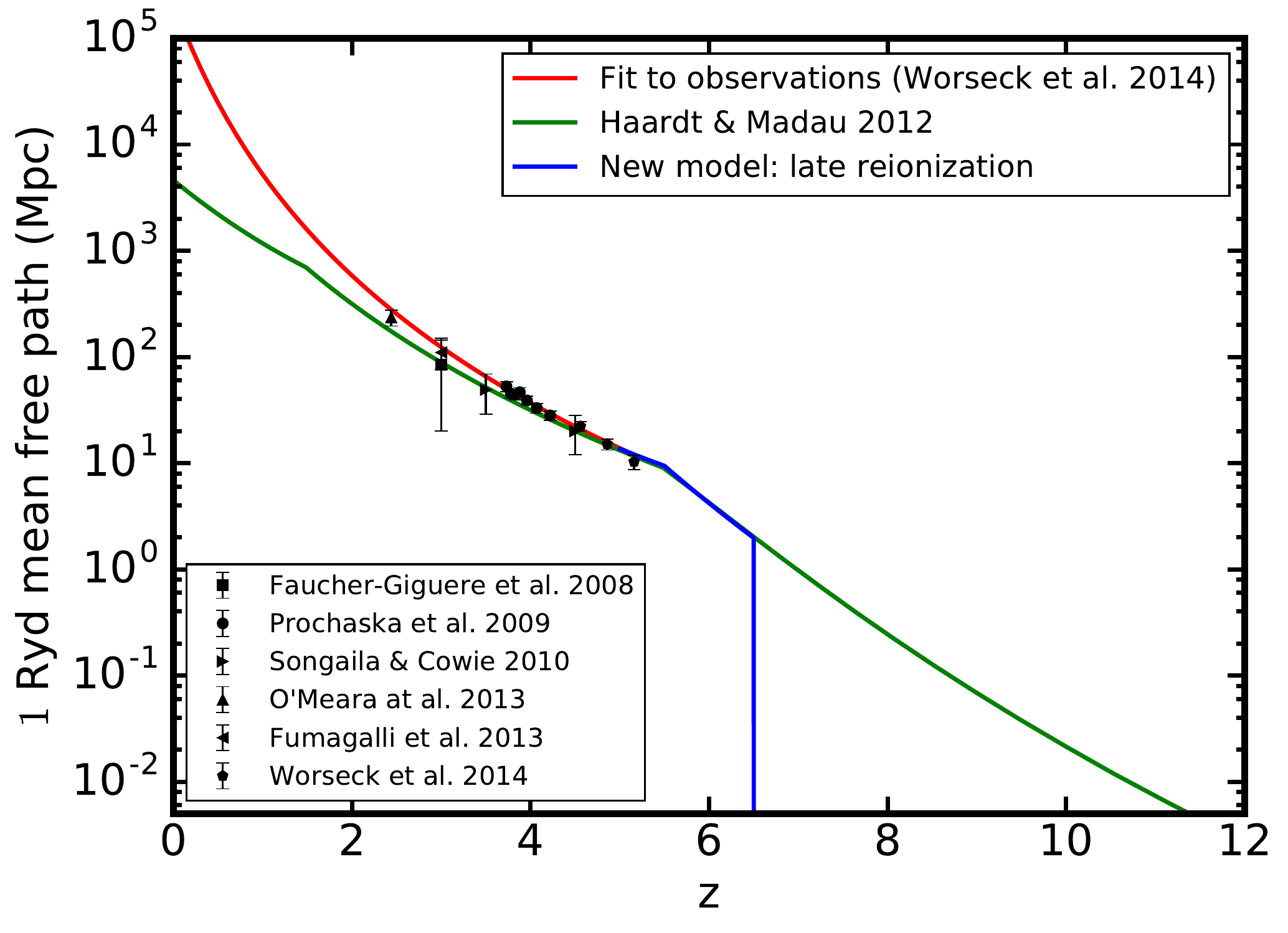}
\end{center}
\caption{
Evolution of the mean free path, $\lambda_{\rm mfp,\nu}$, at 1 Ryd (solid lines)
and 4 Ryd (dot-dashed lines).
Green symbols stand for a compilation of observational
measurements of the mean free path at 1 Ryd
\citep{FaucherGiguere:2008,Prochaska:2009,Songaila:2010,OMeara:2013,Fumagalli:2013,Worseck:2014}.
Red lines: fit from \citep{Worseck:2014} to observations only until $z=5.5$.
Orange lines: tabulated values of the mean free path given by HM12.
Black lines: $\lambda_{\rm mfp,\nu}$ obtained from the tabulated
emissivities $\epsilon_{\nu}$ and intensities $J_{\nu}$ given by HM12 
assuming the source function approximation. 
Blue line: $\lambda_{\rm mfp,\nu}$ obtained from one new reionization model. See text for more details.
\label{fig:mfp}}
\end{figure}

To clarify what is happening at these high redshifts we can use 
the equation of cosmological radiative transfer in its 
``source function'' approximation,
which allows to write the following relation between emissivity,
radiation intensity and mean free path, $\lambda_{\nu}$ by ignoring
photon redshifting effects (minimal at high redshifts):
$4\pi J_{\nu}= \lambda_{\nu}\epsilon_{\nu}$
if only local radiation
sources contribute to the ionizing background intensity 
(HM12). 
From this relation it is now much more easy to see what 
went wrong in the UVB model. 
The $\HI$ photoionization rates, $\Gamma_{\HI}$  given by HM12 
are calculated as
\begin{equation}
\Gamma_{\HI}=\int 4\pi \frac{J_{\nu}}{h\nu} \sigma_{\nu,\HI} d\nu 
\end{equation}
where $J_{\nu}$ is the radiation intensity.
The ionization history, $Q$,
was computed using the emissivity alone, 
using an analytical approximation that does not know anything
about the mean free path assumed in the model.
On the other hand, the photoionization rates depend on both
the emissivity and the mean free path, indicating that the mean free path extrapolation
done at high redshift was wrong and yields values that are systematically too high..
To illustrate this, we show in Figure~\ref{fig:mfp} the mean free path assumed in the model 
(green line) and observational constraints on the mean free path at $1$ Ryd
(black symbols). Notice that at high redshifts $z>5$, there are no
available constraints and the model thus corresponds to a blind extrapolation.

The $\HI$ photoheating rate (energy per time per ion), $\dot{q}_{\HI}$,
given by these models is calculated as
\begin{equation}
\dot{q}_{\HI}=\int 4\pi \frac{J_{\nu}}{\nu} (\nu-\nu_{\HI}) \sigma_{\nu,\HI} d\nu 
\end{equation}
so the photoionization rates are overestimated for the same reason that
the photoheating rates are. 
As noted above,
the total heating rate  
does not depend on the the amplitude of $J_{\nu}$
but only on its shape \citep{Theuns:2002a,McQuinn:2009}.
However, as we argued above, in the standard UVB semianalytic synthesis modeling
approach, $Q$ and $J_{\nu}$ are not required to be internally consistent,
so in practice the actual heating rate per volume approaches a constant
at early times ($z>z_{\rm reion}$), whereas it should go to zero.
Therefore, the total heat produced during $\HI$ and $\HeII$ 
reionization in the simulation was also
applied earlier than when it should be.

We have also tested other widely used UVB models in the literature
\citep[][hereafter HM96, HM01, FG09, respectively]{Haardt:1996,Haardt:2001,FaucherGiguere:2009}.
We present the full results for these other models in Appendix~\ref{app:HMold}
showing that they all share a similar problem,
i.e., for $\HI$ or $\HeII$ (or both)
reionization the gas heating associated with the
reionization event occurs much earlier than one will naively expect
from these models.
Although it was known that these UVB models do not properly model the
reionization process,
it is important that this problem has not been directly confronted 
in the literature \citep[see, however,][]{Puchwein:2015}.
This oversight is likely due to the fact that, for
the ionization and thermal histories of the IGM,
most comparisons between simulations and observations
have been performed in the context of the \lyalpha{} forest and
focused on the redshift range for which such observations are available, i.e. between $2<z<6$,
or even lower redshifts.
The thermal parameters studied so 
far in the literature, $T_{0}$ and $\gamma$, depend on the instantaneous values 
of these rates once we are far enough in time from reionization. So
they quickly forget about early heating
Therefore, what happens at higher redshifts is not that relevant 
for these values.
However, it is expected that this problem will have
more important consequences for correctly modeling the pressure 
smoothing scale, as it depends on the full thermal history.
This is because at IGM densities, the dynamical time that it takes
the gas to respond to temperature changes at the Jeans scale
(i.e., the sound-crossing time) is the Hubble time
\citep{Gnedin:1998}.

To summarize, current UVB models result in different
ionization and thermal histories than those quoted
by the authors.
This is because the low-$z$ mean free path values have been
blindly extrapolated to high-$z$ (see Figure~\ref{fig:mfp}),
where they result in a photoionization rate far too high.
Motivated by these results, we have developed a new method to 
build self-consistent effective ionization and thermal histories
during reionization epochs in optically thin simulations,
which we describe in the next section.

\section{Improved UVB Models} 
\label{sec:newmodel}

As stated in the previous section, current photoionization rates widely used
in optically thin codes do not properly track the desired ionization
histories and lead to incorrect thermal histories for the gas in these simulations, where 
the gas is heated much earlier than it should be. We present here a new way 
of creating self-consistent UVB models 
during reionization events ($\HI$, $\HeI$, and $\HeII$) to be used
in optically thin hydrodynamical simulations.
Different groups in the field have modified
these tables, both ionization and photoheating rates
(especially photoheating,
and more in the context of $\HeII$ reionization where 
more observations are available), 
with different justifications: 
accounting for different physical effects as nonequilibrium ionization or radiative transfer effects,
matching specific observables, or just exploring the parameter space
\citep[e.g.,][]{Haehnelt:1998,Theuns:2002,Bolton:2005,Jena:2005,Wiersma:2009,Pawlik:2009,Puchwein:2015,Lukic:2015}.
However, this has generally been done by just applying
a multiplying factor to the standard models assumed
or by applying different simple cutoffs\footnote{Results 
of simulations using the cutoff approach can be found
in Appendix~\ref{app:HMold}.}.

Our approach will be based on building effective values of the
photoionization and photoheating rates that can be substituted into the
standard
optically thin equations to yield the desired results 
(see FG09 for an early motivation of this approach). 
The main goal is to make sure that the heating
due to reionization in the simulation is consistent 
with the reionization model itself. To enforce this, we have calculated
the volume-averaged values of both the photoionization and photoheating
rates that give us the desired ionization and total heat injection
for an input reionization model.
We give here a global overview of the method and the different 
assumptions made in our models. We explain in full
detail how we derive the photoionization rates in Appendix~\ref{ssec:volave}
and the photoheating rates in Appendix~\ref{ssec:dTdz}. 
We will discuss the different
caveats and limitations of our model in Section~\ref{sec:discuss}.

Each of our reionization models is defined by one free parameter, the
total heat input $\Delta T_{\HI}$ during reionization, and one free
function, the reionization history, which in this context we
define as the volumen-averaged ionization fraction evolution,
$\mean{x_{\HII}}(z)$. 
 The reionization
of $\HeII$ is analogously treated.  Using these parameters, we will
derive effective photoionization and photoheating rates that can be
used in hydrodynamical simulations using the following assumptions:
\begin{enumerate}
\item That all species are in ionization equilibrium at all times. This is done to be fully
  consistent with (most of) the codes that will be using these UVB models but it could be changed in the future.
\item That the gas composition can be approximated as primordial. Therefore, the evolution
of the number density of electrons, $\nel$, is given as $\nel=\nHII+\nHeII+2\nHeIII$.
\item That $\HeI$ reionization is perfectly coupled with $\HI$ reionization.
\item That $\HeII$ reionization is not relevant during $\HI$ reionization and vice versa.
\item That the heating due to reionization is perfectly coupled to the reionization process.
  Therefore, the heating can be written as a function of the total heat injection
  and the ionization history, i.e.,
\begin{equation}
  \frac{dT_{\HI}}{dt} \propto \Delta T_{\HI} \frac{d\mean{x_{\HII}}}{dt}
\end{equation}
\end{enumerate}

Finally, the new effective rates are only used during reionization, i.e., while $\mean{x_{\HII}}<1.0$.
Once the reionization redshift, defined as when the input ionization history is one, $\mean{x_{\HII}}=1$, 
is reached, we can simply use the photoionization and photoheating
rates of common UVB models (in our case HM12). 

Let us first focus on how we obtain the new photoionization rates to be applied
during reionization.
We obtain the new effective photoionization rates 
by volume averaging the ionization equilibrium equations
and using the assumptions enumerated above. In particular, for the $\HI$
photoionization rates we get:
\begin{equation}
\mean{\Gamma_{\gamma, \HI}}(z)= C_{\HII} \mean{\nH}(z) 
\alpha_{\rm r, \HII}(\mean{T})(1+\chi) \frac{\mean{x_{\HII}}^{2}(z)}{\mean{x_{\HI}}(z)} \\[1.5mm]
\label{eq:invphoto}
\end{equation}
where $C_{\HII}$ is a volumen-averaged correction factor
linked with the well known clumping
factor and $\alpha_{\rm r, \HII}(\mean{T})$ is 
the recombination coefficient for which a specific volumen-averaged temperature of the IGM, $\mean{T}$, has to be assumed.
We refer the reader to Appendix~\ref{ssec:volave} for all the 
details\footnote{Based on the same idea used to derive the photoionization rates,
in Appendix~\ref{app:hifrac}, we introduce a numerical method to compute
the expected volumen-averaged ionization history outcome in optically thin
hydrodynamical simulations from a specific mean photoionization rate, $\Gamma_{\gamma,\HI}(z)$.
We recommend this approach to be used in the future when generating
different UVB tabulated models.}.

In addition to affecting abundances, photoionization injects energy into the gas
when a high-energy ($h\nu> h\nu_{\rm T}$) photon transfers more
energy to an electron than what is necessary to unbind it from the atom. 
In order to include this heat transfer to the IGM
by the UVB, models also provide photoheating rates 
that are included in simulations as an effective energy source.
The temperature of the IGM is much more affected by this photoheating and 
different atomic cooling processes than by the gravitational collapse 
of cosmic structure. These processes are accounted for as a global heating and cooling term that
is added to the equation of energy for the gas:
\begin{equation}
\begin{aligned}
  & \frac{\partial E}{\partial t} =
  - \frac{1}{a}\vec{v}\cdot \nabla E
  -\frac{\dot{a}}{a}(3\frac{p}{\rho}+\vec{v}^{2})\\
  & \qquad -\frac{1}{\rho a}\nabla\cdot (p\vec{v})
  +\frac{1}{a}\vec{v}\cdot\nabla\Phi
  +\frac{\Lambda_{\rm HC}}{a \rho}\\  
\end{aligned}
\label{eq:energy}
\end{equation}
where  
$\Lambda_{\rm HC}$ represents the combined heating and cooling terms,\footnote{We have assumed comoving coordinates:
$\vec{r}_{\rm proper}=a\vec{x}$, 
$\rho$ is the comoving baryon density ($\rho=a^{3}\rho_{\rm proper}$),
$p$ comoving pressure ($p=a^{3}p_{\rm proper}$),
$\vec{v}$ is the proper peculiar baryonic velocity ($\vec{v}=\vec{v}_{\rm proper}-\dot{a}\vec{x}$), 
$\Phi$ is the modified gravitational potential, and
$E$ is the total comoving energy ($E=E_{\rm proper}-a\vec{x}\cdot\vec{v}-\frac{1}{2}\dot{a}^{2}\vec{x}^{2}$). Refer
to \citet[e.g.][]{Almgren:2013} for details.}
which can be expanded as
\begin{equation}
\begin{aligned}
  & \Lambda_{\rm HC}= a^{4} \nHI\dot{q}_{\HI}+ a^{4} \nHeI\dot{q}_{\HeI} + a^{4} \nHeII\dot{q}_{\HeII}\\
  & -a^{4}\nel f_{\rm c}(\nHI,\nHII,\nHeI,...)-a^{4}\nel f_{\rm Compton}(T,T_{\rm CMB})+...\\  
\end{aligned}
\label{eq:interenergy}
\end{equation}
where the first three terms represent the photoheating rates
of $\HI$, $\HeI$ and $\HeII$ respectively\footnote{Photoheating
terms in eqn.~(\ref{eq:interenergy}) are written as they are usually defined
in hydrodynamical simulations, a photoheating rate, $\dot{q}_{\HI}$ (in units of energy per
time per ion) multiplied by the ion density, $\nHI$, of 
that resolution element. Therefore, the global heating due to reionization will be determined
also by the amount of ions present, and we cannot expect
a constant temperature increase, independent of density. We will discuss
this in detail in Section~\ref{sec:discuss}.}.
We have also included an atomic cooling term ($f_{\rm c}$)
and an inverse-Compton scattering term ($f_{\rm Compton}$).

We can obtain effective photoheating rates for the reionization heating
by volume-averaging the 
relation between the heat per unit of time produced by 
a certain reionization model, $d\Delta T/dt$, and the one produced by a photoheating rate.
This will allow the use of our new models in standard hydrodynamical codes. 
For the $\HI$ photoheating rate, and using the set of assumptions enumerated above, we obtain
\begin{equation}
\dot{q}_{\HI}=C_{\dot{q},\HI}\frac{ 3 k_{B}}{2 \mean{\mu} X_{\rm p} \mean{x_{\HI}}(z)} \frac{d\Delta T_{\HI}}{dt} 
\label{eq:qdot}
\end{equation}
where $X_{\rm p}$ is the hydrogen mass abundance, $\mean{\mu}$ is the volumen-averaged molecular weight,
$\mean{x_{\HI}}(z)$ the assumed reionization history
of the model, $\Delta T_{\HI}$ is its total heat input, and 
$C_{\dot{q},\HI}$ is a volumen-averaged correction factor that
we set to one at all redshifts. We refer the reader to 
Appendix~\ref{ssec:dTdz} for more details on how this equation was obtained.

\begin{figure*}
\begin{center}
\includegraphics[angle=0,width=0.39\textwidth]{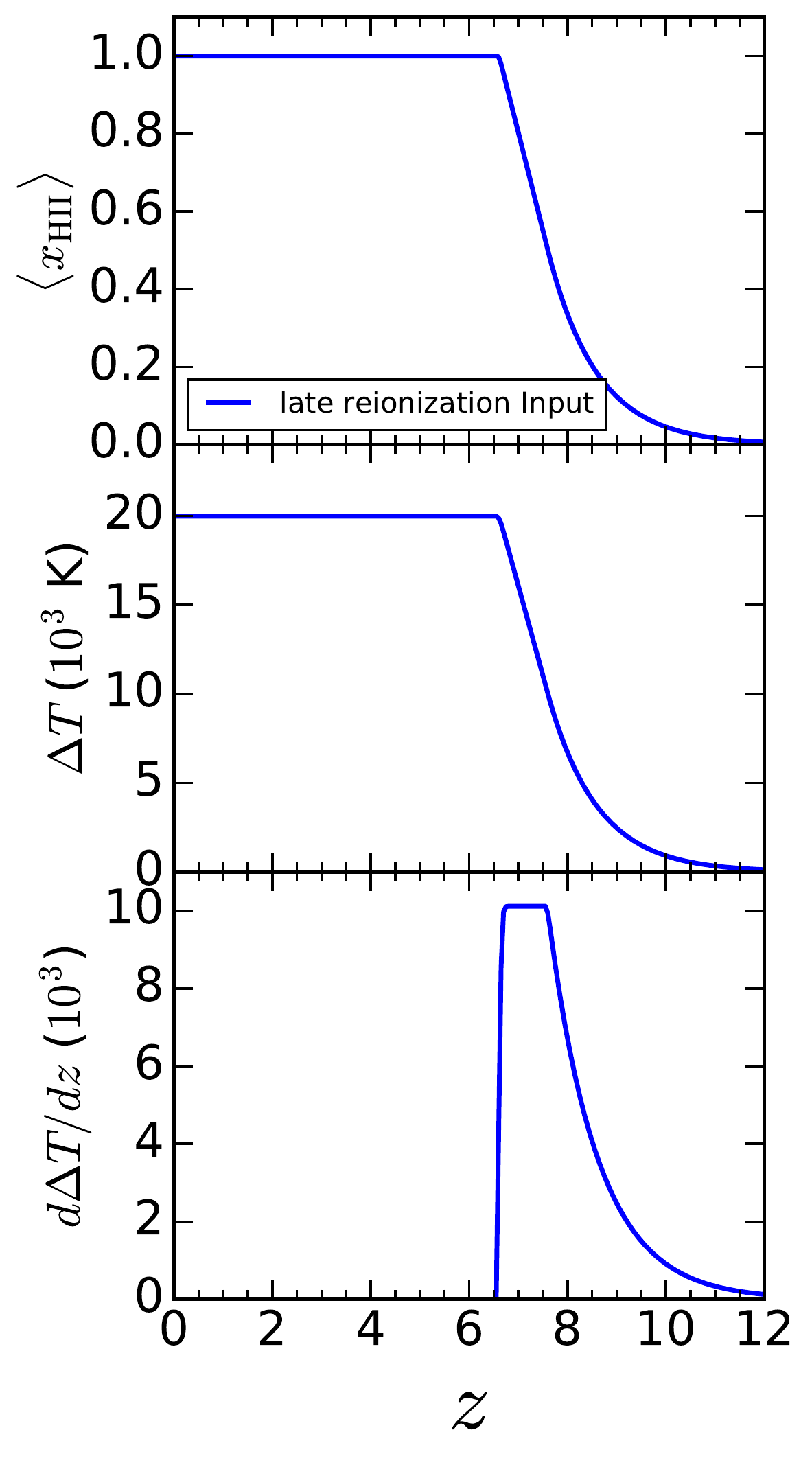}
\includegraphics[angle=0,width=0.6\textwidth]{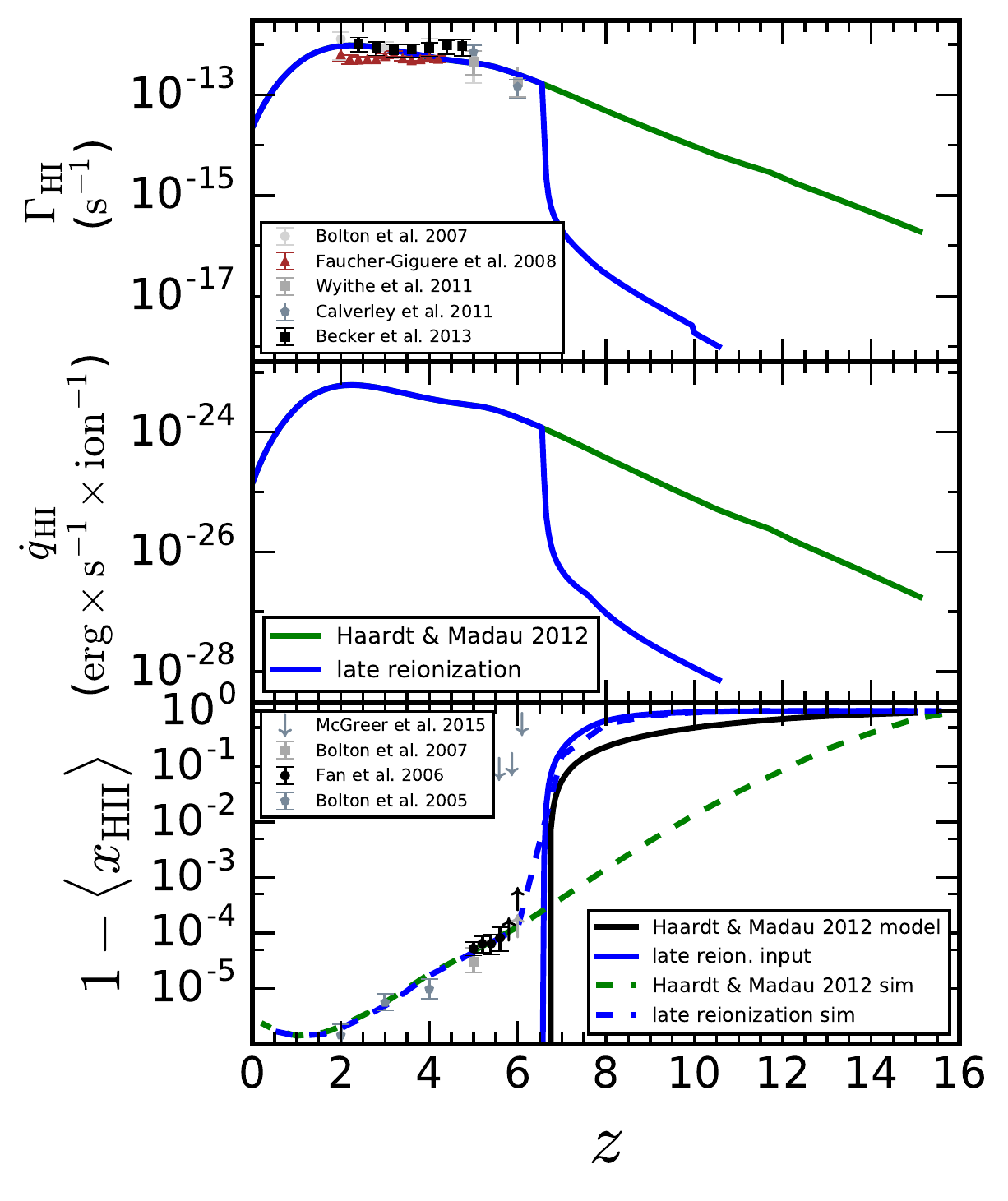}
\caption{New UVB modeling. 
Left panels: show the input parameters in a model that
mimics a late reionization (see upper panel)
and a total heat input of $\Delta T_{\HI}=2E4$ K. The
middle left panel shows the 
thermal history $\Delta T(z)$ and the lower left panel the
heat input $dT/dz$ assumed. 
Right panels: show the new photoionization (upper right panel)
and photoheating values (middle right panel)
that need to be used to run the simulation (blue lines). 
Green lines stand for the standard HM12 rates 
Lower right panel: evolution of the neutral gas fraction
expected from HM12 (black solid line) and the result obtained
using their rates (dashed green lines). We also show the evolution
of the neutral gas fraction that we wanted to impose in our model
during reionization (full blue line) and the result obtained from 
a simulation using the new rates (dashed blue line).
Notice that our proposed model only changes photoionization
and photoheating values above the reionization
redshift as we consider that below this redshift, current models are 
consistent. \label{fig:dTdz}
}
\end{center}
\end{figure*}
  
Figure~\ref{fig:dTdz} shows an example of how a UVB model is built.
In the left panel we show a model that should match 
a specific late $\HI$ reionization history that assumes a reionization
redshift of $z_{\rm rei}=6.55$
(late reionization model; see more details in Section~\ref{sec:reionmodels})
with a total heat input due to $\HI$ reionization of $\Delta
T_{\HI}=2\times10^{4}$ K.
The upper left panel shows the $\HI$
ionization history assumed to build the model.  The middle panel shows
the assumed temperature evolution of the $\HI$ reionization event.
As explained above, we have assumed that the
temperature evolution follows the ionization history. The lower left
panel shows the actual instantaneous heat input, $d \Delta
T_{\HI}/dz$, derived using eqn.~(\ref{eq:dTdz}) for this model. The
integral of this line gives the total heat input, $\Delta T_{\HI}$.

The upper right panel gives the $\HI$ photoionization rate (blue line) obtained
using eqn.~(\ref{eq:invphoto}) and the method described above.
This rate is compared with the equivalent of the HM12 
model (green line).
We also plot observational constraints on the photoionization rates from 
different studies \citep{Bolton:2007,FaucherGiguere:2008,Wyithe:2011,Calverley:2011,Becker:2013}.
Notice that all these constraints are below $z=6$.
The middle right panel shows the photoheating
rates derived using eqn.~(\ref{eq:qdot}) for the $\HI$.
Notice that for the new model,
below its reionization redshift ($z_{\rm reion,\HI}=6.55$)
we apply the same photoionization and photoheating rates 
used in HM12 
as we think that they are reasonable after reionization.
The rates of our model raise abruptly at around
the reionization redshift because in the input $\HI$ reionization model
the transition from $\mean{x_{\HI}}\sim0.1$ to $\mean{x_{\HI}}\sim10^{-4}$
is very fast.
The lower right panel shows the reionization function input
for the model (solid blue line; defined to be $1.0$ at $z=z_{\rm reion,\HI}$).
It also shows the volumen-averaged $\HI$ ionization fraction
obtained running a cosmological hydrodynamical simulation
that uses this new photoionization rate
(dashed blue line).
The solid black line shows the reionization evolution of the
HM12 
model as calculated by their authors, while the dashed green line stands for 
the outcome of a hydrodynamical simulation using
the HM12 photoionization rate.
Current best observational constraints on the $\HI$ fraction at different
redshifts are also plotted using different symbols and errorbars 
\citep{Bolton:2005,Fan:2006,Bolton:2007,McGreer:2015}.
Notice again that all available constraints are below $z=6$ so that
both models are in agreement with these observations.

We can derive the evolution of the mean free path at 1 Rydberg for the new model,
making some simple assumptions, and compare it with the ones
obtained by HM12. Our modeling has no shape information on the intensity, $J_{\nu}$;
however, one can assume the HM12 shape and that all intensities differ by the same constant:
$J_{\rm \nu,new}=C\times J_{\rm \nu,HM12}$.
With this we can approximate the new radiation intensity
at 1 Ryd as the ratio between the two photoionization rates 
$J_{\rm 912,new}=(\Gamma_{\rm \HI,new}/\Gamma_{\rm \HI,HM12})\times J_{\rm 912,HM12}$.
Then we obtain the mean free path using the source function approximation
that has the same emissivity assumed by HM12:
$\lambda_{\rm mfp,912,new}=4\pi J_{\rm 912,new}/\epsilon_{\rm 912,HM12}$.
Figure~\ref{fig:mfp} shows the evolution of the mean free path derived 
for this new model (blue line). The green solid line stands for
the mean free path used in the HM12 model.
This plot illustrates how the volumen-averaged mean free path 
drops significantly above the reionization redshift.

\section{Reionization Models} \label{sec:reionmodels}

We will now discuss the different reionization models that we
want to simulate by using our new approach to compute photoionization
and photoheating during reionization.
In order to define a reionization model, we need to set
the $\HI$ and $\HeII$ reionization history
via the volumen-averaged ionization fractions\footnote{As 
eqn.~(\ref{eq:invphoto}) above shows,
we also need to define the total heat input
expected from each reionization process because there is also
a weak dependence on temperature due to the 
recombination factor.}.

First, we need to define the shape of our reionization histories.
We use the lower incomplete gamma function, $g$:
\begin{equation}
 \mean{x_{\HII}}=\begin{cases}
   0.5+0.5\times g(1/n_{1},|z-z_0|^{n_{1}}), & z <= z_0 \\
   0.5-0.5\times g(1/n_{2},|z-z_0|^{n_{2}}), & z > z_0 \\
 \end{cases}
 \label{eq:Qana}
\end{equation}
where $n_{1}=50$, $n_{2}=1$ 
and $z_0$ is a free parameter that sets the redshift where $x_{\HII}(z_0)=0.5$.
This function defines a slower start for the reionization function but a fast finish. 
This specific shape was motivated by radiative transfer simulation results 
\citep[][]{Ahn:2012,Park:2013,Pawlik:2015} which seem to favor 
rapid and sudden $\HI$ reionization histories in $\sim 20^3$ (Mpc/h)$^3$
cosmological volumes. In particular, $n_{1}$
and $n_{2}$ were fitted to mimic \citet{Pawlik:2015} results.
In this work we use only this shape and experiment with the redshift
of reionization, but our function could be easily modified in the future 
to explore a wider range of models.
We explore the relevant range of reionization parameters 
taking into account the CMB constraints on the integrated electron
scattering optical depth, $\taue=0.078\pm 0.019$ \citep{Planck:2015}\footnote{In
this work we used $\taue$ less constraining results based just
on temperature and polarization Planck data.
Using more data reduces slightly the best value, but it is still in agreement with this best value.
During the making of this paper new constraints
on reionization from Planck were published \citep{Planck:2016a}, moving
this constraints to a lower value and reducing the errors: $\taue=0.058\pm 0.012$.
These results do no change any of the conclusions of this paper and in fact
emphasize the disagreement between standard UVB models and these observational
constraints.}.
We consider three models:
an early, middle, and late $\HI$ reionization history, which have reionization
redshifts\footnote{We define the reionization redshift of the models at the redshift when
$\mean{x_{\HII}}=1$} of $z_{\rm reion,\HI}=9.70$, $8.30$, and $6.55$, respectively, 
all of which give $\taue$ are within $1\sigma$ of the CMB measurements.
These $\HI$ reionization models are plotted as solid lines in the upper panel of Figure~\ref{fig:Qsims}
and their associated $\taue$ values are shown in the lower panel.

For $\HeII$ reionization we consider a history
based on FG09 (last column of their Table $2$)
results which sets the HeII reionization
redshift to finish at $z_{\rm reion,\HeII}=3.0$ (hereafter He\_A).
The specific evolution of the full ionized fraction can be well described
by $\mean{x_{\HeIII}}=1.0-arctan(z-z_{\rm reion})$.
This is in agreement with standard models of $\HeII$ reionization, 
which, as a result of the high energy requirement to double ionize helium 
($E_{\nu}>54.4$ eV),
assume that quasars must be the main drivers of this process
\citep[e.g., ][]{MiraldaEscude:2000,Compostella:2014,Worseck:2015}. 
This $\HeII$ reionization
model is plotted as a dot-dashed line in the upper panel of Figure~\ref{fig:Qsims}.

\subsection{Total Heat Input due to $\HI$ and $\HeII$ Reionization}

The other two parameters that define the reionization events in our
models are the total heat input that happens during $\HI$ and $\HeII$
reionizations.  
There have been several efforts to calculate the specific cumulative
energy increase or total heat input
($\Delta T_{\HI}$,$\Delta T_{\HeII}$) due to reionization events.
Approximate analytical estimates of this energy can be made
given some specific assumptions for the intrinsic 
spectral slopes of the sources responsible for 
reionization, and in general neglecting redshifting effects
\citep{Efstathiou:1992,MiraldaEscude:1994}.
One-dimensional radiative transfer codes have also been used to obtain
a better understanding of the 
radiative  transfer effects on the energy input into the IGM due to the reionization
process \citep[e.g. HM96;][]{Abel:1999}. 
In recent years better radiative transfer codes have been used to 
study this problem,
and current simple analytic models are motivated by 
by these radiative transfer calculations 
\citep{Tittley:2007,McQuinn:2009,McQuinn:2012}.
Most efforts have focused on $\HeII$ reionization 
\citep{Bolton:2004,McQuinn:2009}, 
but the
same set of assumptions have also been applied to $\HI$ reionization
\citep{Bolton:2009}.
The range of values discussed in these works for $\HI$ reionization
center around $\sim2\times10^{4}$ K with up to a factor of two or three 
difference depending on the exact assumptions. For $\HeII$ reionization 
typical values have been around $1.5\times 10^{4}$ K with a similar
range of uncertainty.
In the context of their calculation of self-consistent UVB synthesis models,
FG09 
computed the total
heat input of their model ($\Delta T_{\HeII}=14269$ K)
and its evolution (i.e. $dT/dz$), finding good agreement with
the  heat  input  determined  from  detailed  radiative
transfer simulations \citep{McQuinn:2009}.

We have treated the total heat input of $\HI$ and $\HeII$ reionization as free parameters
in our models and have chosen values based on the aforementioned literature.
In particular, for our default models we will use the standard values
assumed for both $\HI$ and $\HeII$ reionization:  $\Delta T_{\HI}=2\times 10^{4}$ K and 
$\Delta T_{\HeII}=1.5\times 10^{4}$ K.
We also consider other models with a range of heat input
for both $\HI$ reionization ($\Delta T_{\HI}=1.5\times 10^{4}-4\times 10^{4}$ K)
and $\HeII$ reionization ($\Delta T_{\HI}=1\times 10^{4}-3\times 10^{4}$ K), respectively.

\section{Simulations} \label{sec:code}

\begin{figure}
\includegraphics[angle=0,width=0.45\textwidth]{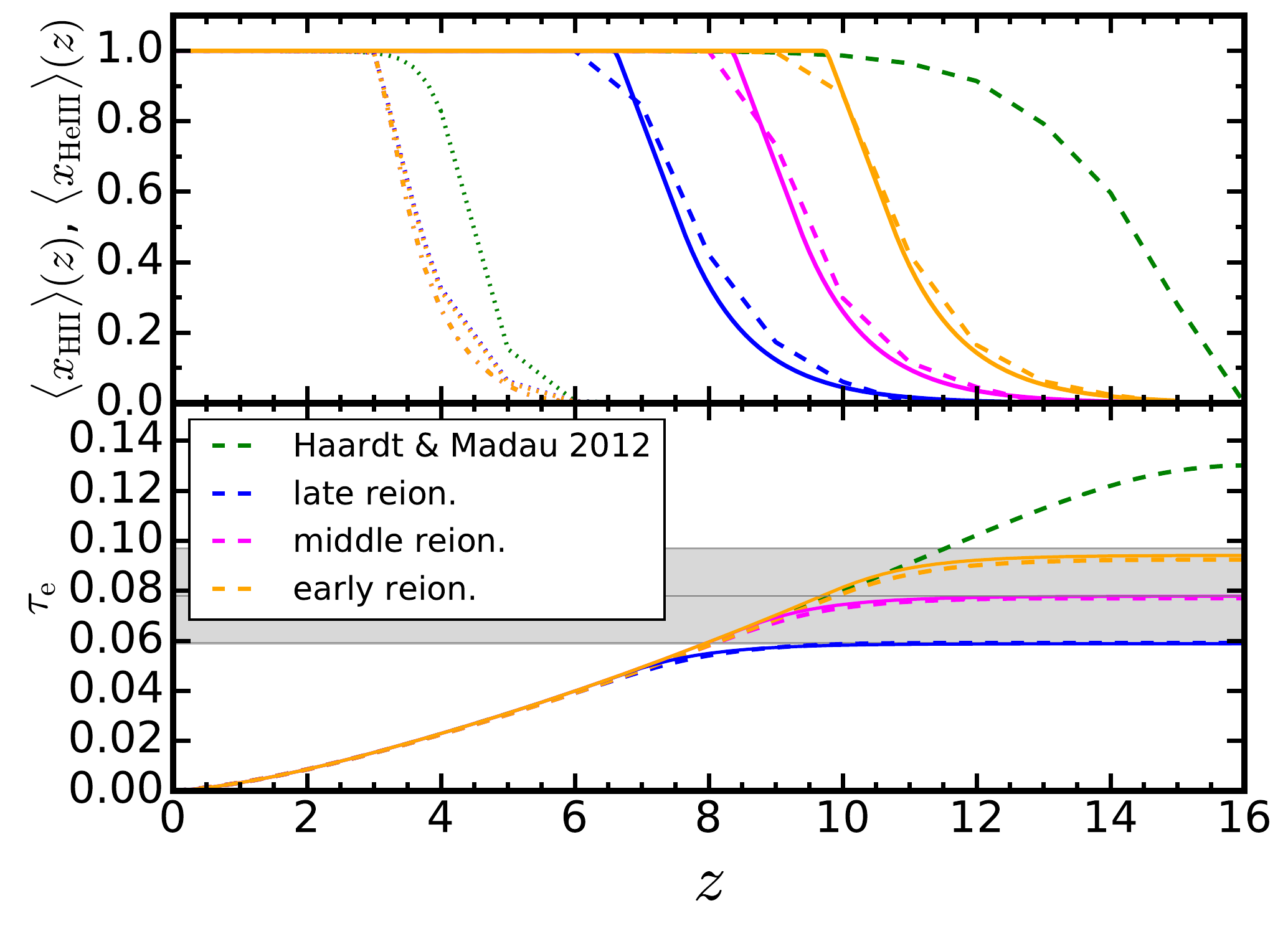}
\caption{Reionization history obtained in simulations 
using different UVB models compared with the input models.
Upper panel: the $\HII$ and $\HeIII$ volume
filling factor evolution calculated in the simulations (dashed and dotted
lines, respectively) compared with the input models (full and
dot-dashed lines respectively). 
Lower panel: integrated electron scattering optical depth, $\taue$
computed from the above volume filling factors. Dashed lines show the results
from the simulations, and solid lines show those for the input models.
Results from a simulation using the HM12 model are also shown 
for direct comparison (green lines, see also Figure~\ref{fig:Qhistgas0}).
The gray band stands for the last constraints on $\taue$ coming from \citet{Planck:2015} data.
\label{fig:Qsims}}
\end{figure}

The simulations used in this work were performed with the Nyx code
\citep{Almgren:2013}. Nyx follows  the  evolution  of  dark  matter
simulated as self-gravitating Lagrangian particles, and baryons
modeled as an ideal gas on a uniform Cartesian grid.
The Eulerian gas dynamics
equations are solved using a second-order-accurate piecewise
parabolic method (PPM) to accurately capture shock waves.
We  do  not  make  use of 
adaptive mesh refinement (AMR) capabilities of Nyx
in the current work, as the 
\lyalpha{} forest signal spans nearly
the entire simulation domain rather than isolated concentrations of
matter,  where  AMR  is  more  effective.
For more details of these numerical methods and scaling behavior tests,
see \citet{Almgren:2013}.

Besides solving for gravity and the Euler equations, we also include
the main physical processes fundamental to modeling the \lyalpha{}
forest. First, we consider the chemistry of the gas as having a primordial composition
with hydrogen and helium mass abundances of $X_{\rm p}$, and $Y_{\rm p}$, respectively.
In addition, we include inverse-Compton cooling off the microwave background
and keep track of the net loss of thermal energy
resulting from atomic collisional processes. 
We used the updated recombination, 
collision ionization, dielectric recombination rates, and cooling rates
given in \citet{Lukic:2015}.
All cells are assumed to be optically thin, and radiative feedback is
accounted for via a spatially uniform but time-varying UVB
radiation field given to the code as a list of photoionization and 
photoheating rates that vary with redshift following the method
described in Section~\ref{sec:newmodel}.

\begin{table*}
\begin{center}
\caption{Summary of Simulations. \label{tab:sims}}
\begin{tabular}{lccccc}
\tableline\tableline
 Sim & $\mean{x_{\HII}}(z)$ & $z_{\rm reion,\HI}$ & $\Delta T_{\HI}$ & $\mean{x_{\HeIII}}(z)$ & $\Delta T_{\HeII}$\\
     & & & (K) & & (K)\\
\tableline
HM12  & HM12\tablenotemark{a} & ... & ... & ... & ... \\
FG09  & FG09\tablenotemark{b} & ... & ... & ... & ... \\
HM01  & HM01\tablenotemark{c} & ... & ... & ... & ... \\
HM96  & HM96\tablenotemark{d} & ... & ... & ... & ... \\
LateR          & Late reionization    & $6.55$ & $2\times10^{4}$  & He\_A & $1.5\times10^{4}$ \\
MiddleR        & Middle reionization  & $8.30$ & $2\times10^{4}$  & He\_A & $1.5\times10^{4}$ \\
EarlyR         & Early reionization   & $9.70$ & $2\times10^{4}$  & He\_A & $1.5\times10^{4}$ \\
MiddleR-Hcold  & Middle reionization  & $8.30$ & $1.5\times10^{4}$& He\_A & $1.5\times10^{4}$ \\
MiddleR-Hwarm  & Middle reionization  & $8.30$ & $3\times10^{4}$  & He\_A & $1.5\times10^{4}$ \\
MiddleR-Hhot   & Middle reionization  & $8.30$ & $4\times10^{4}$  & He\_A & $1.5\times10^{4}$ \\
MiddleR-noHe   & Middle reionization  & $8.30$ & $2\times10^{4}$  & None  & ... \\
MiddleR-Hecold & Middle reionization  & $8.30$ & $2\times10^{4}$  & He\_A & $1\times10^{4}$ \\
MiddleR-Hewarm & Middle reionization  & $8.30$ & $2\times10^{4}$  & He\_A & $2\times10^{4}$ \\
MiddleR-Hehot  & Middle reionization  & $8.30$ & $2\times10^{4}$  & He\_A & $3\times10^{4}$ \\
\tableline
\end{tabular}
\tablecomments{
Column 1: simulation code.
Column 2: $\HI$ ionization history assumed for the model. See Section~\ref{sec:reionmodels} for details.
Column 3: $\HI$ reionization redshift.
Column 4: total heating assumed for $\HI$ reionization.
Column 5: $\HeIII$ ionization history assumed for the model. All models assume a $z_{\rm reion,\HeII}=3.0$.
See Section~\ref{sec:reionmodels} for details.
Column 6: total heating assumed for $\HeII$ reionization.
Unless otherwise stated, all simulations
have a box size of length $L_{\rm box}=20$ ${\rm Mpc}/h$ and $1024^{3}$ resolution elements.
}
\tablenotetext{1}{Using \citet{Haardt:2012} tabulated values.}
\tablenotetext{2}{Using \citet{FaucherGiguere:2009} tabulated values (2011 December update).}
\tablenotetext{3}{Using \citet{Haardt:2001} tabulated values.}
\tablenotetext{4}{Using \citet{Haardt:1996} tabulated values.}
\end{center}
\end{table*}

In order to generate the initial conditions, we have used the
\textsc{music} code \citep{Hahn:2011} and a \textsc{camb}
\citep{Lewis:2000,Howlett:2012} transfer function. All simulations started at
$z_{\rm ini}=159$
to be sure that 
 nonlinear evolution is not compromised \citep[see, e.g.,][for a detailed discussion on 
this issue]{Onorbe:2014}. 
Unless otherwise stated, all the simulations discussed in this paper assumed a
$\Lambda$CDM cosmology with the following fundamental parameters:
$\Omega_{\rm m}=0.3192$, $\Omega_{\Lambda}=0.6808$, $\Omega_{\rm b}=0.04964$,
$h=0.67038$, $\sigma_{8}=0.826$ and $n_{\rm s}=0.9655$. These values are
within $1\sigma$ agreement with last cosmological constraints from the
CMB \citep{Planck:2015}.  The choice of hydrogen and helium mass
abundances ($X_{\rm p}=0.76$ and $Y_{\rm p}=0.24$, therefore
$\chi=0.0789$) is
in agreement with the recent CMB observations and Big Bang
nucleosynthesis \citep{Coc:2013}. Simulations were run down to $z=0.2$,
saving 32 snapshots\footnote{For all simulations we saved an
snapshot at the following redshifts: $20$, $19$, $18$, $17$, $16$, $15$,
$14$, $13$, $12$, $11$, $10$, $9$, $8$, $7$, $6$, $5$, $4$, $3$, $3.8$,
$3.6$, $3.4$, $3.2$, $3.0$, $2.8$, $2.6$, $2.4$, $2.2$, $2.0$, $1.8$, $1.6$,
$1.0$, $0.5$ and $0.2$.}
from $z=20$.
Unless otherwise stated, all simulations presented here have a box size of length
$L_{\rm box}=20$ ${\rm Mpc}/h$ and $1024^{3}$ resolution elements. 
This dynamical range guarantees 
that all the different physical parameters analyzed in this paper 
are converged with enough accuracy ($<5\%$ errors).
We will discuss this issue in more detail
in Section~\ref{ssec:convergence}.

We first run one simulation using photoionization and photoheating
values from the most widely used models
(HM96, HM01, FG09 and HM12). 
We have already presented the reionization and thermal histories of the HM12
model in Section~\ref{sec:typmodels} but below we will further explore
other properties of this simulation.\footnote{Results of the simulations
using HM96, HM01 and FG09 models can be found in Appendix~\ref{app:HMold}}
We also run the three $\HI$ reionization histories presented above,
an early, middle, and late reionization model 
(EarlyR, MiddleR and LateR; see above and Figure~\ref{fig:Qsims}).
All of them share the same 
heat input during $\HI$ reionization,
$\Delta T_{HI}=2\times10^4$ K,
and $\HeIII$ reionization model: $\mean{x_{\HeIII}}(z)$ and 
$\Delta T_{HeIII}=1.5\times10^{4}$ K.

In order to study the effect of different total heat input
during $\HI$ reionization, $\Delta T_{\HI}$, we run three 
more simulations that share 
all parameters with the $\HI$ middle reionization run, MiddleR,
but varying this parameter:
MiddleR-Hcold ($\Delta T_{\HI}=1.5\times10^{4}$ K), MiddleR-Hwarm
($\Delta T_{\HI}=3\times10^{4}$ K) and MiddleR-Hhot ($\Delta T_{\HI}=4\times10^{4}$ K). 
We also explored the effects of different global net heating during
$\HeII$ reionization by running
a set of four more $\HI$ middle reionization simulations (MiddleR, $\Delta T_{\HI}=2\times10^{4}$) in which
we just changed
the heat input during $\HeII$ reionization, $\Delta T_{\HeII}$:
MiddleR-noHe (no $\HeII$ reionization), MiddleR-Hecold ($\Delta T_{\HeII}=1\times10^{4}$ K),
MiddleR-Hewarm ($\Delta T_{\HeII}=2\times10^{4}$ K) and MiddleR-Hehot ($\Delta T_{\HeII}=3\times10^{4}$ K).
A summary of all the relevant parameters used in the runs
presented in this work is shown in Table~\ref{tab:sims} along with
the naming conventions we have adopted.

\subsection{Analysis of the Simulations} \label{ssec:samplesum}

Whenever \lyalpha{} forest spectra are created from the simulation,
we compute the $\HI$ optical depth at a fixed redshift,
which can then be
easily converted into a transmitted flux fraction,
$F=e^{-\tau}$. That is, we do not account for
the speed of light
when we cast rays in the simulation; we use the gas
state at a single cosmic time.
The simulated spectra are not meant
to look like full \lyalpha{} forest spectra, but just recover the statistics of
the flux in a small redshift window.
Our calculation of the spectra accounts for Doppler shifts due to
bulk flows of the gas, as well as for thermal broadening of the
\lyalpha{} line. We refer to \citet{Lukic:2015} for specific details of
these calculations. 
This procedure results in the \lyalpha{}
flux as a function of wavelength  or  equivalently  time  or  distance. 
Following  the standard approach, we then rescale the UV background
intensity so that the mean flux of all the extracted spectra
from the simulation matches the observed mean flux at the respective redshift
(see Section~\ref{sec:meanflux}
for more details on the specific value that we have chosen).
We therefore have neglected noise and metal contamination in our skewers
so far, but this will not be relevant in this paper.

Using these skewers
we have also calculated the curvature flux statistics, $\kurv$,
where,
\begin{equation}
\kappa=\frac{F''}{[1+(F')^{2}]^{3/2}} 
\end{equation}
$F'$ is the first derivative of the flux with respect to the velocity 
separation between pixels, and $F''$ is the second derivative.
We have done this for each simulation following the method
described in \citet[][]{Becker:2011}\footnote{i.e., we renormalize 
the fluxes of each skewer dividing them by its maximum
flux value. Then we only used pixels where the renormalized fluxes
are in the range $0.1\leq F_{\HI}^{R20}\leq0.9$.}.

We measured the thermal parameters of the simulation at each snapshot
by fitting the $\rho_{\rm b}-T$
relation with linear least squares in $\log_{10} \Delta$
and $\log_{10} T$, fitting the range $-0.7< \log_{10} \Delta < 0.0$ and
$\log_{10} T/K< 4.5$ \footnote{We have tested that changing these thresholds 
within reasonable IGM densities
produce differences just at a few per cent level 
\citep[see][for similar conclusions]{Lukic:2015} and 
in any case it does not affect the conclusions presented in this work.
We also found no relevant effects in the main results of this
paper if we employed
a different fitting approach as the one used in \citet{Puchwein:2015}.}. 

To characterize the gas pressure support in all our simulations,
we have followed the recent work by \citet{Kulkarni:2015}
and use the real-space \lyalpha{} flux,
$F_{\HI,\rm real}$.
This quantity is defined as $F_{\HI,\rm real}=\exp(-\tau_{\HI,\rm real})$, where
$\tau_{\HI,\rm real}$ is the real-space \lyalpha{} optical depth 
which is identical to the observed \lyalpha{} optical depth except
that the convolution integral that accounts for the redshift-space 
effects of the peculiar velocity field and thermal line
broadening has not been included. This field 
naturally suppresses dense gas, and is thus robust
against the poorly understood physics of galaxy formation,
revealing pressure smoothing in the diffuse IGM. The
$F_{\HI,\rm real}$ 3D power spectrum
is accurately described 
by a simple fitting function with a gaussian cutoff at
$\lj$, which is then defined as the pressure smoothing
scale. This statistic has the added advantage that it directly relates
to observations of correlated \lyalpha{} forest absorption in
close quasar pairs, proposed as a method to measure this scale,
and enables one to quantify it in simulations
\citep{Rorai:2013,Rorai:2015}.

\section{The Ionization and Thermal History of the IGM} 
\label{sec:results}

\begin{figure*}
\begin{center}
\includegraphics[angle=0,width=0.48\textwidth]{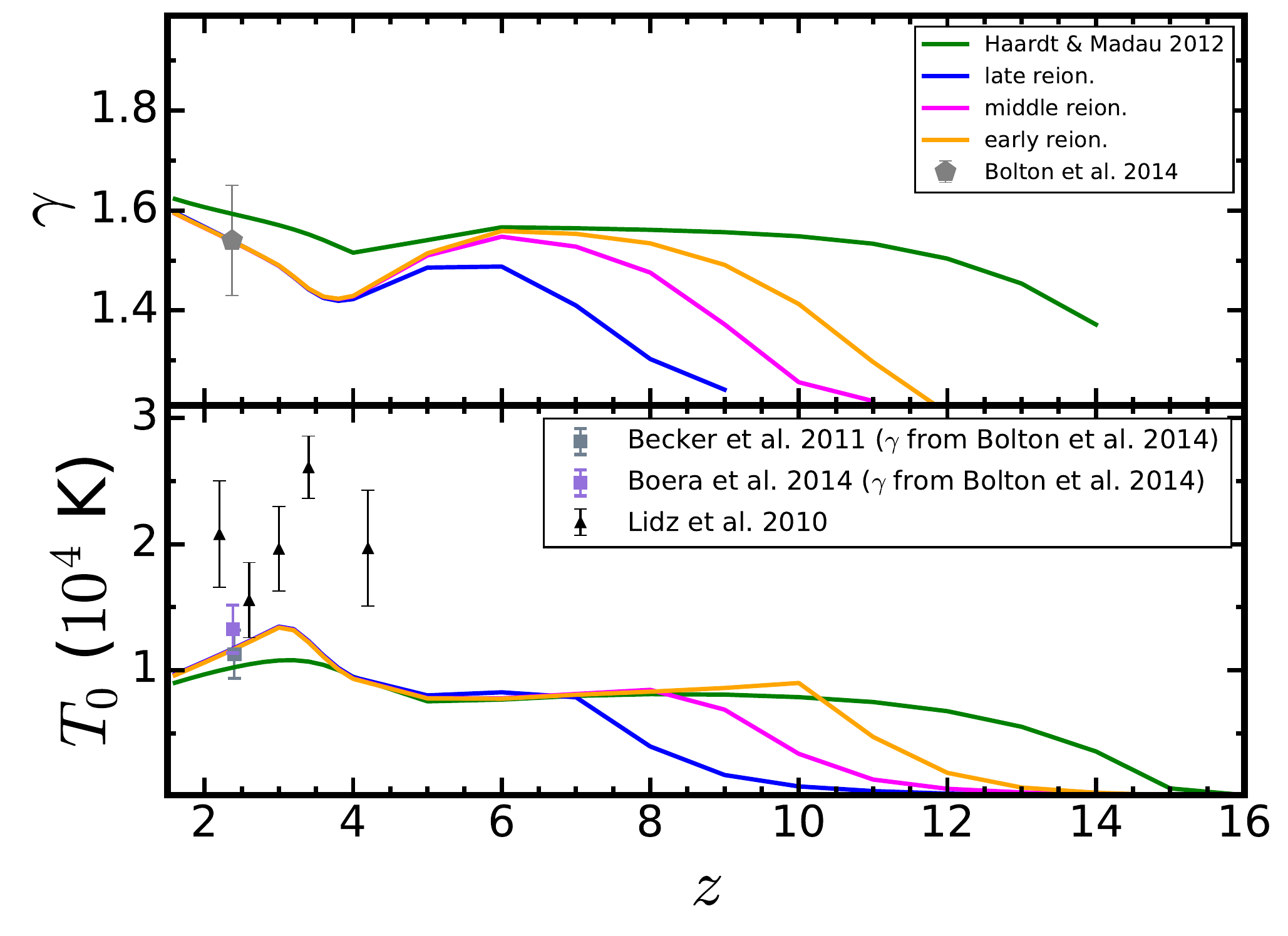}
\includegraphics[angle=0,width=0.48\textwidth]{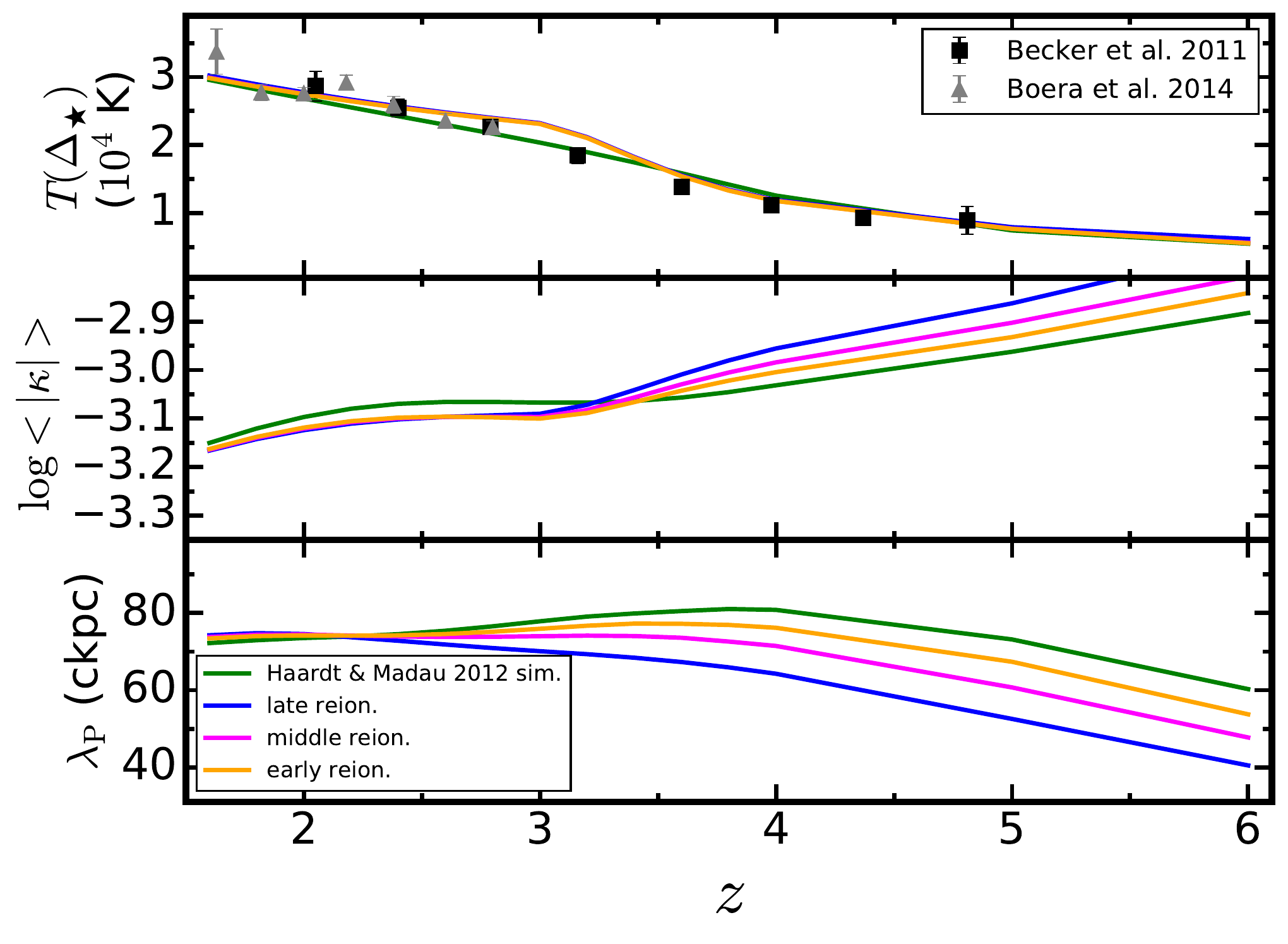}
\end{center}
\caption{
Thermal history obtained in simulations using different
UVB models: HM12 (green line), LateR (blue), MiddleR (magenta) and EarlyR (orange)
compared with observations. 
Left upper panel shows the evolution of the slope of the $\Delta-T$ relation, $\gamma$. 
Right lower panel the evolution of temperature at mean density, $T_{0}$.
The change of slopes at $z\sim3$ are due to $\HeIII$ reionization in both panels.
Right upper panel: evolution of the temperature at the optimal density,
$T(\Deltaopt)$. To compute these values in the models, 
we have assumed the optimal density
values given by \citet[][see text for more details]{Becker:2011}.
Right middle panel: evolution of the curvature flux statistic, $\kurv$, for
each simulation.
Right lower panel: evolution of the gas pressure smoothing scale computed in simulations
using different UVB models. Notice that while the temperature 
is just sensible to the current photoionization and photoheating values,
the actual gas pressure smoothing scale
value depends on the full thermal history of each simulation.
This also explains why the curvature flux statistics differ for
all simulations between $3\lesssim z\lesssim 6$.
Symbols with error bars stand for different observational measurements
and their $1\sigma$ error.
All parameters from the simulations presented in this 
figure are converged within $<5\%$ accuracy (see Section~\ref{ssec:convergence}).
See text for more details.\label{fig:Thistgas1}}
\end{figure*}

We now present thermal properties of the IGM in LateR, MiddleR, and EarlyR simulations, which
only differ in their redshift of $\HI$ reionization.
We first focus on the the evolution with redshift of the temperature at
mean density, $T_{0}$, and the slope of the temperature-density relation,
$\gamma$. 
The left panel of Figure~\ref{fig:Thistgas1} shows the evolution of these parameters for 
the HM12 run (green line). It also shows the thermal history of the 
LateR (blue), MiddleR (magenta) and EarlyR (orange) simulations.
In the upper left panel we plot the evolution of $\gamma$, which
exhibits the expected
convergence to a value close to $\sim 1.6$ after all reionization events for all models, resulting
from the balance of photoheating with adiabatic cooling\citep{Hui:1997,McQuinn:2016}.
The larger decrease of $\gamma$ during $\HeII$ reionization
in the EarlyR, MiddleR, and LateR runs seems to indicate a
temperature increase more independent of density than in the HM12 run.
\citet[][]{Puchwein:2015} have also shown that, for a fixed UVB model, using a
nonequilibrium approach will tend to create a more pronounced feature (we will 
further discuss this in  Section~\ref{sec:discuss}). In our modeling
we use equilibrium photoionization; hence, the flattening occurs for different reasons. 
In fact, this is just because our ionization model injects more heat to
the IGM than in the HM12 model.
One expects this type of effect during reionization
when applying a uniform UVB model through the whole volume,
as in that case we are applying a constant temperature increase
at each resolution element. 
At higher initial temperature this corresponds to a lower increase in the logarithm
of the temperature, so that the temperature-density relation flattens
in log-log space. This effect gets magnified as we increase the amount 
of heat applied to the whole box. Therefore, our results show that the 
exact evolution of $\gamma$ in simulations depends 
on the assumed shape for the $\HI$ and $\HeII$ reionization
histories and their total heat input.
In this panel we also plot the value of $\gamma$ of \citet{Bolton:2014} at $z=2.4$, derived from 
absorption-line profiles in the \lyalpha{} forest. \footnote{We do not directly compare to other measurements
of $\gamma$ \citep{Ricotti:2000,Schaye:2000,McDonald:2001,Garzilli:2012} because they are either 
significantly less precise, employ outdated simulations, do not sufficiently treat degeneracies
between $T_{0}$ and $\lj$, or have other differences in methodology that make direct comparisons between them
challenging.}

The lower left panel of Figure~\ref{fig:Thistgas1} shows the
evolution of $T_{0}$ for the same set of simulations.
As expected, in the new models $\HI$ reionization produces a much later heating than
in the HM12 run. In fact, it can be clearly seen that the temperature at mean density
of the new runs rises following the $\HI$ reionization of each model.
The heating during $\HeII$ reionization also shows significant
differences with the HM12 run. 
Although the heating happens at basically 
the same time as in the HM12 run, the new models also exhibit a larger and sharper
temperature increase at lower redshifts ($3\lesssim z\lesssim4$),
along with the expected decrease of the $\gamma$
value. This is due to the different ionization history
assumed for $\HeII$ which rises steeply at these redshifts.
We compare these models to the 
measurements of the temperature at mean density, $T_{0}$, 
from \citet{Lidz:2010}, based on the wavelet technique \citep{Meiksin:2000,Theuns:2000}.
We also plot constraints obtained by combining the $\gamma$ measurement by 
\citet{Bolton:2014} plotted in the upper panel with 
measurements of the temperature at the optimal density, $T(\Deltaopt)$, by  
\citet{Becker:2011} and \citet{Boera:2014} derived from the observed curvature of the \lyalpha{}
forest transmitted flux.
In this case we propagated errors from both measurements.
Since observations of the \lyalpha{} forest
are only available at $z\lesssim6$,  Figure~\ref{fig:Thistgas1} illustrates
that it will be challenging to constrain
$\HI$ reionization from measurements of $T_{0}$ and $\gamma$ at these
redshifts because the IGM quickly loses thermal memory of HI reionization. 
Our new UVB models provide, by construction, a sharper temperature rise
due to $\HeII$ reionization than the HM12 run. This again illustrates that the exact
evolution of the IGM thermal parameters depends on the the assumed shape
for the $\HI$ and $\HeII$ reionization histories, as well as their total heat input.

In the upper right panel of Figure~\ref{fig:Thistgas1} we plot another interesting property
defining the thermal state of the IGM, which is the temperature at the optimal overdensity, $T(\Deltaopt)$.
This ``optimal'' overdensity at each redshift is defined as the one for which curvature measurements of 
the \lyalpha{} forest
are more sensitive \citep[see][]{Becker:2011,Boera:2014}.
The curvature measurements allow one to determine this parameter
because they are not able to break
the degeneracy between $T_{0}$ and $\gamma$.
To calculate the $T(\Deltaopt)$ of our simulations,
we have used the function fit to these optimal
densities by \citet{Becker:2011} from a suite of hydrodynamical 
simulations ($\log_{10} \Deltaopt=A\times z + B$ where $A=-0.24596$ and
$B=1.22218$)\footnote{Using
optimal densities given by \citet[][$A=-0.21838$ and $B=1.05603$]{Boera:2014}
does not change any of the conclusions of this paper.}.
Inspecting Figure~\ref{fig:Thistgas1}, we see that all simulations give the same 
temperature at the optimal density at $z<6$, which is not surprising as we have already
shown that at these redshifts all of them 
have roughly the same
temperature-density relation (see the left panel of Figure~\ref{fig:Thistgas1}).
We also see the more pronounced
rise in temperature due to $\HeII$ reionization in the new models
compared with HM12 between $3\lesssim z\lesssim4$.

The middle right panel of Figure~\ref{fig:Thistgas1} 
shows the curvature flux statistics, $\kurv$, 
for all these simulations, and it is clear that they do not match
at these redshifts. This is because, as explained above, the flux
statistics depends not only on the temperature density relation of 
the IGM ($\gamma$, $T_{0}$), but also on the pressure smoothing scale
of the IGM, $\lj$.
In fact, the lower right panel of Figure~\ref{fig:Thistgas1} shows the evolution of
the pressure smoothing scale, $\lj$, with redshift for all the simulations.
This panel clearly illustrates the dependence of the pressure smoothing scale on the
full thermal history of the universe and not just on the instantaneous
temperatures.
That is, whereas the temperatures of all models agree at $z<7$, differences in
the pressure smoothing scale persist to much lower redshifts.
This explains the differences between the curvature statistics for all
the simulations and indicates that for fixed values of $T_{0}$
and $\gamma$, the value of the optimal density, $\Deltaopt$, will be  degenerate
with the pressure smoothing scale, $\lj$.

We have confirmed this by recreating the same study 
done by \citet{Becker:2011} to obtain 
the optimal densities using hydrodynamical simulations.
With a similar set of simulations to that of
these authors, we found almost identical results for the
values of the optimal densities\footnote{This grid of simulations
was created by modifying HM12 heating rates using two factors, $A$ and $B$:
$\dot{q}=A\Delta^{B}\dot{q}_{HM12}$}.
However, by including simulations that 
 have identical values of $T_0$ and $\gamma$ but different
 values of $\lj$, complicates the simple unique definition of $\Deltaopt$ by \citet{Becker:2011}, and instead
 adds scatter to the relationship between curvature and $T(\Deltaopt)$.
 The upper panel of Figure~\ref{fig:Thistgas1} also shows the
determinations of the temperature at the optimal density using the curvature of the
\lyalpha{} forest transmitted flux \citep{Becker:2011,Boera:2014}
compared with our simulations.
Given the strong dependence of the curvature on
the pressure smoothing scale $\lj$ resulting from the different
reionization histories, it is clear that the error bars on
$T(\Deltaopt)$ are likely underestimated. These issues
pertaining to the thermal history are discussed in \citet[][see also \citealt{Puchwein:2015}]{Becker:2011} 
but were not included in the error budget. The 
difference in $\log \kurv$ between
our late reionization model (LateR) and early reionization model (EarlyR)
at $z\sim4$ is $\sim0.05$. From the results presented by
\citet[][Figure 1 and 10]{Becker:2011} this difference implies already
$20-25\%$ error in temperature.

The aforementioned issues
related to the pressure smoothing scale and thermal history
can ease the  $2\sigma$ level of disagreement between
\citet{Lidz:2010} measurements of $T_{0}$ and the \citet{Becker:2011}
measurements 
of $T(\Deltaopt)$ at $z > 4$ (Figure~\ref{fig:Thistgas1}) although
this does not seem to be enough to explain it fully.
At lower redshifts
a comparison of the two measurements is more challenging because it depends
on what one assumes for the the temperature
density relation slope, $\gamma$.
Based on our results, it is clear that these conflicting measurements
lead to different interpretations
of the $\HI$ and $\HeII$ reionization events. On the one hand,
the \citet{Becker:2011} results point toward a temperature 
of the IGM at mean density
of $T_{0}\sim 1\times 10^{4}$ K by $z\sim 4.7$,
and a clear heating event later at $z\sim3$, which
they associated with $\HeII$ reionization.
On the other hand, \citet{Lidz:2010} higher measured temperatures 
require a higher energy injection
($\Delta_{T}$) than we assumed in our models ($\Delta T_{\HI}=2\times10^{4}$ K)
and an earlier injection of heat that could be associated with
$\HeII$ reionization at higher redshift than inferred by 
\citet{Becker:2011} and \citet{Boera:2014}.  Based on their
measurements, \citet{Lidz:2010} claim that the $\HeII$ reionization event
should be completed by $z=3.4$ and that the temperature at lower
redshifts is consistent with the fall-off expected from adiabatic
cooling.
Although one can argue about the statistical significance of
these discrepancies, especially given that the error bars are
underestimated because neither study marginalized out the
pressure smoothing scale $\lj$, we see no reason to prefer one set of measurements over the other.
For this reason we will not attempt any further interpretation of these measurements with our
numerical simulations, and we defer detailed data-to-model comparisons to future work.

We now want to discuss the evolution of the pressure smoothing scale in the different
simulations, which,
as can clearly be seen from Figure~\ref{fig:Thistgas1}, retains memory of the reionization events.
As we mentioned in the introduction, this is because, 
at IGM densities, 
the dynamical time that it takes
the gas to respond to temperature changes at the Jeans scale
(i.e., the sound-crossing time) is the Hubble time.
The first thing to notice is that the HM12 model results in a much
larger pressure smoothing scale than that of any our models, even the early
reionization one.  This is a direct result of the premature reionization
of HI and the spurious associated heating (starting at $z\sim15$) produced by this
model. Our new models correct this issue, properly tying reionization heating
to reionization history, resulting in later heating and a smaller overall pressure scale.
In addition, significant differences are also found below $z=6$ among
the simulations with different $\HI$ reionization histories (LateR, MiddleR and EarlyR) even
though these simulations share exactly the same photoionization and photoheating
values at these redshifts and therefore have very similar thermal parameters, i.e. $T_{0}$,
$\gamma$, $T(\Delta)$.
These differences in $\lj$ arise because the IGM has had more time to respond to its hotter temperature
when reionization occurs earlier, resulting in a larger pressure scale.
The sensitivity of the pressure smoothing scale to the reionization history highlights the
importance of constructing self-consistent models of reionization
and applying them to optically thin simulations to better
understand the thermal evolution of the IGM.

It is interesting to discuss the redshift evolution
that we find for the pressure smoothing
scale, using $\lj$. This parameter can be seen as the gas pressure scale
at the density most sensitive
for \lyalpha{} observations. Therefore, there are
two physical effects that contribute to the value of this parameter. First, 
the IGM is heated as the universe evolves, so we expect the pressure smoothing
scale to increase with time. 
On the other hand, there is not just one pressure smoothing scale
in the IGM, but one for each density and, just from linear theory,
we expect it to be higher at lower densities, $\lj \propto \nH^{-1/2}$
\citep{Schaye:2001}.
As we go to lower redshifts,
the neutral hydrogen density is being further diluted by
the expansion of the universe, and observations
start to be more sensitive to higher densities
that have a smaller pressure smoothing scale.
The combination of both processes can produce the 
somewhat surprising behavior of flattening of $\lj$ at $z\lesssim4$
(decrease in the case of the HM12 model due to an $\HeII$ reionization 
with a lower heat injection).
We also found that since all these models share the same UVB
after reionization, they tend to converge to the same
$\lj$ values at lower redshifts. Although the pressure smoothing scale
of the IGM depends on the full thermal history, the 
thermal memory of past reionization events
eventually fades as the gas evolves toward lower redshifts.

\subsection{Heating during Hydrogen Reionization} \label{ssec:hehy}

\begin{figure*}
\begin{center}
\includegraphics[angle=0,width=0.48\textwidth]{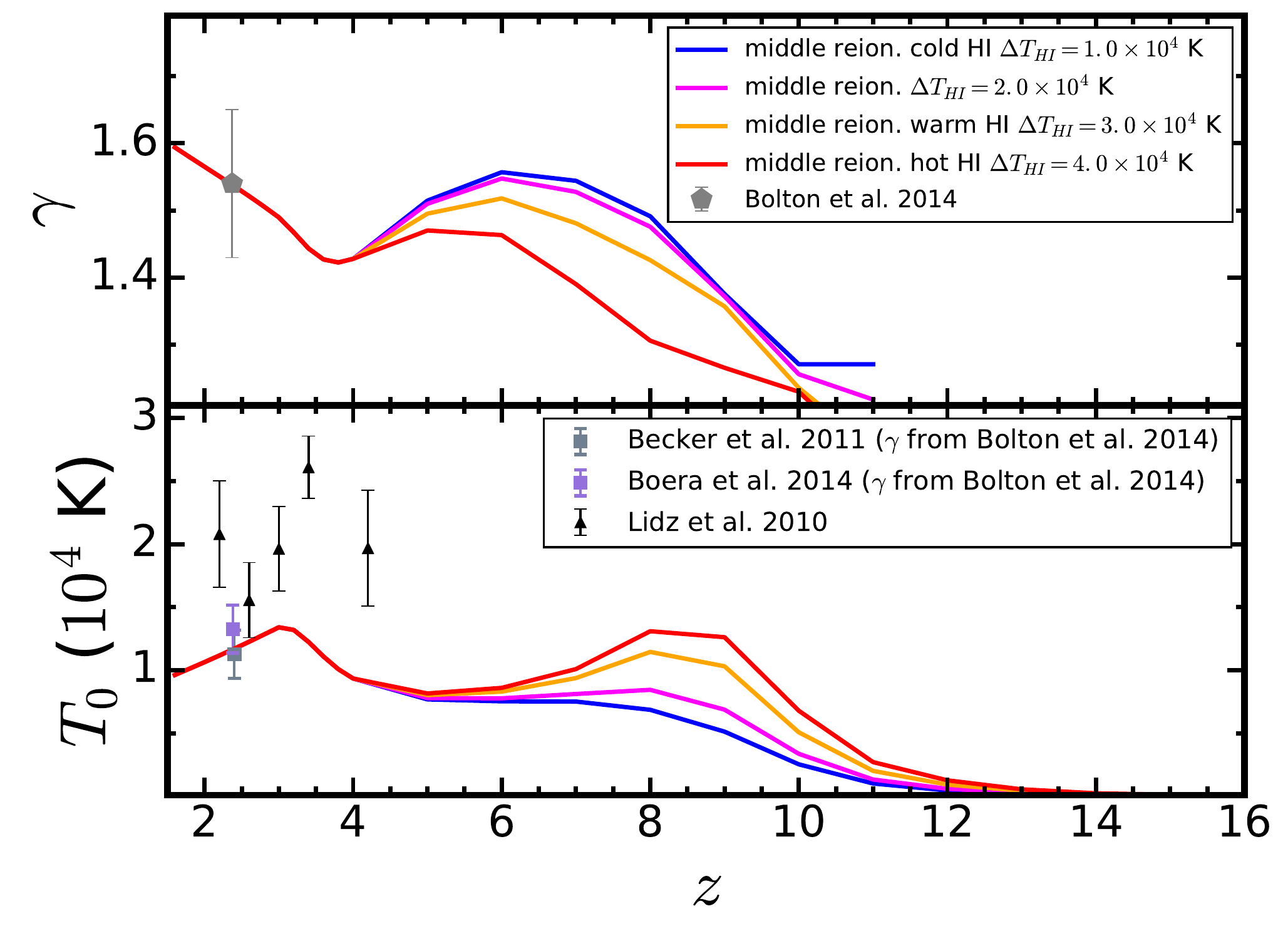}
\includegraphics[angle=0,width=0.48\textwidth]{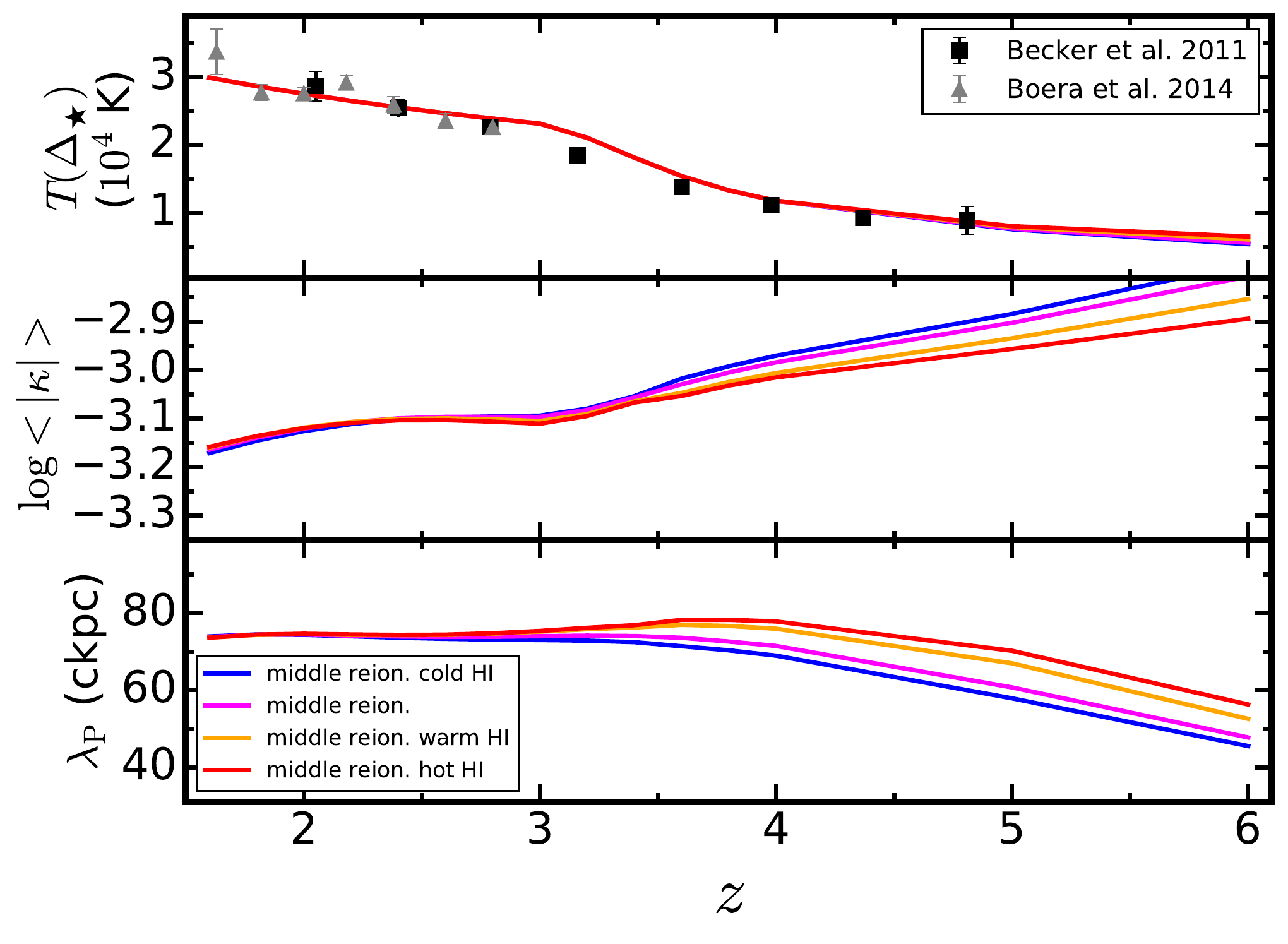}
\end{center}
\caption{Results for simulations using a different total heat input
during $\HI$ reionization, i.e., different $\Delta T_{\HI}$.
Upper left panel: evolution of the slope of the $\Delta-T$ relation, $\gamma$. 
Lower right panel  evolution of temperature at mean density.
Changes of slopes at $z\sim3$ are due to $\HeIII$ reionization.
Upper right panel: evolution of the temperature at the optimal density,
$T(\Deltaopt)$. To compute these values in the models,
we have assumed the optimal density
values given by \citet[][see text for more details]{Becker:2011}.
Lower right panel: evolution of the gas pressure smoothing scale computed in simulations
using different UVB models. Notice that while the temperature 
is just sensible to the current photoionization and photoheating values,
the actual gas pressure smoothing scale
value depends on the full thermal history of each simulation.
Symbols with error bars stand for different observational measurements.
All parameters from the simulations presented in this 
figure are converged within $<5\%$ accuracy (see Section~\ref{ssec:convergence}).
\label{fig:hehymodels}}
\end{figure*}

Figure~\ref{fig:hehymodels} shows the results from simulations
(MiddleR-Hcold, MiddleR, MiddleR-Hwarm and MiddleR-Hhot)
using different total input heat during $\HI$ reionization,
i.e., different $\Delta T_{\HI}$ ($1\times 10^{4}$, $2\times 10^{4}$, 
$3\times 10^{4}$ and $4\times 10^{4}$ K, respectively).
All these simulations share the same $\HI$ ionization history
(middle reionization, $z_{\rm reion,\HI}=8.3$) and exactly the same $\HeII$ ionization
history and heating. 
The left panel of Figure~\ref{fig:hehymodels} shows the evolution of
$\gamma$ and $T_{0}$ for these simulations. As expected, 
they differ significantly during $\HI$ reionization, due to the different heat input
applied in them. 
At  lower redshifts, $z<z_{\rm reion,\HI}$, all these simulations share the same HM12
photoionization and photoheating rates; hence, eventually $T_{0}$ and $\gamma$
thermal parameters tend to converge to the same values. 
This shows again that these thermal parameters depend more strongly on 
the instantaneous value of these rates.
However, it is very interesting to remark that this convergence is not immediate,
but that it takes some time for each simulation to converge after reaching the redshift at which
they all have exactly the same rates \citep{McQuinn:2016}.
In any case, our simulations show that $\gamma$ and $T_{0}$ have little sensitivity to the details
of $\HI$ reionization after $z<5$.

In the right panels of Figure~\ref{fig:hehymodels} we show the evolution of the temperature
at optimal density ($T(\Deltaopt)$, upper panel), the curvature ($\kurv$, middle panel),
and the pressure smoothing scale ($\lj$, lower panel).
As expected, $T(\Deltaopt)$, follows the same trend as $T_{0}$ and $\gamma$
(see discussion above). However, the curvature statistics and 
the pressure smoothing scale for these simulations clearly show a
different behavior at redshifts above $z\gtrsim3$,
while $T_{0}$, $\gamma$ and  $T(\Deltaopt)$ have already forgotten
reionization at $z\sim 5$.
Simulations with a higher heat input at high redshift show higher pressure smoothing scale values
even at lower redshift, due to the dependence of this parameter 
on the full thermal history. 
Comparing this result
with Figure~\ref{fig:Thistgas1}
we can see that the ionization history and the total heat input
are degenerate in terms of the pressure smoothing scale. 
That is, the earlier that $\HI$ reionization injects heat into the IGM, the 
larger the gas pressure scale, $\lambda_{P}$. However,  
a later but hotter (larger heat input) $\HI$ reionization
also results in  a larger pressure smoothing scale. 
This cautions one about the interpretation of curvature-based
measurements of the IGM temperature
\citep{Becker:2011} at $z>3.5$, since the middle right panel of Figure~\ref{fig:hehymodels}
clearly illustrates that the curvature has a strong dependence on thermal history,
even when the instantaneous temperature is the same in all models.

\subsection{Heating during Helium Reionization} \label{ssec:hehe}

\begin{figure*}
\begin{center}
\includegraphics[angle=0,width=0.48\textwidth]{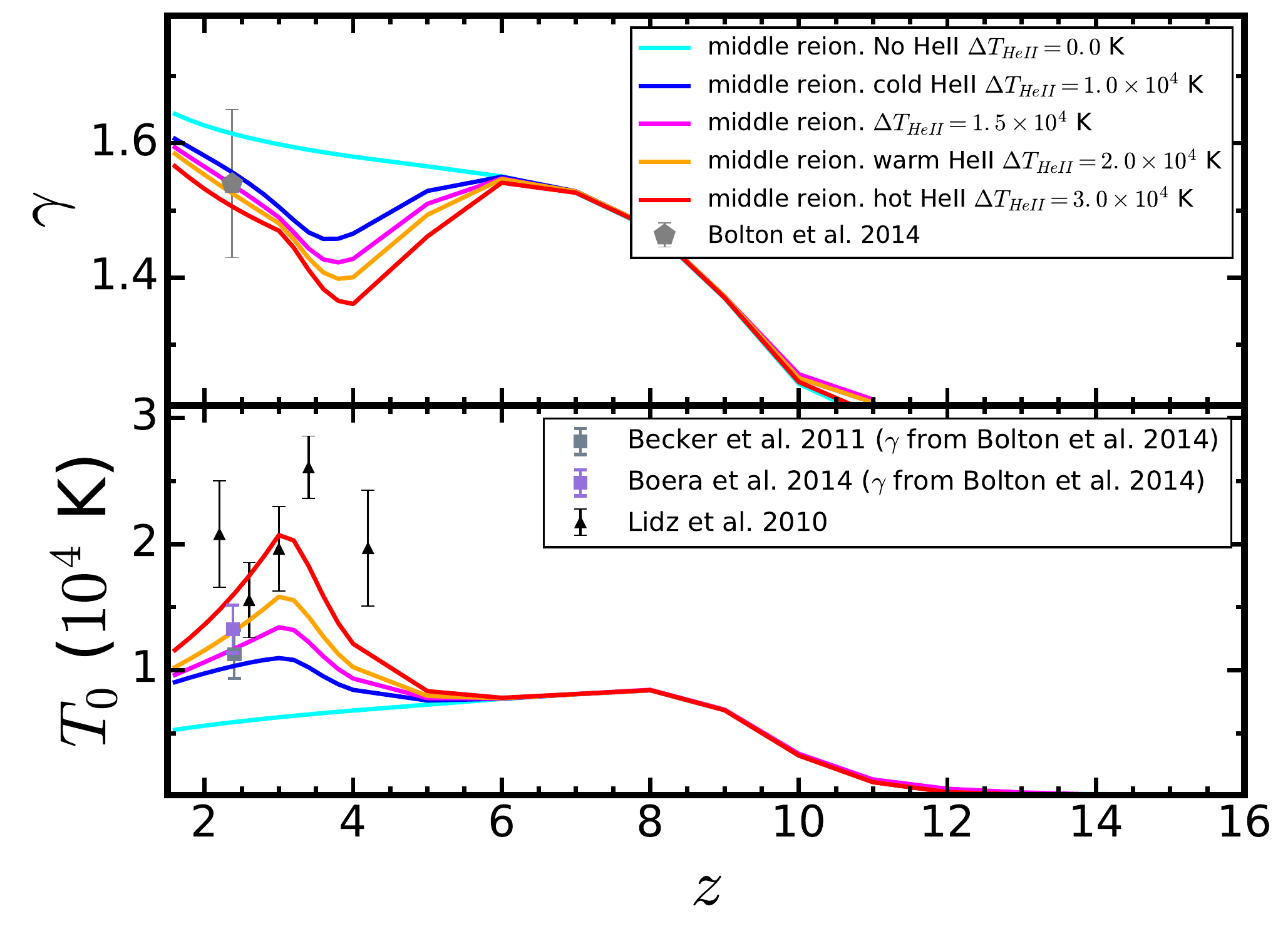}
\includegraphics[angle=0,width=0.48\textwidth]{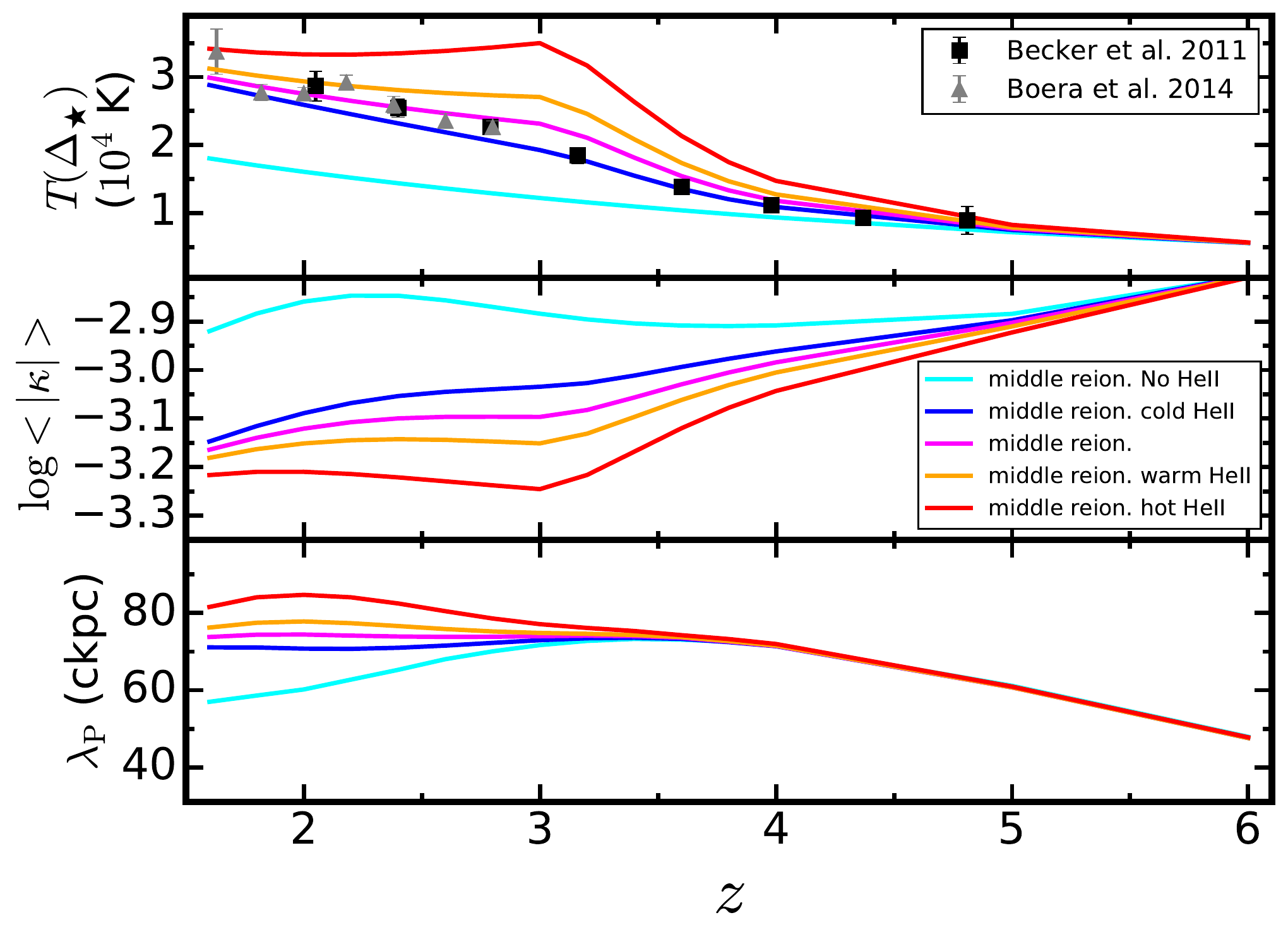}
\end{center}
\caption{Results for simulations using a different total heat input
during $\HeIII$ reionization, i.e., different $\Delta T_{\HeIII}$.
Upper left panel:evolution of the slope of the $\Delta-T$ relation, $\gamma$. 
Lower right panel the evolution of temperature at mean density.
Changes of slopes at $z\sim3$ are due to $\HeIII$ reionization.
Upper right panel: evolution of the temperature at the optimal density,
$T(\Deltaopt)$. To compute these values in the models, 
we have assumed the optimal density
values given by \citet[][see text for more details]{Becker:2011}.
Lower right panel: evolution of the gas pressure smoothing scale computed in simulations
using different UVB models. Notice that while the temperature 
is just sensible to the current photoionization and photoheating values,
the actual gas pressure smoothing scale
value depends on the full thermal history of each simulation.
Symbols with error bars stand for different observational measurements.
All parameters from the simulations presented in this 
figure are converged within $<5\%$ accuracy (see Section~\ref{ssec:convergence}).
\label{fig:hehemodels}}
\end{figure*}

Finally, we discuss simulations for which we modified only the total
heat input during $\HeII$ reionization. These are MiddleR-noHe,
MiddleR-Hecold, MiddleR, MiddleR-Hewarm, MiddleR-Hehot, and the
specific values of $\Delta T_{\HeII}$ used in them were $0$ (no
$\HeII$ reionization), $1\times 10^{4}$, $1.5\times 10^{4}$, $2\times
10^{4}$, and $3\times 10^{4}$ K, respectively. Apart from this, these
simulations share exactly the same $\HeI$ ionization history and also
the same $\HI$ ionization and photoheating rates (middle reionization,
$z_{\rm reion,\HI}=8.30$).  Therefore, it is not surprising that the
evolution of $\gamma$, $T_{0}$, and $T_{\Deltaopt}$ thermal parameters
at redshift above $z\gtrsim 5$ is exactly the same for all simulations
(see upper left, lower left, and upper right panels of
Figure~\ref{fig:hehemodels}).  It is only when $\HeII$ reionization
starts heating the IGM that these simulations differ. In particular,
runs with a higher total heat input result in a much steeper rise
when $\HeII$ reionization commences
in $T_{0}$,  accompanied by a commensurate fall when $\HeII$ reionization is completed. 
With the slope of the temperature-density
relation, $\gamma$, the effect is the opposite. The curvature statistic and
the pressure smoothing scale also illustrate the effect of
the different heating during $\HeII$ reionization.  Simulations with a
larger late heat input due to $\HeII$ reionization give rise to a larger
pressure smoothing scale. Notice that the pressure smoothing scales
of these models begin to diverge at $z<3$, once $\HeII$ reionization has already completed,
as there is a delay before the effect propagates. This delay is due to the dynamical
time that it takes the gas to respond to temperature changes at the Jeans scale
(i.e., the sound crossing time), which, as discussed above for IGM densities, is
close to the Hubble time.

\section{Calibrating the UVB to Yield the Correct Mean Flux} \label{sec:meanflux}

The simplest possible \lyalpha{} flux statistic is the mean transmitted flux
$\mean{F_{\HI}}$, or equivalently, the effective optical depth
$\tau_{\HI}= − \log \mean{F_{\HI}}$.
It  is  commonly  the  case  that  simulations do not recover
the observed mean flux, but that simulated
fluxes are rescaled to match the observed mean.
This rescaling is often understood as equivalent to
adjusting the specific intensity of the $\HI$ photoionization
rate used in the simulation and is justified
based on how poorly constrained the ionizing  background is.
Notice, however,
that this rescaling is generally done directly in redshift space.
\citet{Lukic:2015} conducted a detailed study of  the effect that this 
rescaling can have on different \lyalpha{} statistics. 
They found that for the large rescalings --- those where optical depth has to be rescaled 
by a factor of 2 or more --- the error on flux power spectrum is a few percent.
Also, the larger the rescaling is, the larger the error that is
introduced.  The rescaling error is therefore small, but not
negligible, and most importantly, it is puzzling why one should
continue to run simulations that systematically produce mean flux
values excluded by observations at the few sigma level, and continue
to compensate by rescaling the optical depth by a factor of few as is
currently required with the HM12 or FG09 UVB tables.  For this reason
we wish to correct for this error in our new UVB models by
renormalizing the input $\HI$ photoionization rate so that the
post-processing correction will be minimal at all relevant redshifts.
We want to emphasize here that the goal of this step is not to remove
the need for future rescaling of the optical depth in simulations, but
only to ensure that mean fluxes obtained by simulations are roughly
consistent with current observations, therefore removing the need for
large rescalings.  Trying to do better than that would be
pointless exercise, as the change in cosmological parameters, as well
as having different resolution or box size, will anyway change the mean
flux at a few percent level. The resolution of the simulations
discussed here is thus sufficient for obtaining mean flux 
converged at a few percent level
\citep{Lukic:2015}. In fact, we have run simulations of our
LateR, MiddleR and EarlyR UVB models using
a larger box size, $L_{box}=40$ $Mpc/h$ and $2048^3$ resolution elements to
confirm that this is indeed the case.

\begin{figure} 
\begin{center}
\includegraphics[width=0.45\textwidth]{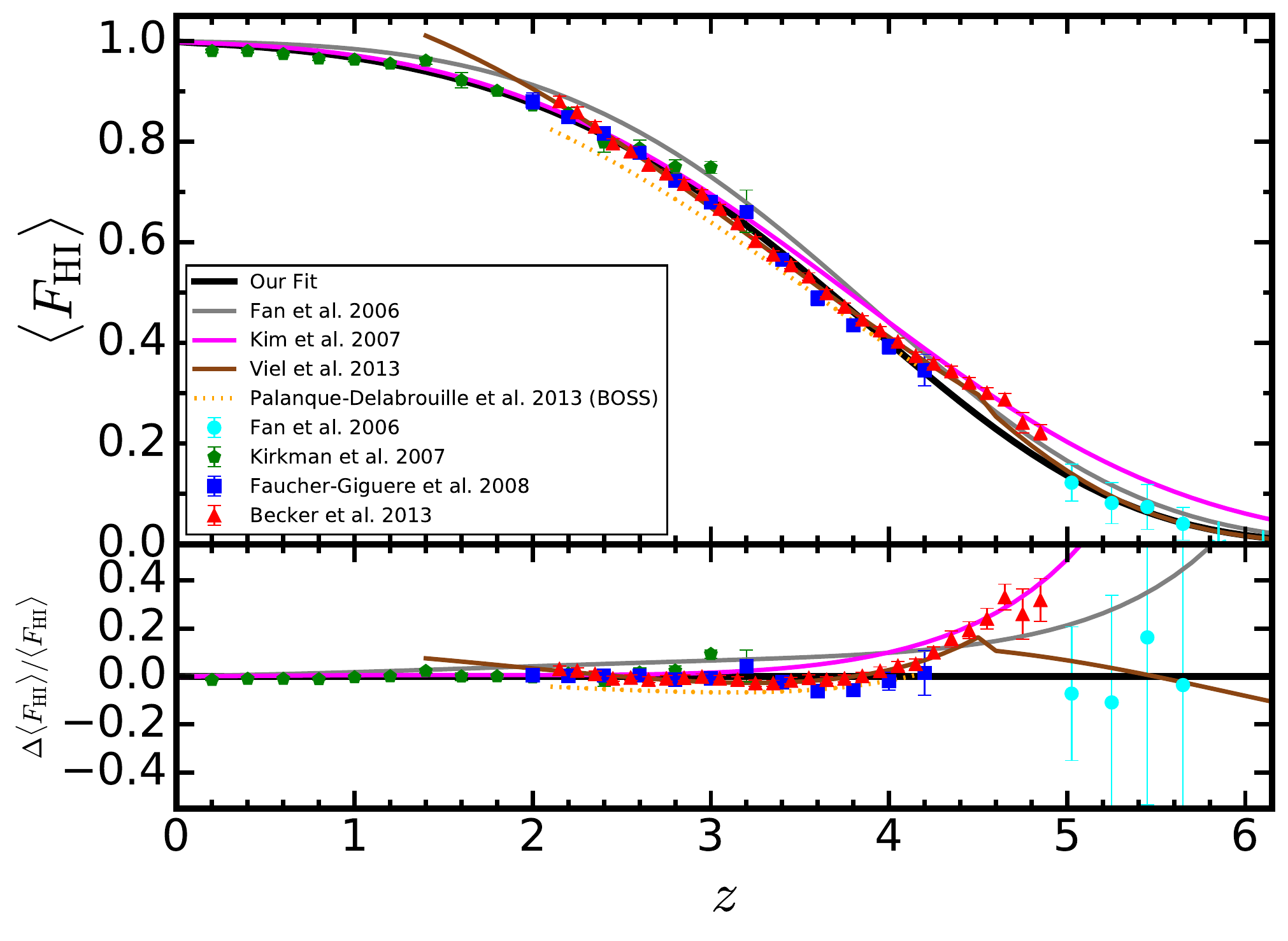}
\end{center}
\caption{Different observation sets of the \lyalpha{} mean flux
 evolution from high resolution quasar spectra
 \citep{Fan:2006,Kirkman:2007,FaucherGiguere:2008,Becker:2013}.
 The orange dotted line stands for the fit obtained using $\sim 13000$ quasar spectra from the BOSS collaboration \citep{PalanqueDelabrouille:2013}.
 We also plotted some suggested fits in the literature 
 \citep{Fan:2006,Kim:2007,Viel:2013b} as well as our suggested fit the the whole high resolution data set $0.2<z<5.85$
 (solid black line). See text for more details.}
\label{fig:meanflux_fit}
\end{figure}

Observational constraints coming from quasar absorption lines
\citep{Fan:2006,Becker:2007,Kirkman:2007,FaucherGiguere:2008,Becker:2013} 
show that the mean flux 
smoothly evolves from about $0.2$ at $z=5$, to about $0.9$ at
$z=2$, as expansion gradually lowers the density
and the UVB intensity
slowly increases. Figure~\ref{fig:meanflux_fit}
shows a compilation of these observations
using different symbols with $1\sigma$ error bars. We also plot suggested fits
to the mean flux evolution by various authors  \citep[][]{Fan:2006,Kim:2007,Viel:2013b}.
However, none of these fits do a particularly good job of describing the full
evolution of the mean flux, and we therefore opt for our own
fit using all observational data points between $0.2<z<5.85$. 
We found that the functional form
\begin{equation}
 \tau_{\HI}= A \times e^{(B\times \sqrt{z})}
\end{equation}
provides an optimal fit, with $A=0.00126$ and $B=3.294$ as the best-fit parameters.
This fit is also shown in Figure~\ref{fig:meanflux_fit} as a solid black line.

\begin{figure*}
\begin{center}
\includegraphics[angle=0,width=0.45\textwidth]{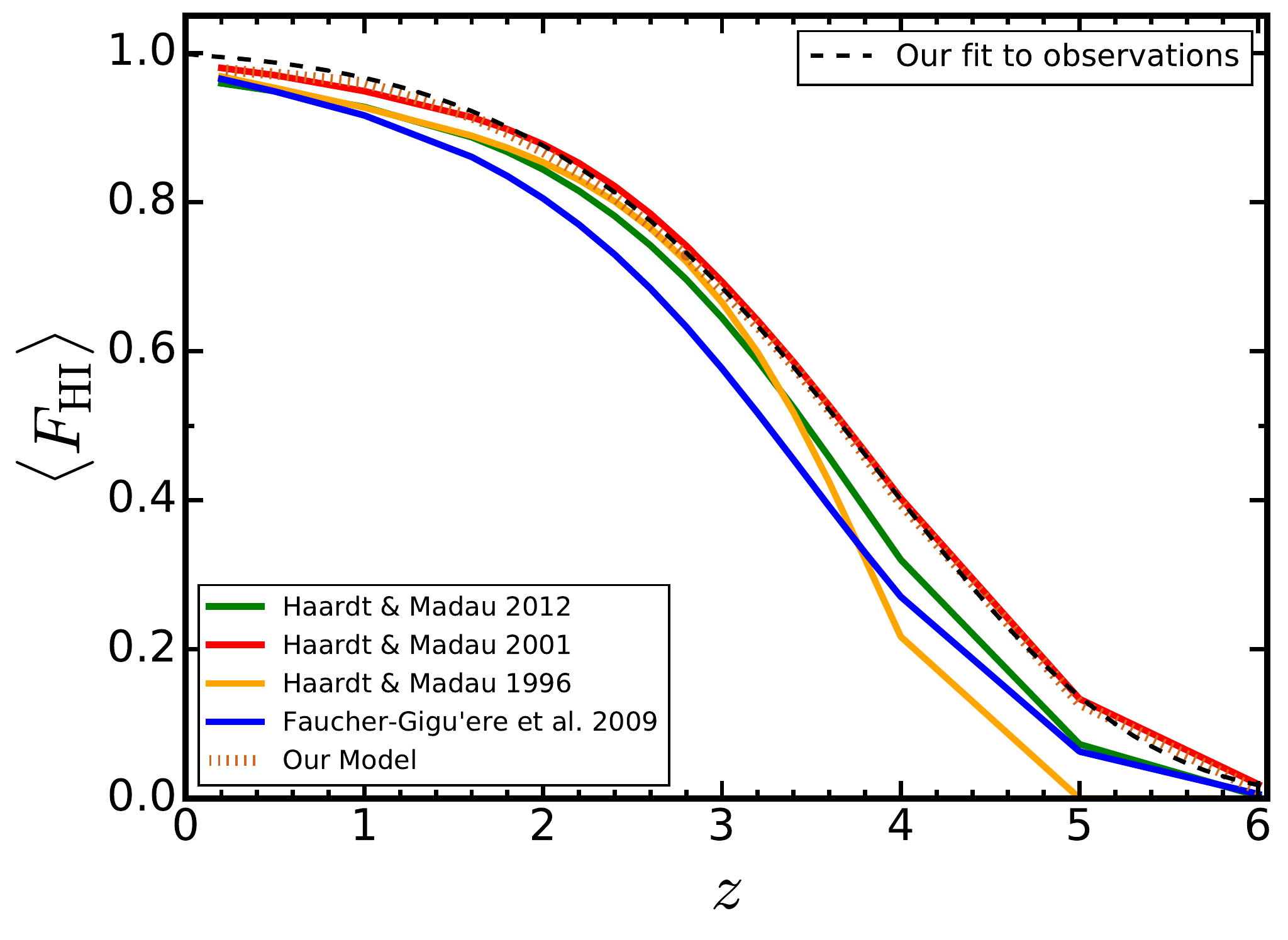}
\includegraphics[angle=0,width=0.45\textwidth]{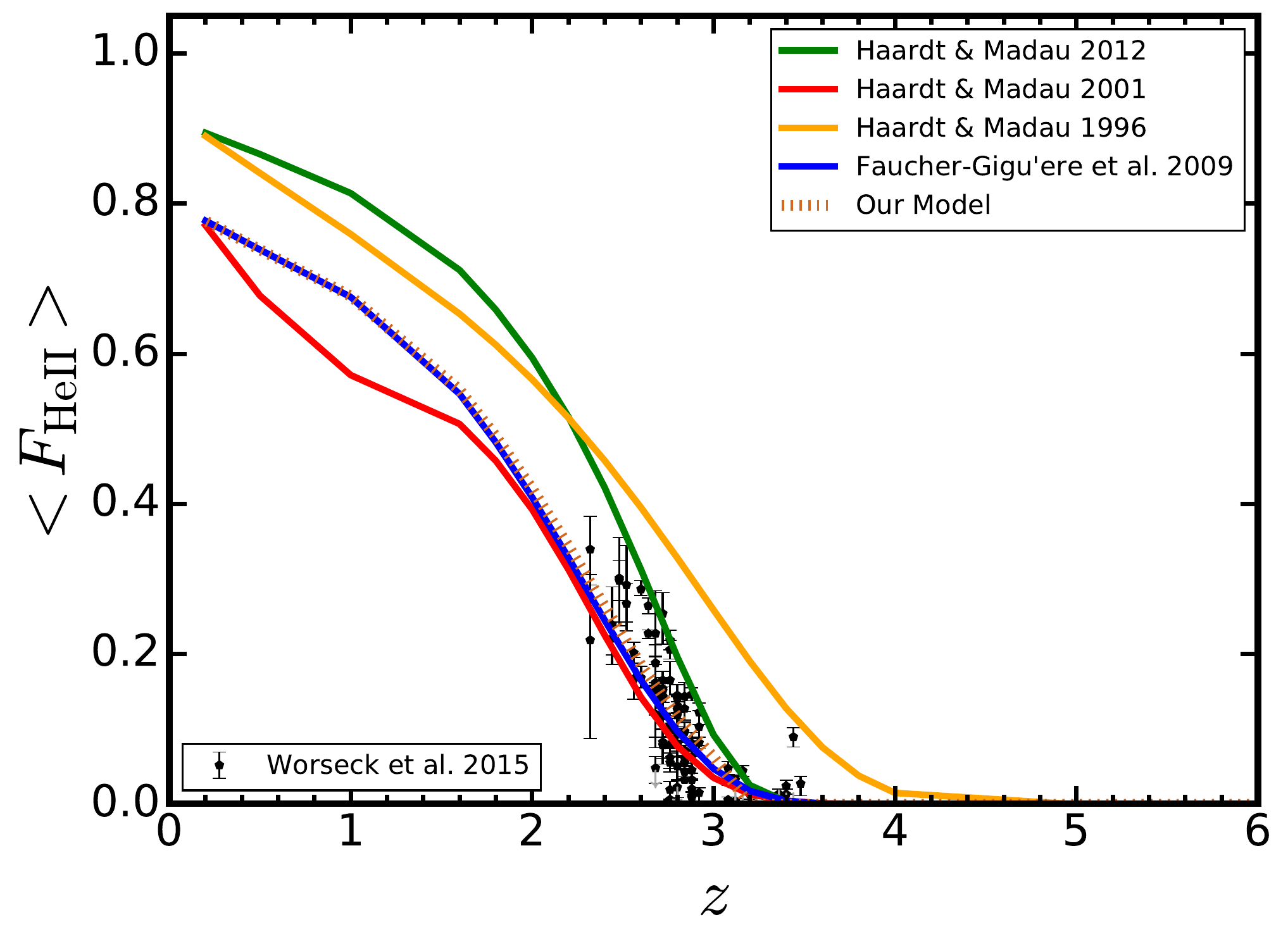}
\caption{Left panel: $\HI$ mean flux evolution obtained using standard UVB
models (HM96, HM01, FG09, HM12) 
compared with our best fit to observations. 
Right panel: $\HeII$ mean flux evolution obtained using the same standard UVB
models as in the left panel compared with observations \citep{Worseck:2015}.
The dotted line stands for the mean flux
evolution obtained in our models. See text for more details.
\label{fig:meanfsims}}
\end{center}
\end{figure*}

The left panel of Figure~\ref{fig:meanfsims} shows
the $\HI$ mean flux evolution
in simulations using different well-known UVB models. It is clear
that several of them significantly underpredict the observed mean flux at all redshifts. 
It is also worth pointing out that, somewhat coincidentally,
the HM01 
UVB model is doing a good job in recovering the observed mean flux.
Notice, however,
that since our thermal history is different from theirs, we cannot just
assume their photoionization rates.
Hence, we have modified the $\HI$ photoionization rate in our models after reionization
so that the $\HI$ mean flux in the simulations matches the fit to current observational 
constraints. Before $\HI$ reionization, when we apply our new methodology, this does not apply.
We also modified the $\HI$ and $\HeI$ photoheating rate by the same factor so that
the heat input at these redshifts is conserved, i.e.,
$n_{\HI,\rm old}\dot{q}_{\HI,\rm old}=n_{\HI,\rm new}\dot{q}_{\HI,\rm new}$
and therefore we get exactly the same thermal histories.
We have confirmed that this is the case by comparing the evolution 
of the thermal parameters in the new models versus the old ones,
and we show the mean flux evolution of the MiddleR simulation 
as a dashed line in Figure~\ref{fig:meanfsims}. 

Regarding the effect of changing the thermal history, we
have compared the differences between the simulations in which we change
the heat input due to $\HeII$ reionization. We found differences in the 
$\HI$ mean flux up to $\sim 10\%$ and $\sim 15\%$ 
between our our fiducial MiddleR model 
and the two most extreme simulations MiddleR-Hehot and MiddleR-noHe, 
respectively. This is because through the recombination factor ($\alpha \propto T_{0}^{-0.7}$)
the neutral fraction has a sensitivity
to the gas temperature, not just the photoionization:
$n_{\HI}\propto \Gamma_{\HI}^{-1} T_{0}^{-0.7} \Delta^{0.7*(\gamma-1)}$.
This maximum difference corresponds to $z\sim 3$, the redshift
at which the thermal parameters between these simulations are most different.
Of course, simulations using thermal histories that deviate even further 
from these would increase these differences.
We have confirmed that mean flux differences due to current uncertainties in the cosmological
parameters have a much weaker effect on the mean flux than any of the above systematics
in the UVB models described above (see Appendix~\ref{app:cosmo} for a
full discussion on cosmological parameters).

Finally, in the right panel of Figure~\ref{fig:meanfsims} 
we show the most recent observations of the $\HeII$ transmission
\citep{Worseck:2015} and compare them again with the mean HeII flux 
derived from our hydrodynamical simulations 
using standard UVB models 
(HM96, HM01, FG09, HM12)\footnote{Note that in this plot observations compute the mean flux
 averaging over a much smaller window than in the simulations;
however this would only change the variance, but not the mean.}.
We show the mean fluxes, $\mean{F_{\HeII}}=e^{-\tau_{\HeII}}$, and
not the optical depths in order to focus on the low redshift results,
$z\lesssim2.7$, where observations seem to indicate that $\HeII$ reionization
has already finished.
Results at higher redshifts are not that conclusive 
and may indicate that this reionization happened much more slowly 
than has been assumed \citep[see][for a detailed discussion]{Worseck:2015}.
Therefore, we want to focus here just on the low redshift values, where
reionization is completed and our method should be valid.
For these it seems that FG09 
$\HeII$ photoionization rates are doing
the best job in reproducing the observations. For this reason we
decided to use the $\HeII$ photoionization and
photoheating rates of this model in our new UVB models
after $\HeII$ reionization.

\section{Discussion} \label{sec:discuss}

In this section we elaborate on the convergence of the results
presented in this work, we compare them with recent efforts
done in the field,
and finally discuss their implications for galaxy formation
simulations.

\subsection{Resolution and Convergence}
\label{ssec:convergence}

We have also run a set of simulations to explore resolution and box size 
effects on the different methods and parameters discussed in this paper.
Some of these results for the mean flux, $T_{0}$, $\gamma$ 
and flux power spectrum relevant for this work have already been presented in \citet{Lukic:2015}.
We refer to this work for more details of the accuracy of these simulations.
In this regard we are confident that the thermal parameters discussed
in this paper, $T_{0}$, $\gamma$, and $T(\Deltaopt)$
as well as the pressure smoothing scale, $\lj$, are converged at least at the 5\% level
for $z\lesssim6$. This is also the case for the curvature statistic, $\kurv$. 
Although we have used Nyx, an Eulerian code, to run all the tests of the new
models created in this work, they will produce the same ionization and thermal
histories in any other optically thin hydrodynamical code available.
We have explicitly confirmed that Nyx and Gadget (which uses the SPH method for the hydrodynamics) 
agree well in their values for the mean flux and $\rho$-$T$ relation.
Some differences could arise at lower redshifts in some observables,
depending on the specific galaxy formation sub-grid model implementation \citep[see e.g.,][]{Viel:2013a}.
However, it is hard to think of a realistic galaxy formation feedback model
that will significantly affect the global ionization
and thermal histories of the IGM \citep[see, e.g.,][but see Figure 10 of \citealt{Meiksin:2014}]{Kollmeier:2006,Desjacques:2006,Shull:2015}.

\subsection{Comparison to Previous Work}

Recently, \citet{Puchwein:2015} tried to solve some of the discrepancies
between the HM12 model and $2<z<4$ observations of thermal parameters
by including a nonequilibrium ionization solver in their
hydrodynamical simulations.
This approach goes in the same
direction as this work, in the sense that they both try to improve how
things are currently done during reionization events. 
As was expected, for a fixed UVB model they showed that using a non-equilibrium solver
will produce a bigger temperature increase of the IGM during reionization.
There is no doubt that a nonequilibrium approach is more physically relevant,
as during reionization events the equilibration
timescale, which is the time it takes the ionized fraction to change in 
response to a change in the photoionization rate $\Gamma$,
can be comparable to the Hubble time, $t_{\rm eq} \simeq \Gamma_{\rm UVB}^{-1} \simeq t_{\rm Hubble}$.
In a time-dependent (nonequilibrium) ionization calculation
the neutral fraction will thus be elevated relative to the equilibrium value, and this results
in more photoionization heating, $\sim \nHI \times \dot{q}_\HI$,
i.e. $\nHI$ is higher in a nonequilibrium calculation.
\citet{Puchwein:2015} found that this effect brings the HM12 model much more in agreement
with the \citet{Becker:2011} curvature measurements.

\citet{Puchwein:2015} also showed that the change of the
slope in the temperature density relation of the IGM, $\gamma$, is in
fact significantly smaller
in the ionization equilibrium approximation by running
the same UVB model using ionization equilibrium and nonequilibrium
algorithms.
However, the different thermal histories 
that we found using our new UVB
models indicate that this in fact degenerated with the ionization
history and total heat input of the reionization event assumed to
build the UVB model. The \citet{Puchwein:2015} calculations use the HM12
heating rates, which are based on \textsc{cloudy} 1D slab
calculations.  The validity of the various approximations is dubious,
as the heating during reionization is a complicated physical process
that depends not only on the shape of the spectrum but also on the
local density field and how fast the ionizing front travels
\citep{McQuinn:2012,Davies:2016}.
Due to the present lack of knowledge about how
much and when the reionization heats the IGM, we prefer to simply
parameterize our ignorance of the details of reionization heat
injection with a free parameter $\Delta T$. The differences in the IGM
thermal properties between equilibrium and nonequilibrium codes thus
seem moot given the large uncertainty in this $\Delta T$ parameter.
However, as observational constraints improve and begin to constrain $\Delta T$,
an improvement of our calculation would be to implement a nonequilibrium calculation
along the lines of \citet{Puchwein:2015}, but from the perspective of
our thermal history. This would amount to modification of the value of
the $\Delta T$ that we choose or infer from data. That said, most
cosmological hydrodynamical and galaxy formation codes use equilibrium
solvers, and thus our current tables have wider applicability in their
present assumption of equilibrium.

Recently, \citet{UptonSanderbeck:2015} have also analyzed the thermal
histories of different reionization models in a similar spirit as our
approach in this work.
However, they adopt a fast semianalytical approach
that allows them to study how intergalactic gas is heated and cooled during and after 
reionization processes in a multiple-zone scenario, as opposite
to the one-zone model assumption currently used in hydrodynamical simulations.
Using this approach, the authors also explored
reionization models with different ionization histories and heat
injection. 
This method allows them more freedom in the types of models and parameters
that can be explored. It probes to be a very useful tool to build intuition on 
the possible effects of a wide variety of reionization scenarios.
However, a full numerical hydrodynamical method is required to generate
simulations of different reionization models that can be directly 
compared to the observations, and perhaps most importantly, it is not possible to simulate
the pressure smoothing effects resulting from different thermal/reionization history
analytically. So in fact, the authors must interpret their results on the thermal parameters
derived from observations using hydrodynamical simulations.

Finally, we want to point out that both the \citet{Puchwein:2015} 
and \citet{UptonSanderbeck:2015} final conclusions on possible
$\HI$ reionization scenarios are based only on \citet{Becker:2011} measurements
at  high redshift ($z>4$), ignoring the constraints
on $T_{0}$ given by \citet{Lidz:2010}, which
point to a much hotter IGM.
As discussed above, these two measurements seem to be in
disagreement at the 2$\sigma$ level at high redshift, but
there is no clear reason to us why any of them should be ignored.

\subsection{Implications for Galaxy Formation}

As mentioned in the introduction, 
an ionizing UVB inhibits gas accretion and photoevaporates gas
from the shallow potential wells of low-mass
dwarf galaxies. This effect due to gas being heated up by photoionization 
can result in negative feedback: suppressing star formation inside
reionized regions, 
thus impeding their continued growth \citep[see, e.g.][]{Rees:1986,Efstathiou:1992}.
In our current picture of galaxy formation
this feedback is considered to be a good candidate for
resolving the "missing satellite" problem \citep{Moore:1999,Klypin:1999}
which arises because in the cold dark matter 
($\Lambda$CDM) framework 
the number of simulated dark matter subhalos is much larger than the number
of observed dwarf galaxies \citep{Babul:1992,Bullock:2001}.
This picture seems consistent with recent observations
that have seen uniformly old stellar populations
in ultrafaint galaxies \citep{Brown:2014}.

This topic has been a very active area
of research in the past years, and
using cosmological simulations is key to trying to understand
these effects in their full context
\citep[][]{Quinn:1996,Thoul:1996,Gnedin:2000,Hoeft:2006,McQuinn:2007,Okamoto:2008,Noh:2014,BenitezLlambay:2015}.
More recent simulations with much
better resolution and a more complete feedback model
also seem to point in this 
direction \citep[][]{Onorbe:2015,Wheeler:2015a}.
In this context, the work by \citet{Simpson:2013} is
particularly relevant. Using an \textsc{enzo} high-resolution cosmological hydrodynamical simulation,
they found that turning on the UVB at $z=7.0$ versus $z=8.9$ resulted in
an order-of-magnitude change in the final stellar mass of a $10^{9}$ $\Msun$
dark matter halo.
In addition, some very interesting constraints are starting
to come from radiative transfer hydrodynamical simulations. 
\citet{Wise:2014} find that very
faint galaxies ($M_{UV}\sim-6$, $M_{*}\sim10^{3.59}$ 
in halos of $M_{h}=1\times 10^{7}$) will still form at high redshift
and contribute a significant
amount to the ionizing photon budget during cosmic reionization.
However, in order to avoid overproducing
the observed abundance of classical satellites
of the Milky Way, studies based on dark-matter-only simulations
argue for a critical mass closer to
$\sim 10^9$ $\Msun$
\citep[][]{Madau:2008,BoylanKolchin:2014}.

\begin{figure}
\begin{center}
\includegraphics[angle=0,width=0.48\textwidth]{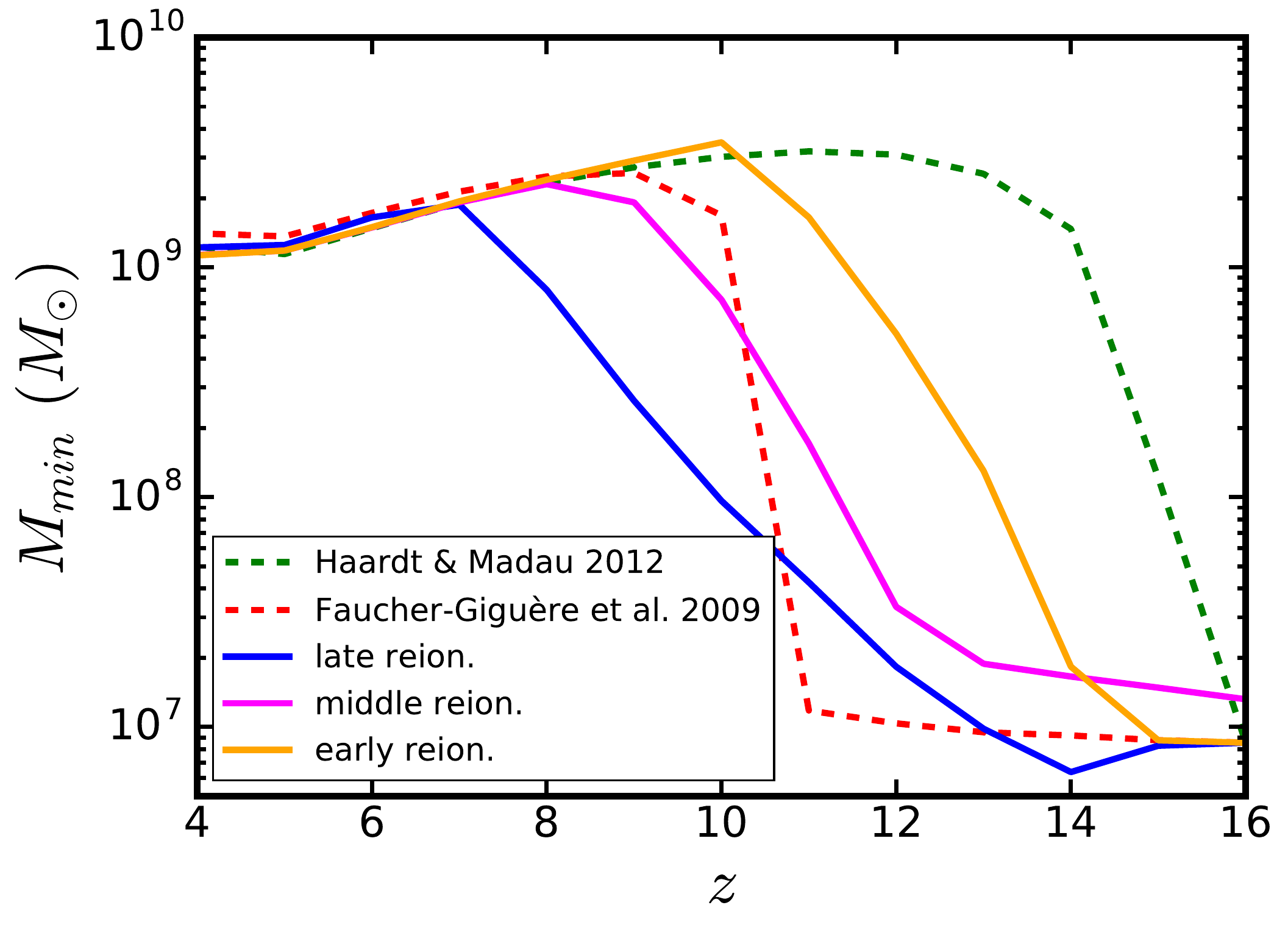} 
\end{center}
\caption{Expected cosmological Jeans mass of the neutral IGM
for the different thermal histories of HM12, FG09, LateR, MiddleR, and EarlyR simulations.
See text for details.
\label{fig:MJ}}
\end{figure}

For all this it is clear that the use of current standard UVB models that reionize and,
more importantly here, heat the IGM at a much higher redshift than was desired
will have a strong impact on the results of galaxy formation hydrodynamical simulations.
For example, simulations that use the HM12 UVB background start spuriously heating up
the IGM at $z\sim15$ (see Figure~\ref{fig:Qhistgas0}).
We can now perform a simple first analytical calculation to get an approximate idea of
this effect. In linear theory, the instantaneous cosmological mass cooling threshold 
of the neutral IGM  (sometimes also referred to as the Jeans mass) is
given by \citep[see, e.g.,][]{Iliev:2007}:
\begin{equation}
\begin{aligned}
& M_{\rm min}=3.9\times 10^{9} \Msun  \left(\frac{T_{\rm IGM}}{10^{4}K}\right)^{3/2}
\left(\frac{1+z}{10}\right)^{3/2} \\
& \qquad \left(\frac{\Omega_{\rm b}h^{2}}{0.0223}\right)^{-3/5}
\left(\frac{\Omega_{\rm m}h^{2}}{0.15}\right)^{-1/2}
\label{eq:Mcool}
\end{aligned}
\end{equation}
where $T_{\rm IGM}$ is the temperature of the IGM.
As has been found in the simulations mentioned above, the actual cooling mass 
differs somewhat from this instantaneous Jeans mass
since the mass scale on which baryons succeed in collapsing out of the IGM
along with the dark matter must be determined, even in linear theory, 
by integrating the differential equation for perturbation growth over
time for the evolving IGM \citep{Gnedin:1998}.
In fact, there is no single mass above which a collapsing halo retains all
its gas and below which the gas does not collapse with the dark matter.
Instead, simulations show that the cooled gas fraction in halos decreases 
gradually with decreasing halo mass \citep[e.g.][]{Okamoto:2008}.
Using eqn.~(\ref{eq:Mcool}) can give us a first estimate of the possible
effects of using different UVB models. Figure~\ref{fig:MJ} shows the results 
of this equation when we use the different thermal histories of the
HM12, FG09 as well as LateR, MiddleR, and EarlyR models.
The differences between the models are quite significant.

Different studies of the reionization redshift of collapse structures 
have shown that the median reionization redshift of halos
moves to lower redshift and the scatter increases
\citep{Weinmann:2007,Alvarez:2009,Li:2014}.
Although both parameters depend substantially on the details of reionization
\citep[see, e.g.][]{Ocvirk:2013}, massive halos, $>10^{15}$ $\Msun$,
can be reionized significantly earlier 
than the average region in the universe
\citep[$\Delta z_{\rm reion,50\%}\sim 2$, see Figure 2 of][]{Li:2014}.
For Milky Way halos ($\sim10^{12}$ $\Msun$, ), or dwarf 
galaxies ($\sim10^{10}$ $\Msun$)
the typical reionization redshift is expected to be much
lower ($\Delta z_{\rm reion,50\%} \sim 0.5$ and
$\Delta z_{\rm reion,50\%} \lesssim 0$ respectively),
making FG09 and HM12 models not optimal for these studies. This is 
even more true if one considers the last constraints on $\taue$ from \citet{Planck:2016a}.
By overestimating the heat at high $z$ we are not only
overestimating the pressure smoothing scale at lower redshifts but also overestimating
the effect of the UVB in galaxy formation and evolution\footnote{Notice that 
as mentioned above, some authors have tried
to solve for this issue by applying different simple redshift cutoffs
to the standard UVB models. Results of simulations using the cutoff approach can
be found in Appendix~\ref{app:HMold}.}.
For all these reasons, a detailed review of the results of galaxy formation simulations
using new consistent UVB models that fulfill all
observational constraints, like the ones developed here, is certainly needed. 
This is another reason why we make
the models presented in this work 
publicly available to the community (see Appendix~\ref{app:tables}) and encourage colleagues
to adopt them in future galaxy formation work and revisit previous calculations.

\section{Conclusions} \label{sec:conc}

%
In this paper we have presented results from optically thin 
cosmological hydrodynamical simulations using the Nyx code
\citep{Almgren:2013,Lukic:2015}. As commonly done in multiple IGM and galaxy
formation studies, the UV background
is modeled as a uniform and isotropic field that evolves with redshift.
Operationally, the UVB determines the photoioinization and photoheating
rates of $\HI$, $\HeI$ and $\HeII$, which are inputs to the code.


We have demonstrated that when canonical models of the UVB,
like that of HM12, are used in hydrodynamical simulations, the ionization of the IGM and,
more importantly, the concomitant heating occur far too early, inconsistent
with the reionization histories calculated by the respective authors, and in violation of
current observational constraints on reionization. We argue that this results from the
fact that these models dramatically overestimated the mean free path of ionizing photons at
root at $z > 5$, resulting from the blind extrapolation
of a model fit to lower redshift ($z < 5$) measurements.
As a result,
the amplitudes of the photoionization and photoionization heating
rates are far too high at $z > 6$.
This premature heating spuriously heats the IGM to $\sim 10^4$ K by $z \sim 13$,
and because the IGM gas pressure smoothing scale depends on the full thermal history,
it produces an erroneously large pressure smoothing scale at nearly all redshifts.
We argue that a correct and consistent model of the reionization and
thermal history is crucial for obtaining the correct pressure scale in
simulations, which is necessary for interpreting \lyalpha{} forest
statistics at $z<6$ -- not doing so can bias estimates of the
thermal state of the IGM. We also discussed the implications of this spurious
early heating on galaxy formation simulations.

Motivated by these issues, we have developed a new method to generate
UVB models for hydrodynamical simulations that allow one to
self-consistently simulate different reionization models. We implement
this by volume-averaging the photoionization and energy equations. In
the model, each reionization event is defined by the ionization
history with redshift and the total heat input of the reionization
event.  In this sense our new models provide a very promising tool to
explore the parameter space of possible ionization and thermal
histories.  In this work, we investigated models in which we changed
the redshift at which $\HI$ reionization ends and the amount of heat
input associated with both $\HI$ and $\HeII$ reionization.  We studied
the effect of these changes on the thermal history of the IGM, in
particular the temperature at mean density, $T_{0}$, the slope of the
temperature-density relation, $\gamma$, the temperature at the
optimal density probed by curvature measurements, $T(\Deltaopt)$, and
the gas pressure smoothing scale, $\lambda_{P}$.  We have shown how important
the degeneracies between these parameters can be in order to derive
the thermal parameters of the IGM using the curvature \lyalpha{}
statistic. 
These rates have also been corrected to
improve the agreement with measurements of the average $\HI$ and $\HeII$ transmission
after reionization.

The UVB plays a fundamental role in 
determining the star formation of the first galaxies and their evolution
by not only setting the minimum halo mass able to form stars
but also regulating the gas accretion from the IGM into more massive halos.
Previous studies utilizing UVB models that suffer from the spurious early heating
described in this paper have thus overestimated the effect of this photoheating feedback and
the resulting suppression of star-formation at high redshift. 
We therefore argue that galaxy formation simulations should be revisited using our
new UVB models.

We make our new UVB models publicly available so that the 
community can better explore the consequences and effects
of different ionization and thermal histories in all types
of hydrodynamical cosmological simulations. 
Tables with the photoionization and photoheating rates of the new
models can be found in Appendix~\ref{app:tables}, which are in the
the standard ``TREECOOL'' file format, ready and easy to use with most cosmological codes, including
\textsc{gadget}, \textsc{arepo}, and \textsc{gizmo}.
We encourage anyone interested in implementing
some other specific UVB model to contact the authors.
We will also be happy to provide help incorporating
these models into other codes by request.

Like all optically thin simulations,
our approach misses UVB fluctuations that could produce scatter
in reionization times and temperatures between different regions of the
universe
\citep[e.g.][]{Abel:1999,Meiksin:2004,Pontzen:2014,GontchoaGontcho:2014,Davies:2014,Malloy:2015,DAloisio:2015,Davies:2016b}.
Of course, an immediate solution to this problem
in optically thin simulations
will be to add
some extra dependence on a specific property of each resolution element in the simulation
(e.g. density, distance from a halo that could host a galaxy/quasar, etc).
However, the computational challenge is to try to do this without making the simulation
prohibitively expensive. In this sense, an interesting solution that it
is worth exploring will be to assign a specific reionization redshift to
each resolution element of the simulation using, for example, 
an excursion set formalism \citep[e.g.][]{Furlanetto:2004}.
In this picture, a resolution element will not see the UVB background until its reionization
redshift. We plan to pursue this idea in the near future.
Another very valuable piece of information that would improve current
optically thin hydrodynamical simulations
would be if future radiative transfer simulations
\citep[e.g.][]{So:2014,Gnedin:2014,Pawlik:2015,Norman:2015,Ocvirk:2015}
were to make the probability distribution function
of their photoionization rates publicly available (and perhaps
also its dependence on density), and not just the
evolution of the mean and/or median values.

Finally, our new parameterization for the heating and ionization
produced by the UVB
allows us to explore a broader range of reionization models,
as well as any other physical scenarios that could alter the thermal
history of the IGM. This will allow us to better test
the effect of such models in simulations of galaxy formation and
the IGM. 
These models could include Population III
stars \citep{Manrique:2015}, X-ray pre-heating
coming from from starburst 
galaxies, supernova remnants, or miniquasars 
\citep{Oh:2001,Glover:2003,Madau:2004,Furlanetto:2006},
dark matter annihilation or decay \citep[][]{Liu:2016}
or cosmic rays \citep{Samui:2005}, from the
intergalactic absorption of blazar TeV photons \citep{Chang:2012,Puchwein:2012},
or from broadband intergalactic dust absorption \citep{Inoue:2008}.
We expect more detailed studies on these physically
motivated models in the future.

\acknowledgments

We thank the members of the ENIGMA group
at the Max Planck Institute for Astronomy (MPIA) for helpful
discussions.
J.F.H. acknowledges generous support from the Alexander von Humboldt
foundation in the context of the Sofja Kovalevskaja Award. The 
Humboldt foundation is funded by the German Federal Ministry for
Education and Research.
Z.L. was supported by the Scientific Discovery through Advanced
Computing (SciDAC) program funded by U.S.\ Department of Energy Office of
Advanced Scientific Computing Research and the Office of High Energy Physics.
Calculations presented in this paper used the \textsc{hydra} cluster of the
Max Planck Computing and Data Facility (MPCDF, formerly known as RZG)
MPCDF is a competence center of the Max Planck Society located 
in Garching (Germany). 
We also used resources of the National Energy Research
Scientific Computing Center (NERSC), which is supported by the
Office of Science of the U.S. Department of Energy under 
Contract no. DE-AC02-05CH11231.
The ASCR Leadership Computing Challenge (ALCC) program
has provided NERSC allocation under the project
``Cosmic Frontier Computational End-Station''.
This work made extensive use of
the NASA Astrophysics Data System and of the astro-ph preprint
archive at arXiv.org.

\appendix

\section{Volume-averaged Ionization and Heating Equations} \label{app:volave} 

In this paper we have presented
a new way of obtaining effective photoionization and photoheating rates
for different reionization models.
These can be used in optically thin hydrodynamical simulations to account for
emission of the galaxies and quasars.
Here we provide detailed derivation of these rates for future reference.

We remind the reader that the new effective
rates are only computed during reionization, i.e.,
while $\mean{x_{\HII}}<1$. After reionization, our models use the rates from
common UVB models (e.g. FG09, HM12, see Section~\ref{sec:meanflux}).
New values of the photoionization rates during reionization are forced
to never exceed the values of the model plugged after reionization.
This is done to guarantee no numerical
artifacts in the limit where reionization is almost complete, $\mean{x_{\HII}}\sim1$,
where our new rates are not well defined.

\subsection{Volume Average Optically Thin Ionization Equations} \label{ssec:volave} 

In the context of optically thin hydrodynamical codes, 
it is often assumed that the gas is of the primordial chemical composition, 
where the resulting reaction network includes six atomic species:
$\HI$, $\HII$, $\HeI$, $\HeII$, $\HeIII$ and e$^-$.  Codes generally evolve those species under the
assumption of ionization equilibrium
\citep[see, however][for nonequilibrium treatments]{Gnat:2007,Vasiliev:2011,Oppenheimer:2013,Richings:2014a,Richings:2014b,Puchwein:2015}.
The resulting system of algebraic equations is:
\begin{equation}
  \begin{aligned}
     & \left( \Gamma_{\rm e, \HI} \nel + \Gamma_{\gamma, \HI} \right) \nHI
      = \alpha_{\rm r, \HII} \nel \nHII \\[1.5mm]
     & \left( \Gamma_{\rm e, \HeI} \nel + \Gamma_{\gamma, \HeI} \right) \nHeI
      = \left( \alpha_{\rm r, \HeII} + \alpha_{\rm d, \HeII} \right)
      \nel \nHeII \\[1.5mm]
     & \left[ \Gamma_{\gamma, \HeII} + \left(\Gamma_{\rm e, \HeII}
      + \alpha_{\rm r, \HeII} + \alpha_{\rm d, \HeII} \right)
      \nel \right] \nHeII \\[1.5mm]
     & \qquad = \alpha_{\rm r, \HeIII} \nel \nHeIII
      + \left( \Gamma_{\rm e, \HeI} \nel + \Gamma_{\gamma, \HeI} \right)
      \nHeI
  \end{aligned}
  \label{eq:photo}
\end{equation}
In addition, there are three closure equations for the conservation of charge
and hydrogen and helium abundances. Radiative recombination
($\alpha_{\rm r, X}$), dielectronic recombination ($\alpha_{\rm d, X}$), and
collisional ionization ($\Gamma_{\rm e, {\rm X}}$) rates are strongly dependent on
the temperature, which itself depends on the ionization state through the mean
mass per particle $\mu$
\begin{equation}
  T = \frac{2}{3} \frac{m_p}{k_{\rm B}} \mu\ e_{\rm int}
\end{equation}
where $m_p$ is the mass of a proton, $k_{\rm B}$ is the Boltzmann
constant and $e_{\rm int}$ is the internal thermal energy per mass of
the gas. For a gas composed of only hydrogen and helium, $\mu$ is related to the
number density of free electrons relative to hydrogen by 
$\mu=(1+4\chi)/[1+\chi+(\nel/\nH)]$.
The reaction network equations are iteratively solved 
together with the
ideal gas equation of state, $p = 2\rho e_{\rm int}/3$, to
determine the temperature and equilibrium distribution of species.
Above and throughout this paper we have assumed an adiabatic index of $5/3$.

In order to produce consistent UVB models that reliably reproduce
different reionization histories, we have derived the
volume-averaged version of the ionization equilibrium equations
presented in eqn.~(\ref{eq:photo}).
We start with the $\HI$ reionization
and derive here in detail the equation for $\HI$ photoionization.
We will address the $\He$ single
and double reionization afterward, as the method is very similar.
We start by doing the volume-averages of eqn.~(\ref{eq:photo}) for $\HI$,
 \begin{equation}
   \begin{aligned}
      & \mean{\Gamma_{\rm e, \HI} \nel \nHI} + \mean{\Gamma_{\gamma, \HI} \nHI}
       = \mean{\alpha_{\rm r, \HII} \nel \nHII} 
   \end{aligned}
   \label{eq:photomean1}
 \end{equation}
The first thing is that collisional ionization terms, $\Gamma_{\rm e}$, are mainly relevant in shocks
at high temperatures and densities. They will have a negligible effect 
on this volume-averaged calculation, so we can discard them.
The important point here is that each term in this equation is nonlinear;
hence, in principle, it is not possible to compute their volume
averages unless the cross-correlation of the abundances
of $\nHI$ and $\nHII$ with each other, as well as with the radiation
and temperature fields are known (since $\Gamma_{\rm e,\HI}$ and 
$\alpha_{\rm r, \HII}$ are temperature-dependent). 
A convenient way to encapsulate this
unknown information is using correction factors.
With all this we can write the volume-averaged equivalent of
eqn.~(\ref{eq:photo}) as
\begin{equation}
  \begin{aligned}
     & C_{\gamma,\HI} \mean{\Gamma_{\gamma, \HI}} \mean{\nHI} 
     = C_{\rm r,\HII}\mean{\alpha_{\rm r, \HII}} \mean{\nel} \mean{\nHII} 
  \end{aligned}
  \label{eq:photomean2}
\end{equation}
where the correction factors are defined as
\begin{equation}
  \begin{aligned}
     & C_{\gamma,\HI}=\frac{\mean{\Gamma_{\gamma, \HI} \nHI}}{\mean{\Gamma_{\gamma, \HI}} \mean{\nHI} } \\[1.5mm]
     & C_{\rm r,\HII} =\frac{\mean{\alpha_{\rm r, \HII} \nel \nHII}}{\mean{\alpha_{\rm r, \HII}} \mean{\nel} \mean{\nHII}} 
  \end{aligned}
  \label{eq:clumping}
\end{equation}
Then, we can rewrite eqn.~(\ref{eq:photomean2}) as
\begin{equation} 
  \begin{aligned}
     & \mean{\Gamma_{\gamma, \HI}}  
      = \frac{C_{\rm r,\HII}}{C_{\gamma,\HI}}
      \frac{\mean{\alpha_{\rm r, \HII}} \mean{\nel} \mean{\nHII}}{\mean{\nHI}}  \\[1.5mm]
  \end{aligned}
  \label{eq:photomean3}
\end{equation}

Now, in order to obtain an $\HI$ photoionization rate, we will
make use of the following assumptions:
1) We assume that the $\HeI$ reionization occurs perfectly coupled
with the $\HI$ reionization process. This is a very common assumption in
reionization models because ionization of hydrogen and the first
ionization of helium require photons with similar energies 
(13.6 eV and 24.6 eV respectively). Therefore, the same physical
process responsible for $\HI$ reionization must be also liable for
the $\HeI$ reionization.
We also consider that the  the helium second ionized state number 
density is negligible during $\HI$ and $\HeI$ reionization, 
$\nHeIII\sim 0$, so  $\mean{x_{HeIII}}=0$.
This is a correct assumption for all the models discussed in this work 
but could be easily modified in the future if needed.
2) The evolution of number density of free electrons, $\nel$,
can be approximated by
$\nel=\nHII +\nHeII + 2\nHeIII$. 
We can rewrite this equation
as a function of the hydrogen density and the ionized fractions:
$\nel=\nH\left[(1+\chi)x_{\HII} +\chi x_{\HeIII}\right]$
where $\chi=Y_{\rm p}/4X_{\rm p}$ and we have assumed again that 
$\HeII$ reionization follows that of $\HII$ one.
The volume-averaged value can be written as
\begin{equation}
\begin{aligned}
   \mean{\nel}=\mean{\nH}\left[C_{x_{\HII}} \left(1+\chi\right) \mean{x_{\HII}} +
  C_{x_{\HeIII}}\chi \mean{x_{\HeIII}}\right] 
  \label{eq:elecden}
\end{aligned}
\end{equation}
where $C_{x}$ factors stand for the correction factors defined as
$C_{x_{\HI}}\mean{\nH}\mean{x_{\HI}}=\mean{\nHI}$, $C_{x_{\HII}}\mean{\nH}\mean{x_{\HII}}=\mean{\nHII}$, etc.
As we are assuming that $\HeII$ is not relevant during $\HI$ reionization,
we can discard the second term inside parentheses for our current 
derivation.
Finally, putting together eqn.~(\ref{eq:photomean3}) with eqn.~(\ref{eq:elecden}), we arrive at
\begin{equation}
\mean{\Gamma_{\gamma, \HI}}(z)= C_{\HII} \mean{\nH}(z) 
\alpha_{\rm r, \HII}(\mean{T})(1+\chi) \frac{\mean{x_{\HII}}^{2}(z)}{\mean{x_{\HI}}(z)} \\[1.5mm]
\label{eq:invphotoA}
\end{equation}
where $C_{\HII}$ encapsulates all correction parameters described 
above and can be written as
\begin{equation}
 C_{\HII}=\frac{C_{\alpha_{\rm r,\HII}}\times C_{\rm r,\HII}\times C_{x_{\HII}}^{2}}{C_{\gamma,\HI} \times C_{x_{\HI}}}
 \label{eq:clumpingfinal}
\end{equation}
where $C_{\alpha_{\rm r,\HII}}$ is defined as 
$C_{\alpha_{\rm r,\HII}}=\mean{\alpha_{\rm r, \HII}}/\alpha_{\rm r, \HII}(\mean{T})$
to clarify that in general the volume-averaged recombination factor 
is redefined as the recombination factor at a certain mean temperature value.
It is customary to use the temperature at mean density,
$T_{0}$ as this value.
In the case of optically thin hydrodynamical simulations a constant photoionization
rate is used throughout the whole volume, so $C_{\gamma}=1$.

In general, the unknown ratio of the IGM's true recombination rate to its
hypothetical rate under the assumption of uniform density 
and temperature 
is often referred to as the IGM clumping
factor ($C_{\rm IGM}=C_{\rm r,\HII}$).
Notice also that, if for eqn.~(\ref{eq:clumping}) we 
assume that $\nel=\nHII$ and that $\alpha_{\rm r}$ is a constant, it is easy to redefine the IGM clumping in
a much more familiar form: 
$C_{\rm IGM}=\mean{\nHII^2}/\mean{\nHII}^{2}$
\citep{Kohler:2007,Pawlik:2009,Finlator:2012,Kaurov:2014}.

We can also obtain analogous equations for the 
volume-averaged helium photoionization rates 
from  eqn.~(\ref{eq:photo})
to the one we obtained above, eqn.~(\ref{eq:invphotoA}),
for the hydrogen. We have just made one extra assumption,
which is that $\nHeIII$ is negligible during the single reionization of helium
and that during the double reionization of helium $\nHeI$ is negligible.
In the context of all the $\HeII$ reionization models discussed in this paper
this is a fair assumption.
For the $\He$ first ionization we obtain
\begin{equation}
\mean{\Gamma_{\gamma, \HeI}}(z)= C_{\HeII} \mean{\nH}(z) 
\alpha_{\rm r, \HeII}(\mean{T})(1+\chi) \frac{\mean{x_{\HeII}}^{2}(z)}{\mean{x_{\HeI}}(z)} \\[1.5mm]
\label{eq:invphotoHeI}
\end{equation}
and $C_{\HeII}$ is analogous to $C_{\HII}$,
\begin{equation}
C_{\HeII}=\frac{C_{\alpha_{\rm r,\HeII}}\times C_{\rm r,\HeII}\times C_{x_{\HeII}}^{2}}{C_{\gamma,\HeI} \times C_{x_{\HeI}}}.
\label{eq:clumpingHeII}
\end{equation}
For the $\He$ double ionization,
\begin{equation}
\begin{aligned}
& \mean{\Gamma_{\gamma, \HeII}}(z)=\frac{1}{C_{\gamma,\HeI}} \mean{\nH}(z) 
[C_{x_{\HII}}(1+\chi) + \chi C_{{x_{\HeIII}}}\mean{x_{\HeIII}}(z)]\\
& \quad \left[C_{\alpha_{\rm r,\HeIII}}\times C_{\rm r,\HeIII}\frac{\alpha_{\rm r, \HeIII}(\mean{T})\mean{x_{\HeIII}}(z)}{\mean{x_{\HeII}}(z)} 
-C_{\alpha_{\rm r,\HeII}}\times C_{\rm r,\HeII} \times \alpha_{\rm r, \HeII}(\mean{T})\right] \\[1.5mm]
\end{aligned}
\label{eq:invphotoHeII}
\end{equation}
In order to describe the reionization history of one model,we will 
need to specify two functions,
the evolution of $\HII$ and $\HeIII$ volumen-averaged ionization
fractions, $\mean{x_{\HII}}(z)$ and $\mean{x_{\HeIII}}(z)$.
As $\mean{x_{\HI}} + \mean{x_{\HII}}=1$  and 
$\mean{x_{\HeI}} + \mean{x_{\HeII}} +\mean{x_{\HeIII}}=1$ and
we assume that $\HeI$ reionization is totally coupled 
with $\HI$ reionization.

To calculate the average values of the
radiative recombination rates ($\alpha_{\rm r, i}$)
which depend on temperature, we need to specify the 
evolution of the volume-averaged temperature, $\mean{T}(z)$ in our model.
To do this, we first need to define the total heat input 
produced by each reionization event, 
$\Delta T_{\HI}$ and $\Delta T_{\HeII}$
which will be two free parameters in our reionization models.
We describe in the next paragraph the recent efforts to calculate these
parameter from theoretical models.
Thus, we will make another assumption in our model
to describe the evolution of the volumen-averaged temperature.
3) That the evolution of the volumen-averaged temperature 
can be well approximated with the evolution of the reionization history times
the total heat input, i.e,
$\mean{T}_{\HI}(z)=\Delta T_{\HI}\times \mean{x_{\HII}}(z)$ and  
 $\mean{T}_{\HeII}(z)=10^{4}+\Delta T_{\HeII}\times x_{\HeIII}(z)$.
This is a very rough estimation of the thermal history, as we
are neglecting all cooling, but it serves
as a first-order approximation of the gas temperature during reionization
in order to compute the
different rates that we need to compute eqn.~(\ref{eq:invphotoA}).
We will show below that this assumption is accurate enough for 
our purposes and for the range of models covered in this work. 
More elaborate assumptions, including
analytic estimations of the Compton cooling, 
could be implemented in the future.

Finally, to generate the models of this work, we used the following correction factors. For
the $\HI$ and $\HeI$ photoionization rates we used $C_{\HII}=C_{\HeII}=1.5$ for $z\geq10$
and $C_{\HII}=C_{\HeII}=2.0$ for $6<z<10$. Notice that we never go below redshift $z=6$
when we compute the $\HI$ rates, and that these values are in good agreement
with results from radiative transfer simulations \citep[clumping factor, $C_{100}$, of][]{Pawlik:2009}
in this range of redshifts.
For the $\HeII$ photoionization rates we used $C_{\HeIII}=1.5$ at all redshifts, which seems
to allow us to accurately recover the input ionization models.

\subsubsection{First-order Estimation of the Ionization History in Hydrodynamical Simulations from the Photoionization Rates} 
\label{app:hifrac}

Using the relation between volume-averaged quantities derived above,   
one can obtain an analytical expectation of the volume-averaged $\HI$ ionization fraction, $\mean{x_{\HI}}$, 
evolution in an optically thin simulations once a certain photoionization rate is assumed.
This can be useful to predict the approximate ionization history that a certain photoionization rate model
will produce in a optically thin hydrodynamical simulation.
In particular, for the $\HI$ photoionization rate we can rewrite eqn.~(\ref{eq:invphotoA}) above as
\begin{equation}
 C_{\HII} \mean{\nH}(z) \alpha_{\rm r, \HII}(\mean{T})(1+\chi) (1-C_{x,\HI}\mean{x_{\HI}}(z))^{2}
 - C_{x,\HII}^{2}\mean{\Gamma_{\gamma, \HI}}(z)\mean{x_{\HI}}(z)=0 \\[1.5mm]
\end{equation}
where now we are using that $C_{x,\HII}\mean{x_{\HII}}=1-C_{x,\HI}\mean{x_{\HI}}$.
If we define $A=C_{\HII} \mean{\nH}(z) \alpha_{\rm r, \HII}(\mean{T})(1+\chi)$, we can rewrite this equation as
\begin{equation}
A\mean{x_{\HI}}^{2}(z) - [2A+\mean{\Gamma_{\gamma, \HI}}(z)]\mean{x_{\HI}}+A=0
\end{equation}
Solving for the volume-averaged $\HI$ ionization fraction, we get
\begin{equation}
\mean{x_{\HI}}=1-\mean{x_{\HII}}=\frac{ 2A+\Gamma_{\gamma,\HI} \pm \sqrt{\Gamma_{\gamma,\HI}^{2}+4A\Gamma_{\gamma,\HI}} } {2A}
\label{eq:hifrac}
\end{equation}
which gives us an estimation of the expected hydrogen neutral fraction once we discard the nonphysical solution.
It is expected that $C_{\HI}$ changes between simulations that implement very different physical processes.
Therefore, this formula could also be used to obtain the value of $C_{\HI}$ in optically thin hydrodynamical simulations
using one run and then compute analytically the expected results with different UVB models.

\begin{figure}
\begin{center}
\includegraphics[angle=0,width=0.48\textwidth]{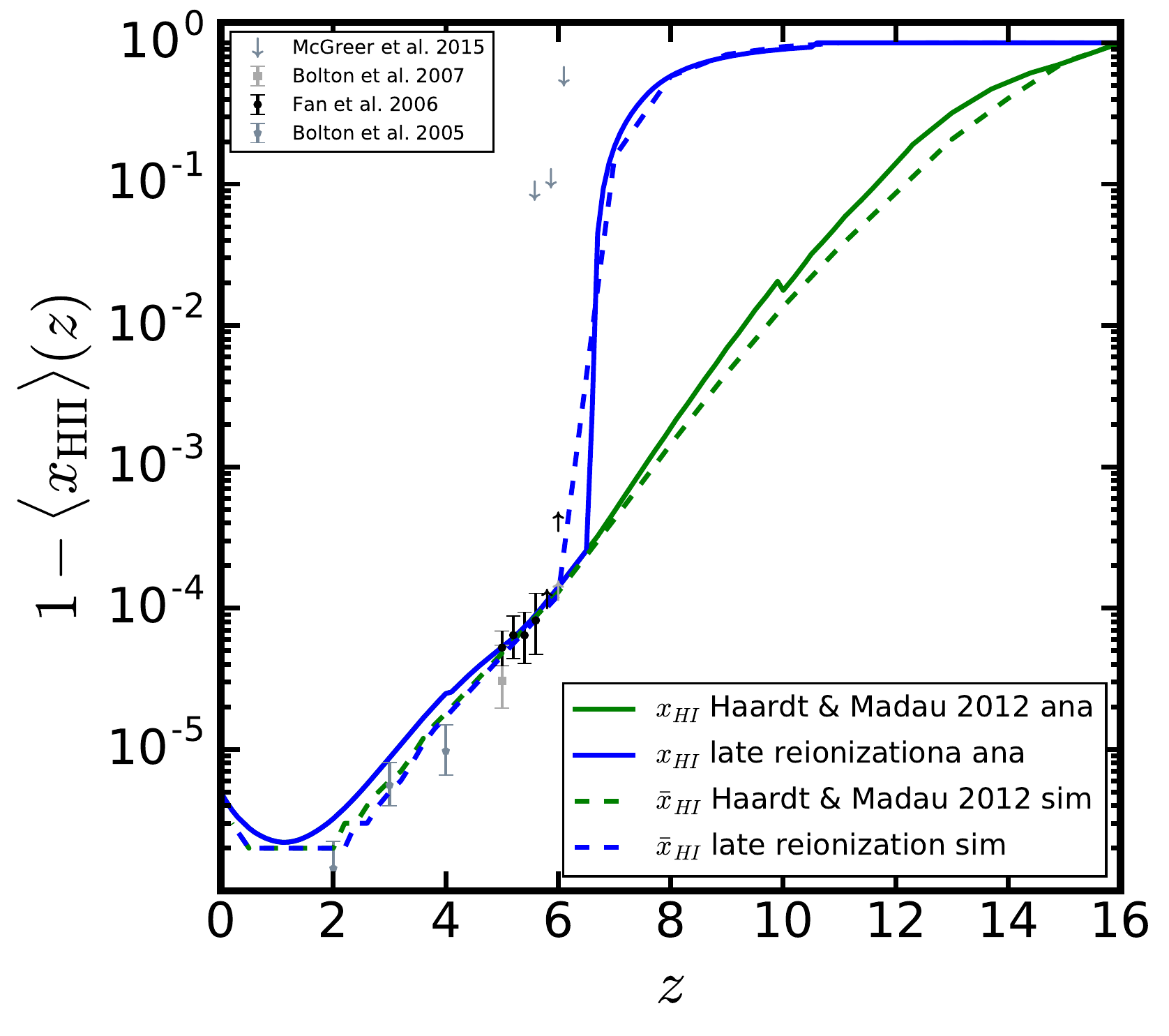}
\end{center}
\caption{Evolution of the volume-averaged $\HI$ fraction, $\mean{x_{\HI}}=1-\mean{x_{\HII}}$,
obtained in the hydrodynamical simulations (dashed lines) vs.
the expected evolution using the volume-averaged analytical prediction
given by
eqn.~(\ref{eq:hifrac}). See text for more details.
\label{fig:xhiana}}
\end{figure}

In Figure~\ref{fig:xhiana}
we plot the result obtained from eqn.~(\ref{eq:hifrac}) using
the photoionization rate from the HM12 
model 
(green line) and the late reionization model (blue line). We assumed $T=2\times10^{4}$ K at all redshifts, and
we used the clumping given by \citet[][their fit to $C_{100}$]{Pawlik:2009}.

\subsection{Volume-averaged Heating Equations} \label{ssec:dTdz}

The next step is to include in a consistent way the heating occurring
during different reionization epochs. To do this, we have 
implemented a variation of an the idea suggested
by FG09.
In this work the authors also raised the problems pointed out 
in the previous section of incorporating the effects
of a prescribed UVB in cosmological hydrodynamical simulations
under the assumption of an optically thin plasma. 
Their work was motivated by $\HeII$ reionization, but the idea
can be well applied for any reionization event.
They proposed as a more physically motivated approach to
increase the temperature of each  gas  element  by  an  amount
$d \Delta T_{\HI} (z) / dz$, (which  is subsequently allowed to cool)
at each time step $\Delta z$ in the simulation.\footnote{
This method injects the same amount of heat regarding of 
the density at each resolution element and in fact, we 
have also implemented it in our code and did several tests.
This creates a very different evolution
of the temperature-density relation slope, $\gamma$, with redshift.
The evolution for the temperature at
mean density, $T_{0}$, was the same regardless of the method. 
This method has the advantage of only requiring an additional 
heating term and adding negligible computational 
overhead while capturing the timescale and magnitude of  
the heat input more realistically. 
Although we might explore it in more detail in the future, we decided
to change to our current approach because it
implied changing the standard algorithm
used in several state-of-the-art cosmological hydrodynamical codes.
Our current method produces new UVB models 
that can be directly plugged into any of these codes.}
Here $\Delta T_{\HI} (z)$ is the total heat input or 
cumulative temperature increase evolution
owing to reionization and is precomputed given
the desired $\HI$ reionization history.
Both the total heat input from $\HI$ reionization and from $\HeII$
reionization will be free parameters in our model. We discuss the specific
values used in this work in Section~\ref{sec:reionmodels}.

Now, we want to go one step further from the
$d\Delta T/dt$ idea, and relate this change in temperature
to a specific photoheating rate, $\dot{q}$.
This will allow the use of our new models in 
standard hydrodynamical codes.
To do this we can, again, volumen-averaged the relation between the heat per unit
of time in one cell produced by the $dT/dz$ model and the one produced by a photoheating rate.
The change of internal energy density due to a certain change in temperature of the gas can be related
to a new photoheating rate in the following way:
\begin{equation}
 \frac{de_{int}}{dt}=\frac{3 k_{\rm B}}{2 m_{\rm p}\mu}\frac{d\Delta T_{\HI}}{dt}=\frac{\nHI\dot{q}_{\HI}}{\rho_{\rm proper} }
 \label{eq:dtdzq}
\end{equation}
Using the same assumptions considered to obtain the new photoionization
rates, 
we can get a volume-averaged value for the heating rate. We need also
to assume that the volume-averaged molecular weight is 
$\mean{\mu}=(1+ 4\chi)/\left(1+\chi + \mean{\nel/\nH}\right)$.
Then we obtain
\begin{equation}
\dot{q}_{\HI}=C_{\dot{q},\HI}\frac{ 3 k_{\rm B}}{2 \mean{\mu} X_{\rm p} \mean{x_{\HI}}(z)} \frac{d\Delta T_{\HI}}{dt} 
\label{eq:qdotA}
\end{equation}
where $C_{\dot{q},\HI}$ is a correction factor defined as
\begin{equation}
\langle \frac{\mu}{x_{\HI}} \frac{d\Delta T_{\HI}}{dt} \rangle = C_{\dot{q},\HI} \frac{\mean{\mu}}{\mean{x_{\HI}}}\mean{ \frac{d\Delta T_{\HI}}{dt}} 
\label{eq:Cqdot}
\end{equation}
Note that this approach sets $\dot{q}_{\HeI}=0.0$ as we include
the heating produced by $\HeI$ reionization in the $\HI$ heating rate.
An identical approach is used to obtain the \HeII{} heating 
rate, $\dot{q}_{\HeII}$. To compute these values in our models, we have used a
correction factor, $C_{\dot{q}}=1$.

Therefore, in our model, once a total heat input due to reionization ($\Delta T_{\HI}$)
is chosen, the exact photoheating rates will depend on the assumption 
made on $d\Delta T_{\HI}/dz$.
In order to simplify this, we assume that the evolution of the total heat input can be well
approximated by the volume-averaged ionization fraction
evolution: $d\Delta T_{\HI}/dt\sim d\mean{x_{HII}}/dt$.
From here we can derive the derivative of the total heat input 
evolution with redshift, which is used in eqn.~(\ref{eq:qdotA}) to
obtain the tabulated photoheating rates that will
go into the code. For this reason in our
reionization models, the thermal history is defined based on one 
free parameter, the total heat input $\Delta T$, and one free
function, the ionization history, which in this context
we define as the volumen-averaged ionization fraction 
evolution, $\mean{x_{\HII}}(z)$.
In the case of $\HI$ reionization the heating term is defined as
\begin{equation}
\left|\frac{d\Delta T_{\HI}(z)}{dz}\right| =  \frac{\Delta T_{\HI} \left| 
\frac{d\mean{x_{\HII}}(z)}{dz} \right|}{\int_{z_{\rm reion}}^{z_{\rm recomb}} \left| \frac{d\mean{x_{\HII}}(z)}{dz} \right| dz}
\label{eq:dTdz}
\end{equation}
The denominator factor is a normalization to guarantee 
that the integrated amount of heating injected at each time step corresponds
to the total heat input.

\section{Ionization and Thermal Histories for Other UVB Models} \label{app:HMold} 

\begin{figure*}
\begin{center}
\includegraphics[angle=0,width=0.32\textwidth]{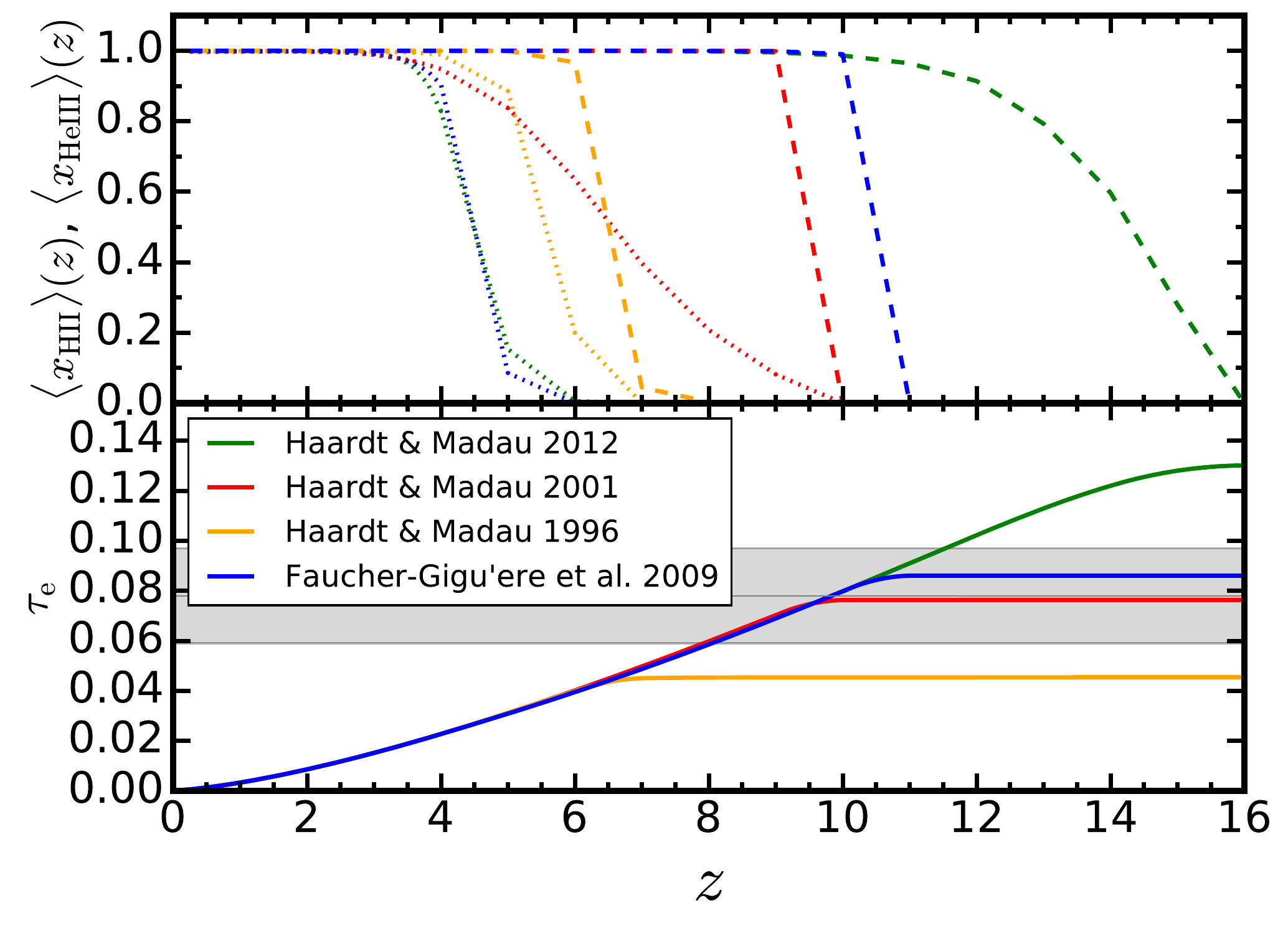}
\includegraphics[angle=0,width=0.32\textwidth]{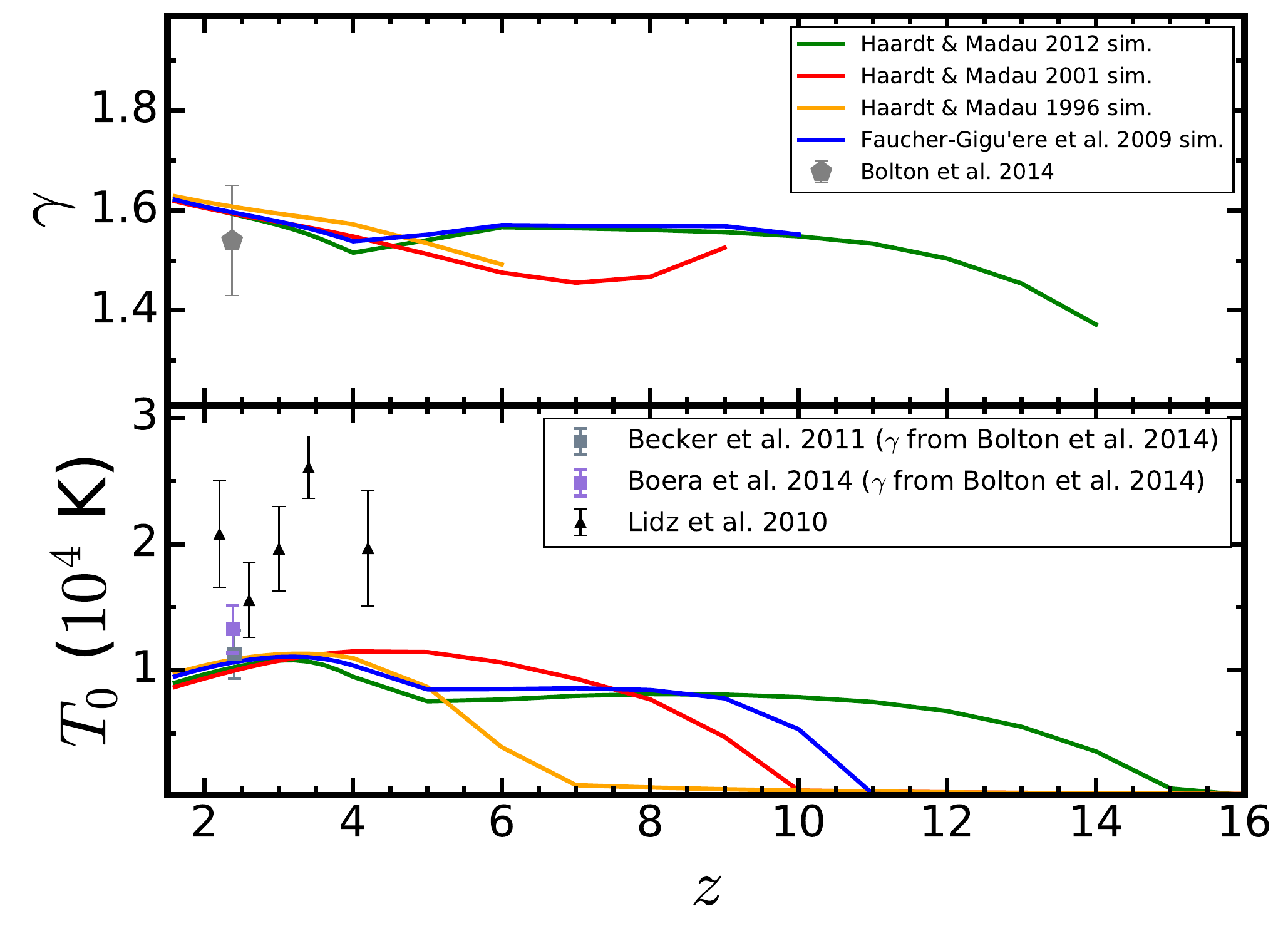}
\includegraphics[angle=0,width=0.32\textwidth]{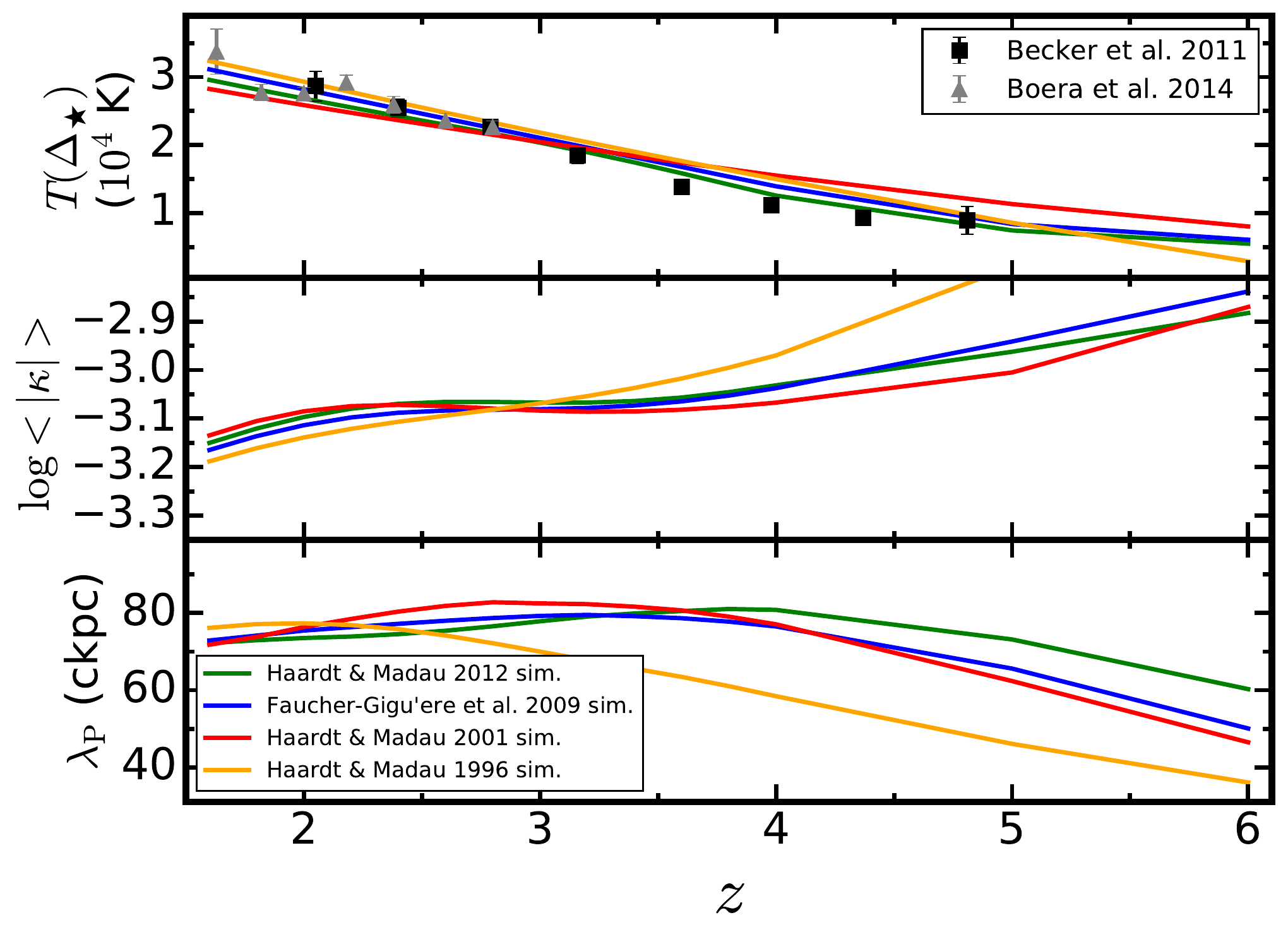}
\end{center}
\caption{Results for simulations using different common UVB models:
HM96, HM01, FG09, HM12. Left panel: ionization history.
Middle panel: the thermal history parameterized by the slope
of the density-temperature relation, $\gamma$, and the temperature
at mean density, $T_{0}$.
Right panel: evolution of the temperature at the optimal density
(upper), the curvature (middle) and the gas pressure smoothing scale (lower).
\label{fig:hmold}}
\end{figure*}

Here we present the ionization and thermal histories
produced by other UVB models, widely used in the literature.
We have run simulations using the following UVB models
(HM96, HM01, FG09),
in addition to the HM12 
model largely discussed in this work.
The left panel of the Figure~\ref{fig:hmold} shows the ionization history for all these models.
It is particularly interesting to first focus on the results
of the FG09 
which use the same approximation as HM12 (Eq.~\ref{eq:Q})
to calculate the reionization redshift of their model
that obtained complete $\HeII$ reionization by $z\sim3$,
and $\HI$ reionization by $z=6$. 
However, it is clear that this model is producing a much higher $\HI$ reionization,
more close to $z\sim10$, 
indicating that they have the same problems as the HM12
prescription. In fact, the method to compute their expected ionization history is the same 
as the one used by HM12. 
In this case the effect is not as extreme, only because tabulated values start at 
a lower redshift and, although they produce what seems to be a very high redshift
$\HI$ reionization, the final outcome of the model is still within CMB
observational constraints \citep{Planck:2015}\footnote{During the making of
this paper, new constraints
on reionization from Planck were published \citep{Planck:2016a}, moving
these constraints to a lower value and reducing the errors: $\taue=0.058\pm 0.012$.
The HM01 and FG09 models are in clear disagreement at $1\times\sigma$ with these new
constraints.}.
We do not have any information
on what were the expected ionization histories for the other two older Haardt \&
Madau models, but it is still very interesting to show their reionization histories,
as they are still used in the literature.
First of all, notice that in these models, as well as in the FG09 run, the $\HI$ reionization
is described as a simple step function. 
The HM01 model assumes a much more shallower $\HeII$ reionization history
than any other model considered in this work, starting as early as $z\sim10$.
This will allow us to 
see the effect of these types of models on the thermal history.
HM96 $\HI$ reionization happens very late, and therefore this model
does not fulfill \citet{Planck:2015} CMB constraints\footnote{Notice, however, that
it is in good agreement with the new \citet{Planck:2016a} constraints.}.

The middle and right panels of Figure~\ref{fig:hmold} illustrate the thermal histories
of these models. The first thing that we want to point out is the effect of a shallower
$\HeII$ reionization model on the HM01 
model. This basically produces
a smoother evolution of the temperature at mean density, $T_{0}$, eliminating any
sharp behavior at lower redshifts. 
At lower redshifts, the instantaneous thermal parameters $\gamma$ and $T(\Delta)$ are
converging to the same values, but pressure smoothing scale and curvature are not.

\begin{figure*}
\begin{center}
\includegraphics[angle=0,width=0.32\textwidth]{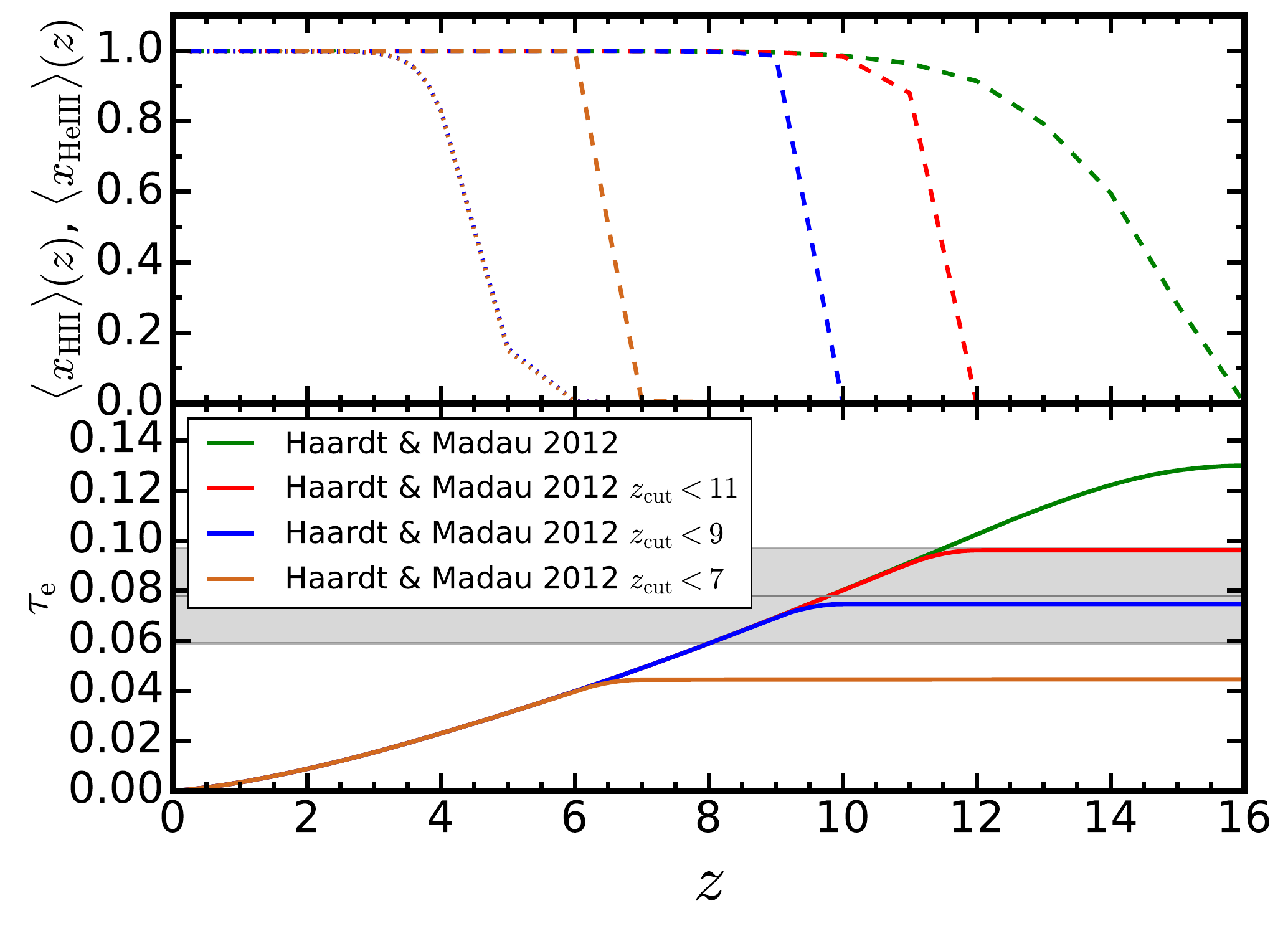}
\includegraphics[angle=0,width=0.32\textwidth]{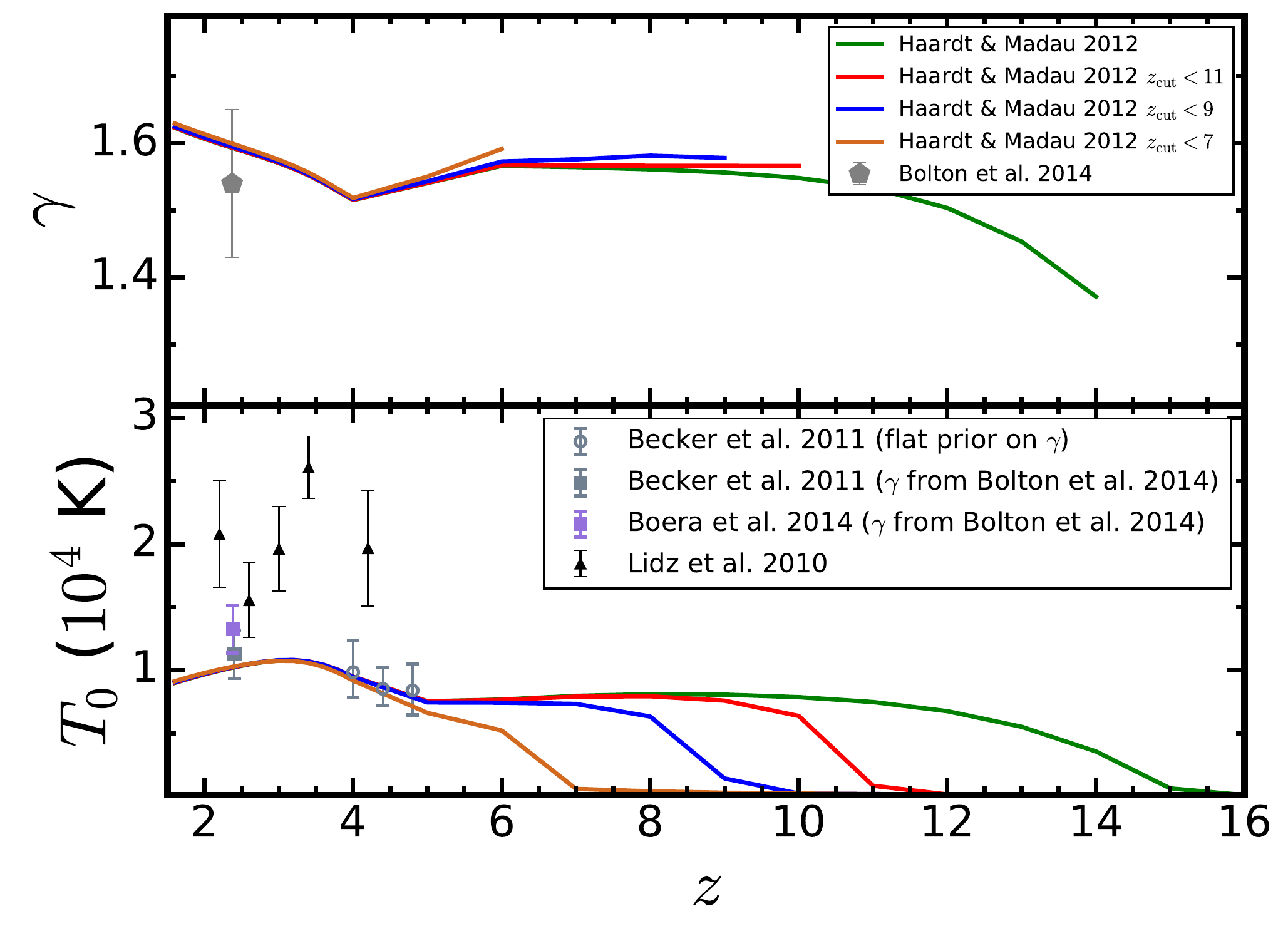}
\includegraphics[angle=0,width=0.32\textwidth]{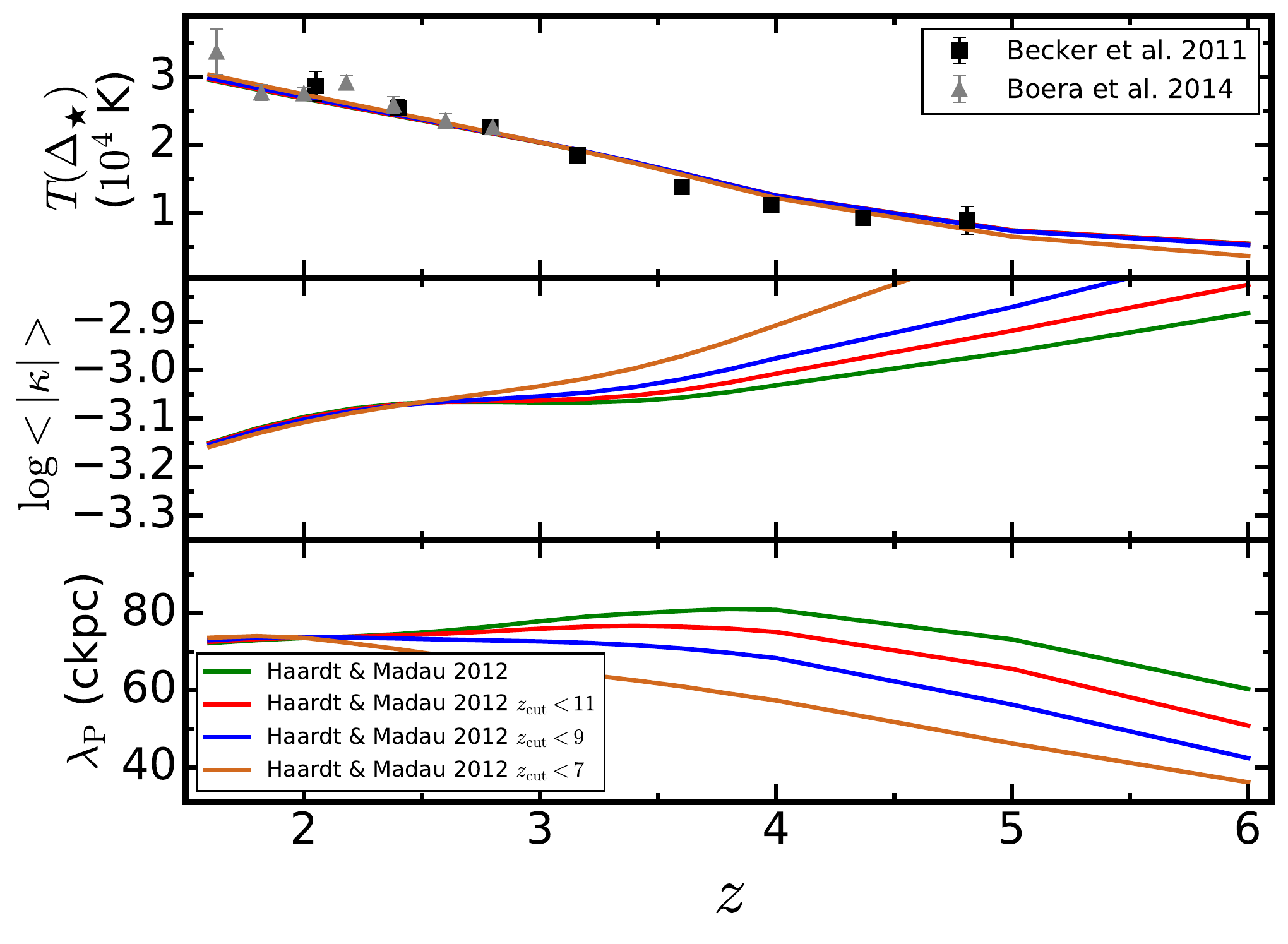}
\end{center}
\caption{Results for simulations using the HM12 model
cut at different redshifts:
$z=11$ (red), $z=9$ (blue), $z=7$ (brown), compared
with the original HM12 (green).
Left panel: ionization history.
Middle panel: the thermal history parameterized by the slope
of the density-temperature relation, $\gamma$, and the temperature
at mean density, $T_{0}$.
Right panel: evolution of the temperature at the optimal density
(upper), the curvature (middle) and the gas pressure smoothing scale (lower).
\label{fig:hmcut}}
\end{figure*}

Figure~\ref{fig:hmcut} shows results
for hydrodynamical simulations in which a simple redshift cutoff was applied
to the HM12 model.
In this approach the $\HI$ and $\HeI$ UVB rates are set to zero above
a certain redshift: $z=11$ (red), $z=9$ (blue), and $z=7$ (brown). 
We also show the original HM12 model (green lines) for comparison.
All the runs share the same $\HeII$ rates.
The left panel of Figure~\ref{fig:hmcut} shows the ionization histories 
of all these models. The middle and right panels present their thermal histories.
As expected, by applying a cutoff to the UVB rates, the reionization redshift
and its thermal signatures move down to the cutoff redshift.
The gas pressure scale shows very clearly the effect of producing
the heating due to $\HI$ reionization at lower redshift.
As discussed in Section~\ref{sec:results}, this is because
the pressure smoothing scale 
depends on the full thermal history of the universe and not just on
the instantaneous temperatures.

\section{Cosmological Effects on the Ionization and Thermal Properties of the IGM} \label{app:cosmo}

In this appendix we present the results of simulations that only differ in the cosmological parameters and have the
same UVB model, HM12. The random seeds in the initial condition are also the same.
Table~\ref{tab:cosmo} summarizes the cosmological parameters used in each run.
Cosmological model A is the default cosmological model used in this work.
Models B and C were selected from the posterior distribution of the Planck results \citep{Planck:2015}
in order to differ as much as possible in the expected matter power spectrum. These should maximize the difference between the models
while keeping them within the limits allowed by CMB observations. Model D was used in \citet{Lukic:2015} and is also
in agreement with the last CMB results.

\begin{figure}
\begin{center}
\includegraphics[angle=0,width=0.45\textwidth]{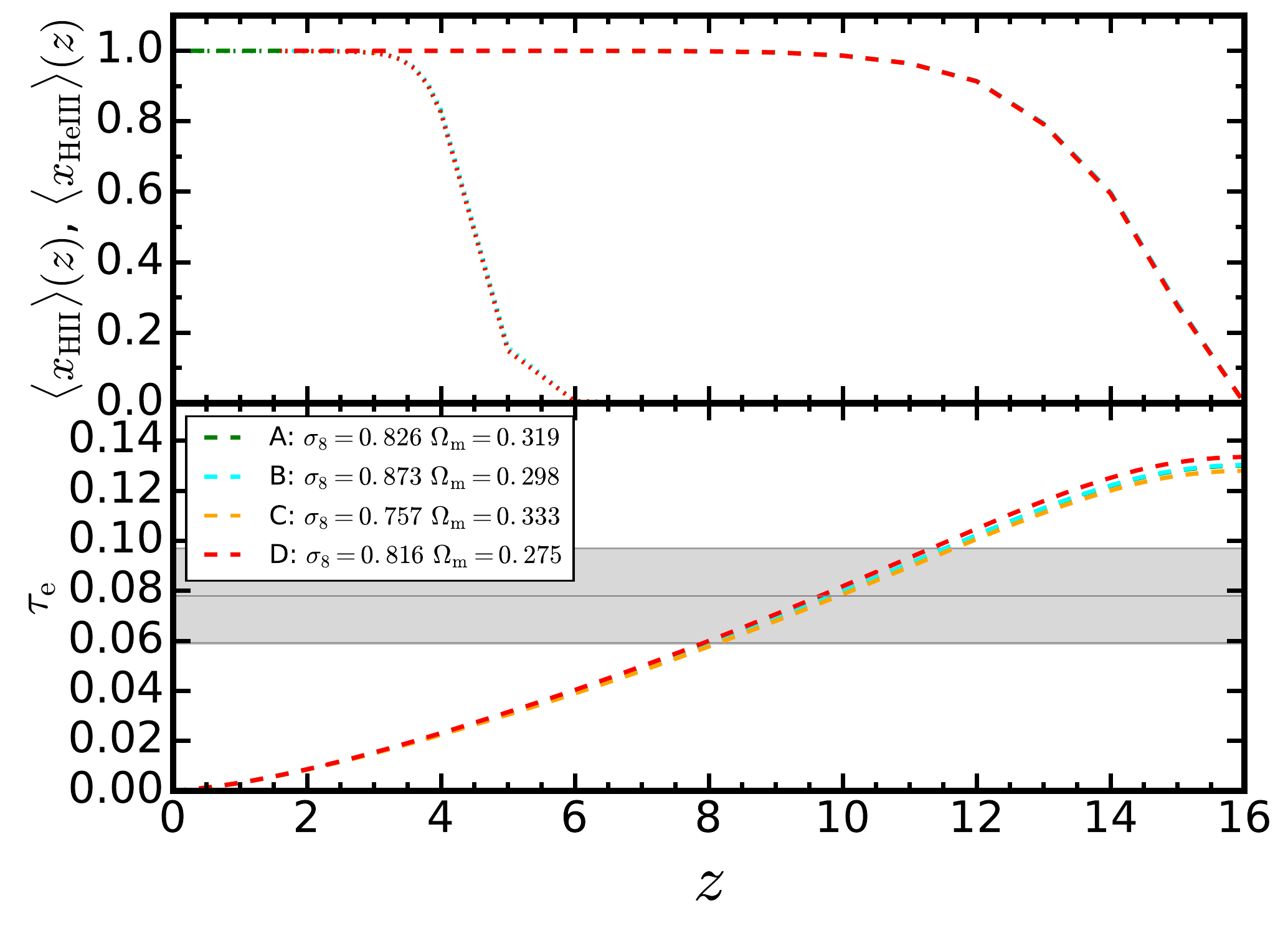}
\includegraphics[angle=0,width=0.45\textwidth]{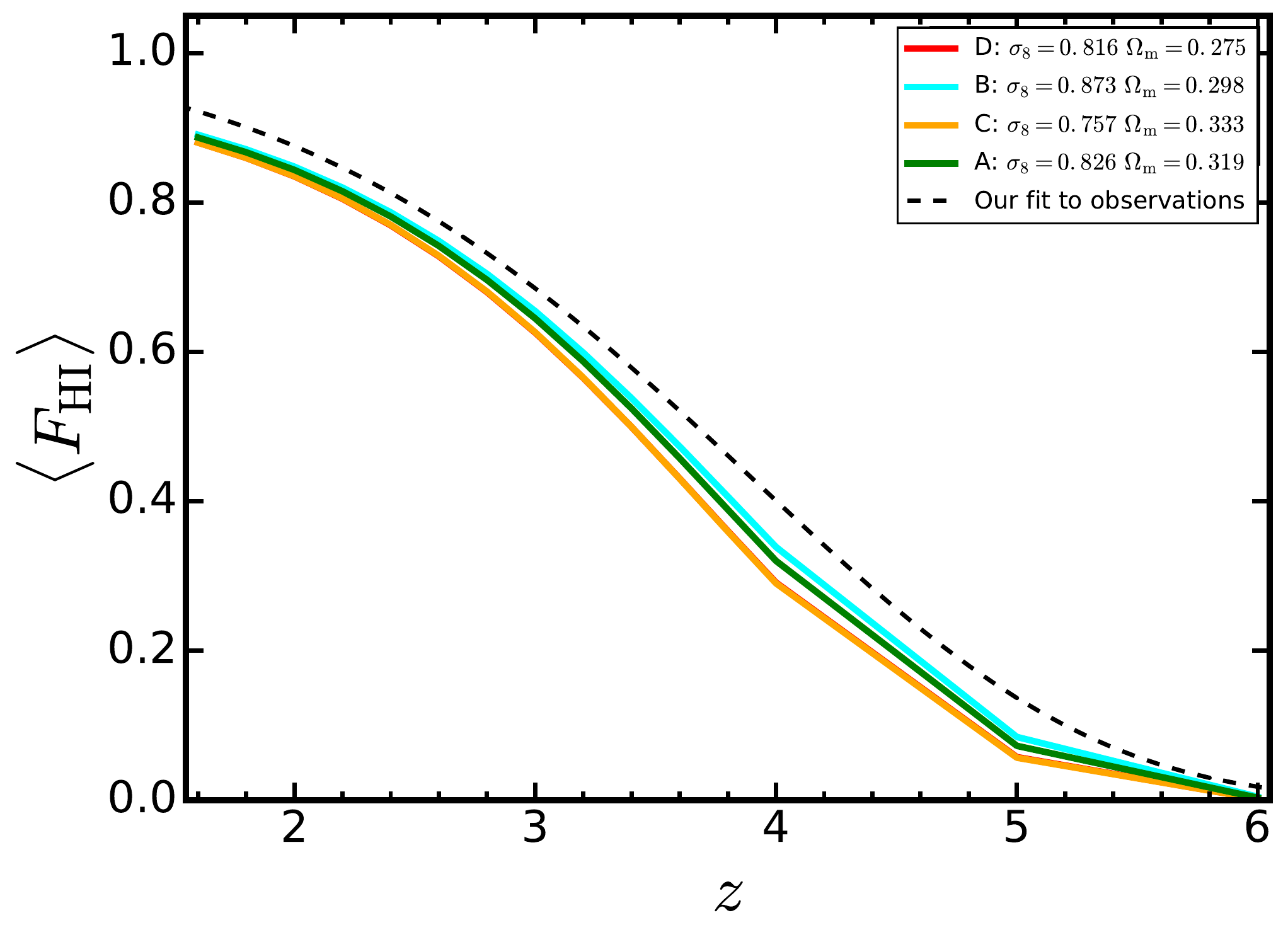}\\
\includegraphics[angle=0,width=0.45\textwidth]{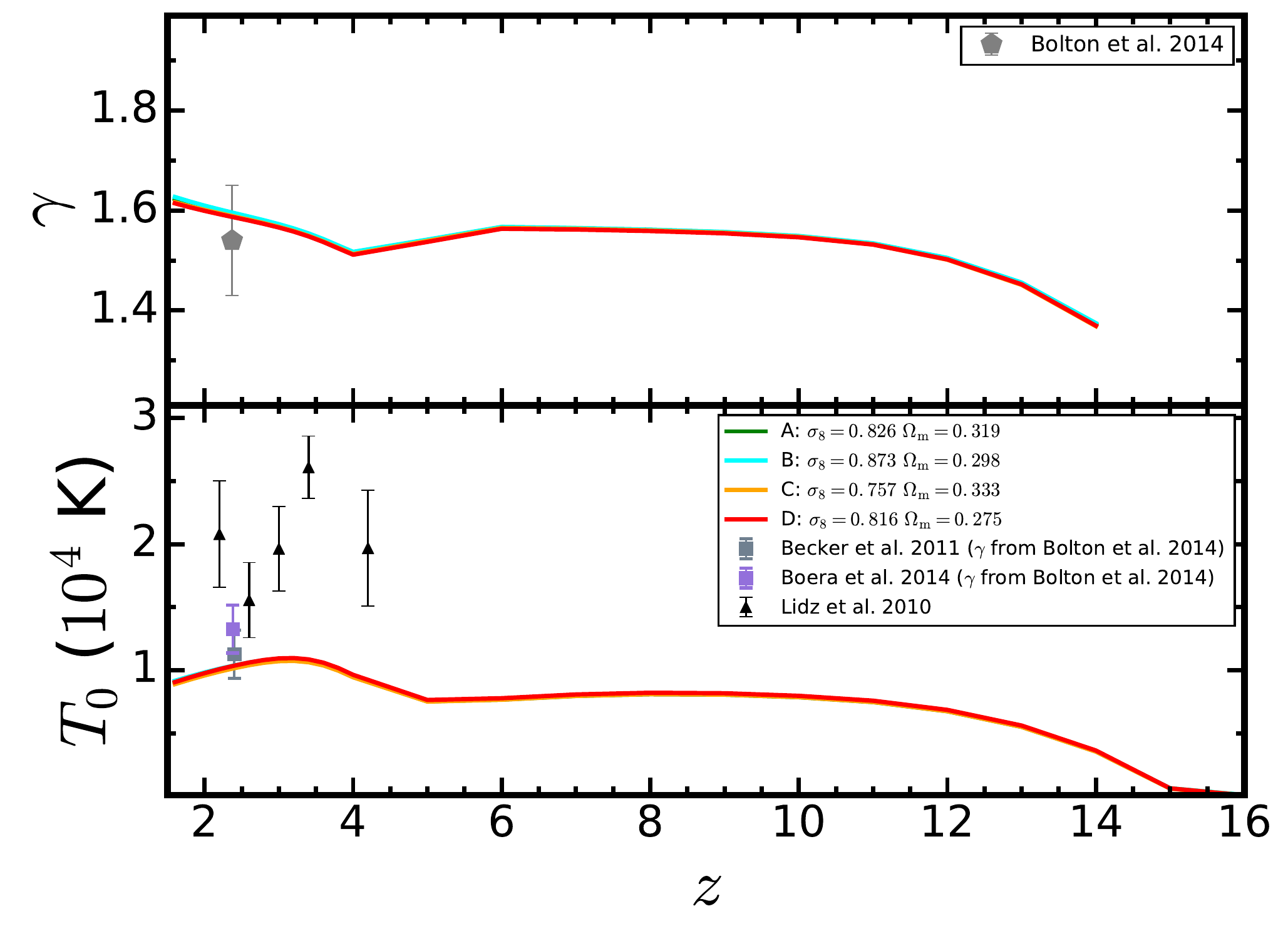}
\includegraphics[angle=0,width=0.45\textwidth]{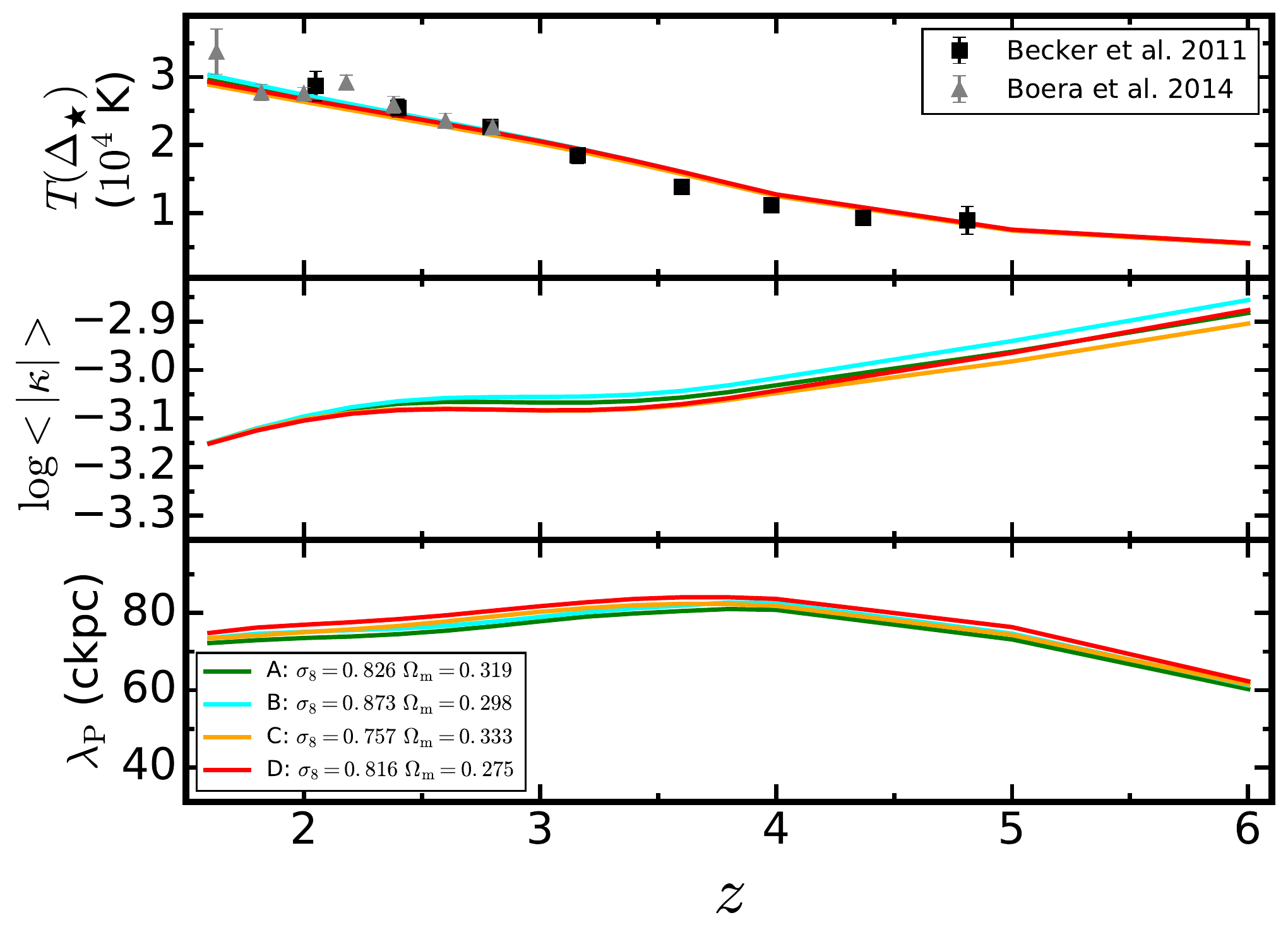}
\end{center}
\caption{Cosmological dependence of the IGM properties from simulations 
using different cosmological parameters but the same UVB (HM12). 
Upper left panel: $\HI$ and $\HeII$ ionization histories computed from the simulations.
Upper right panel: evolution of the $\HI$ mean flux.
Lower left panel: evolution of thermal parameters. slope of the temperature density relation, $\gamma$ (upper), and the temperature at mean density, $T_{0}$ (lower). 
Lower right panel: evolution of the temperature at the characteristic density, $T(\Delta)$ (upper), the curvature statistics, $\kurv$ (middle), and
the pressure smoothing scale, $\lj$ (lower).
\label{fig:cosmology}}
\end{figure}

Figure~\ref{fig:cosmology} shows the ionization and thermal histories for these four simulations. Both ionization histories, as well 
as the evolution of thermal parameters, are almost identical for all the runs. 
This means that within current observational constraints, the cosmological structure formation does not change
significantly the thermal evolution of the IGM, which is determined by the UVB model used.
In fact, this result is expected, as from our analytical calculation of the photoionization and photoheating
rates we can easily calculate how these rates will change with some cosmological parameters.
The difference in photoionization rate values for
different cosmologies will be the difference between $\Omega_{\rm b}h^{2}$ in the models. Difference in effective 
photoheating values for different cosmologies will be driven by the difference between $H(z)$\footnote{In all cases
we are assuming the same abundance values for hydrogen, $X_{\rm p}$, and helium, $Y_{\rm p}$.}. 

We want to emphasize here that even when two simulations/models share the same ionization and thermal evolution, 
that does not mean that the \lyalpha{} observables (probability density function, flux power spectrum, etc.) from these runs will not show differences
between them as the observables could have their own dependence on cosmological parameter or other
parameters.

\begin{table}
\begin{center}
\caption{Cosmological Parameters Used in Simulations\label{tab:cosmo}}
\begin{tabular}{lcccccc}
\tableline\tableline
 Model & $\Omega_{\rm m}$ & $\Omega_{\rm b}$ & $\Omega_{\Lambda}$ & $h$ & $\sigma_{8}$ & $n_{\rm s}$\\
\tableline
A & 0.320 &0.0496&0.681&0.670&0.826&0.966\\
B & 0.298 &0.0477&0.702&0.686&0.873&0.974\\
C & 0.333 &0.0517&0.667&0.658&0.757&0.971\\
D & 0.275 &0.0460&0.725&0.702&0.816&0.960\\
\tableline
\end{tabular}
\tablecomments{
Column 1: model name.
Column 2: total matter content.
Column 3: total baryonic matter content.
Column 4: cosmological constant.
Column 5: normalized Hubble constant ($0-1$).
Column 6: normalization of the power spectrum.
Column 7: spectral index.}
\end{center}
\end{table}

\section{New Optically Thin Photoionization and Photoheating Rates} \label{app:tables}

\begin{figure}
\begin{center}
\includegraphics[angle=0,width=0.33\textwidth]{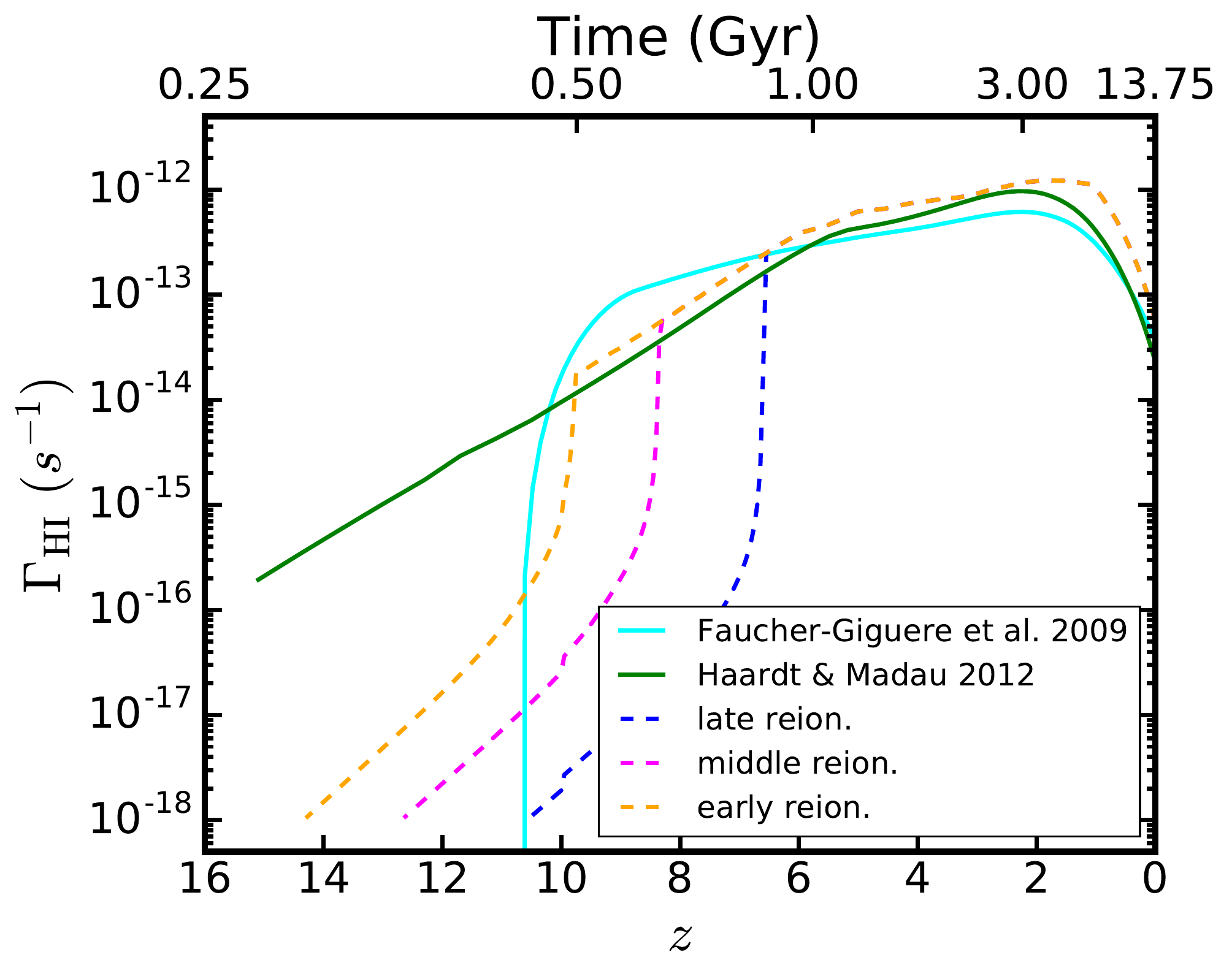}
\includegraphics[angle=0,width=0.33\textwidth]{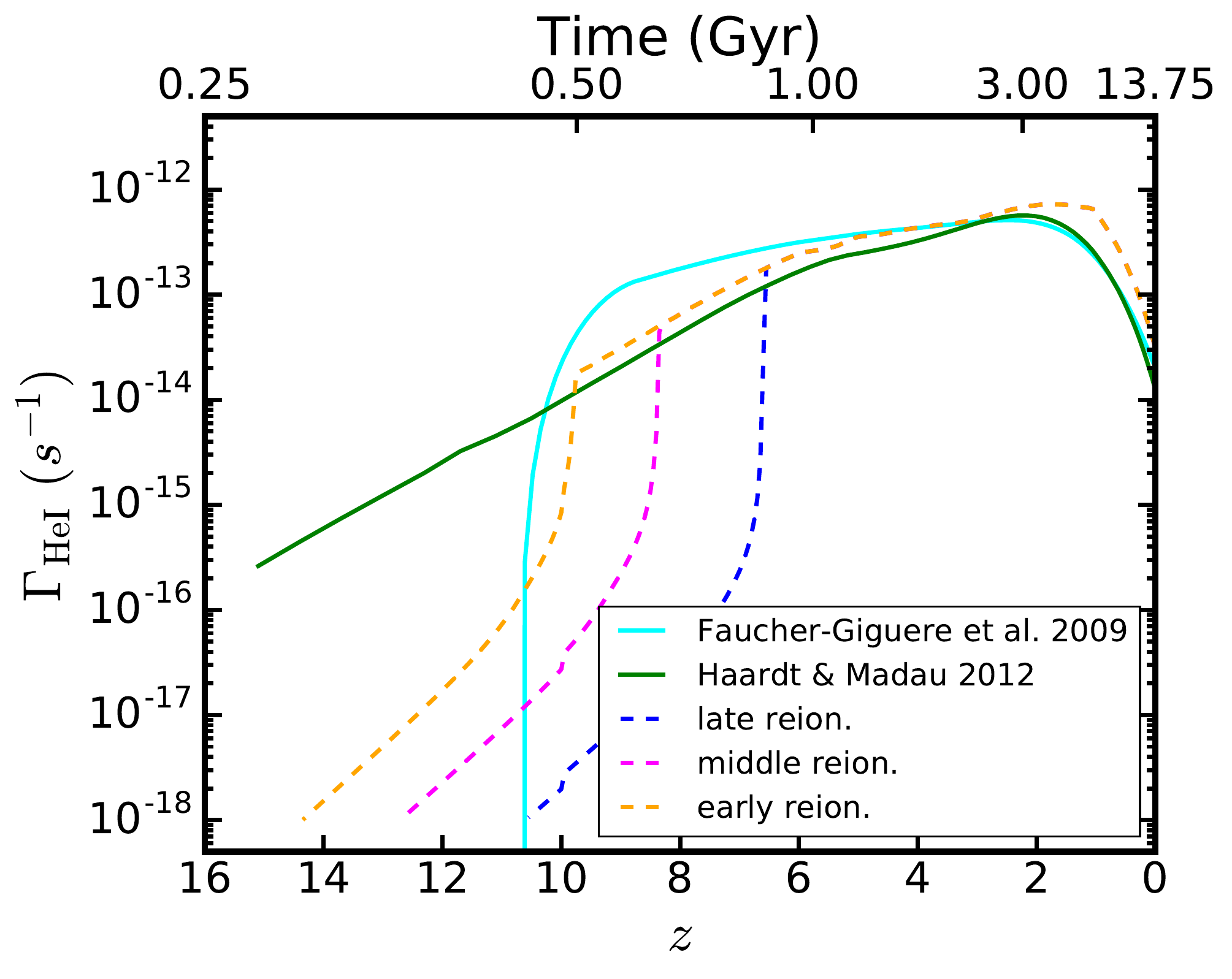}
\includegraphics[angle=0,width=0.33\textwidth]{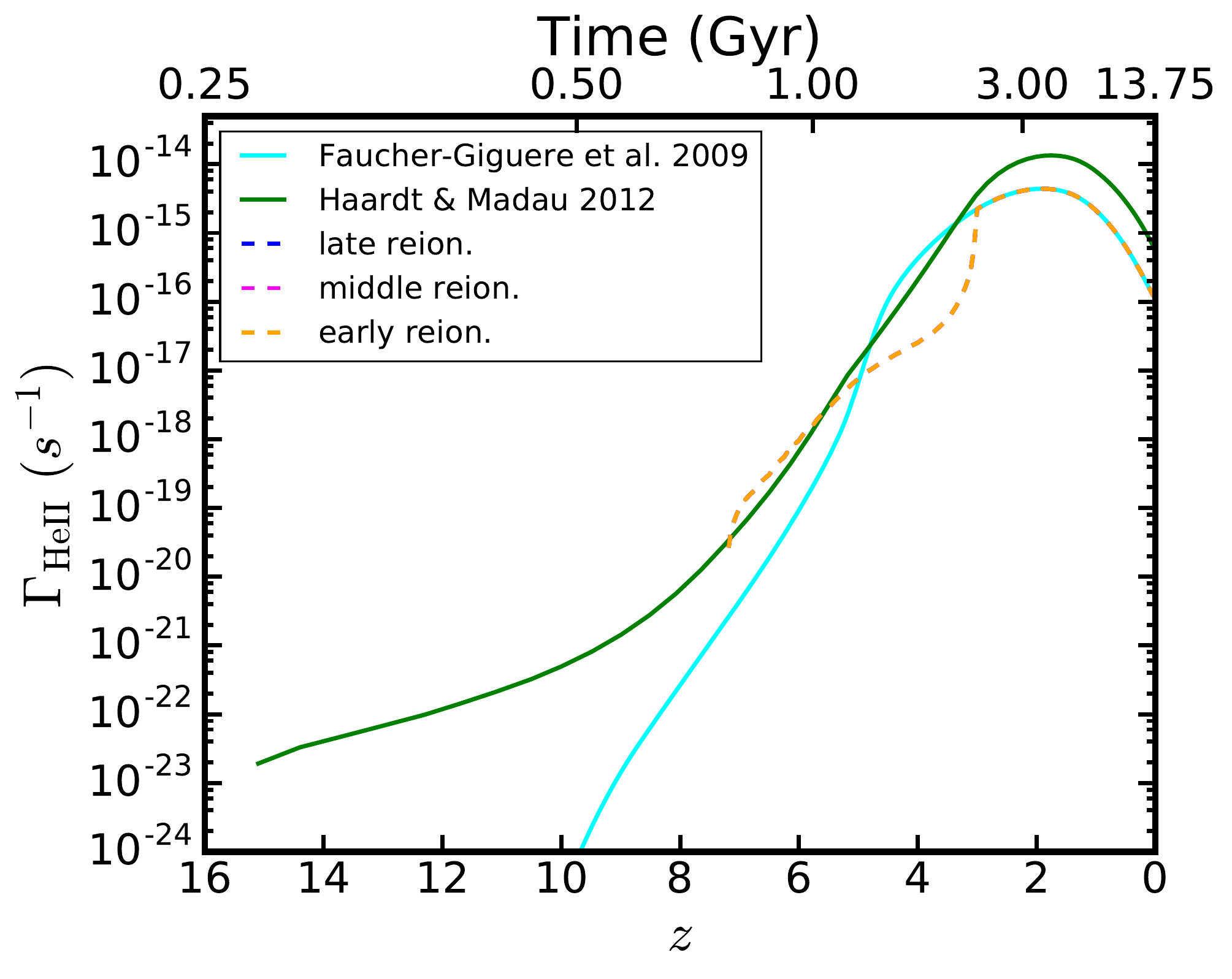}\\
\includegraphics[angle=0,width=0.33\textwidth]{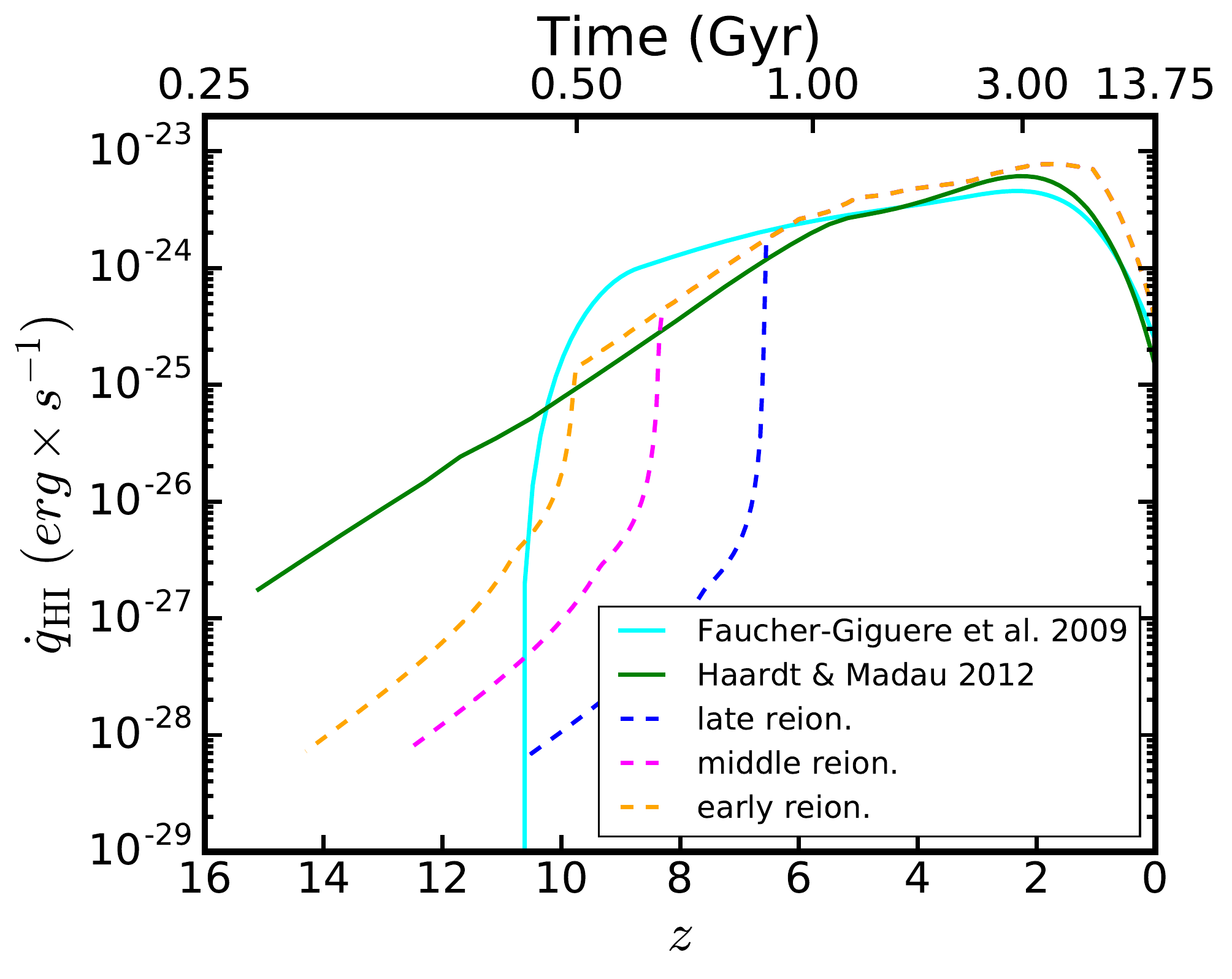}
\includegraphics[angle=0,width=0.33\textwidth]{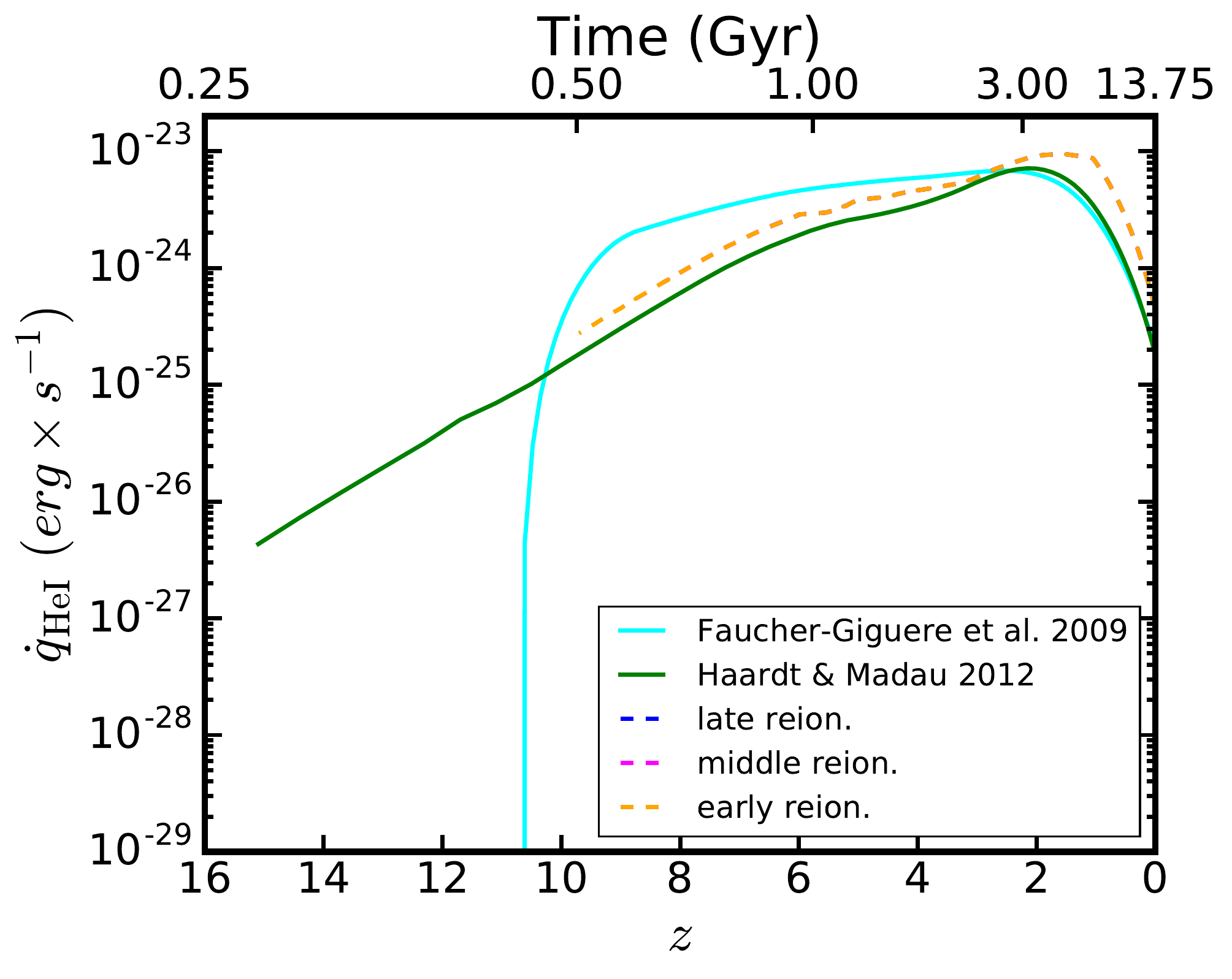}
\includegraphics[angle=0,width=0.33\textwidth]{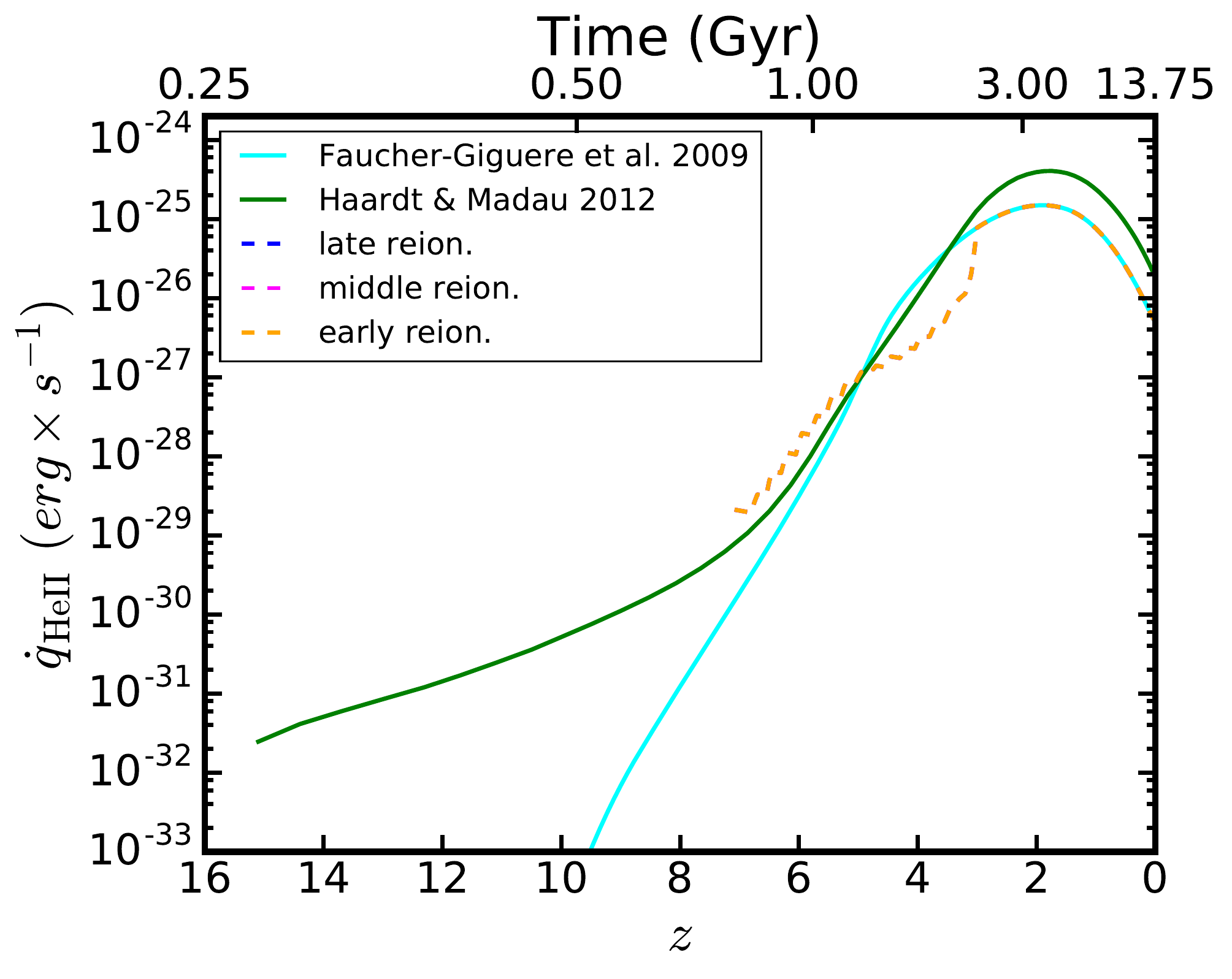}
\end{center}
\caption{Evolution of photoionization and photoheating rates with redshift for our new late reionization,
middle reionization, and early reionization UVB models. See text for more details on how these models
were computed. The HM12 
and FG09 
models are also shown for comparison.
\label{fig:newuv}}
\end{figure}

We are  making  our new UVB models freely  available  for  public  use,
so that they can be used by the whole community in future hydrodynamical
simulations.
We provide here the photoionization and photoheating rates 
for the late $\HI$ reionization (table~\ref{tab:uv1}), 
middle $\HI$ reionization (table~\ref{tab:uv2}), and 
early $\HI$ reionization (table~\ref{tab:uv3})
models, assuming a total heat input during reionization of
$\Delta T_{\HI}=2\times10^{4}$ K for $\HI$ and 
$\Delta T_{\HeII}=1.5\times10^{4}$ K for $\HeII$.
The rates for these models are also shown in Figure~\ref{fig:newuv}.
We refer to Section~\ref{sec:newmodel} for
a careful explanation on how these rates were obtained.
In our new models we have also applied a small correction to the $\HI$ 
and $\HeI$ photoionization rates of HM12 
once reionization is completed 
to ensure that the simulations match current best observations
of the $\HI$ mean flux at different redshifts.
The goal is to reduce the effect of the current standard post-process rescaling
approach done with simulations that aim to reproduce \lyalpha{} statistics.
The photoheating rates have been corrected by the same factor in order
to keep the heat input per volume element constant in the models and therefore
keep exactly the same thermal histories.
Section~\ref{sec:discuss} elaborates on the limitations
and applicability of these models.
Table~\ref{tab:uv1}, table~\ref{tab:uv2} and table~\ref{tab:uv3}
are published in their entirety in a machine-readable
format.
Just a portion is shown here, as a guidance to
its format and content.
We encourage anyone interested in running some other specific model to 
contact the authors.

\begin{table}
\begin{center}
\caption{Tabulated UV Background for Late \HI{} Reionization ($z_{\rm reion,\HI}=6.55$), $\Delta T_{\HI}=2\times10^{4}$ K;
He\_A Reionization, $\Delta T_{\HeII}=1.5\times10^{4}$ K. \label{tab:uv1}}
\begin{tabular}{lcccccc}
\tableline\tableline
 $\log_{10}(z+1)$ & $\Gamma_{\HI}$ & $\Gamma_{\HeI}$  & $\Gamma_{\HeII}$ & $\dot{q}_{\HI}$ & $\dot{q}_{\HeI}$ & $\dot{q}_{\HeII}$\\
   & (s$^{-1}$) & (s$^{-1}$) & (s$^{-1}$) & (erg s$^{-1}$) & (erg s$^{-1}$) & (erg s$^{-1}$)\\
\tableline
0.0000 & 5.700e-14 & 3.100e-14 & 1.122e-16 & 3.561e-25 & 4.486e-25 & 5.008e-27\\
0.0212 & 7.131e-14 & 3.942e-14 & 1.291e-16 & 4.466e-25 & 5.632e-25 & 5.729e-27\\
0.0414 & 8.817e-14 & 4.882e-14 & 1.564e-16 & 5.546e-25 & 6.944e-25 & 6.874e-27\\
0.0607 & 1.081e-13 & 6.037e-14 & 1.892e-16 & 6.806e-25 & 8.499e-25 & 8.215e-27\\
...\\
\tableline
\end{tabular}
\tablecomments{
This table is published in its entirety in a machine readable
format. A portion is shown here for guidance regarding 
its form and content.
Column 1: redshift (logarithm).
Column 2: $\HI$ photoionization rate.
Column 3: $\HeI$ photoionization rate.
Column 4: $\HeII$ photoionization rate.
Column 5: $\HI$ photoheating rate.
Column 6: $\HeI$ photoheating rate.
Column 7: $\HeII$ photoheating rate.}
\end{center}
\end{table}

\begin{table}
\begin{center}
\caption{Tabulated UV Background for Middle \HI{} Reionization ($z_{\rm reion,\HI}=8.30$), $\Delta T_{\HI}=2\times10^{4}$ K;
He\_A Reionization, $\Delta T_{\HeII}=1.5\times10^{4}$ K. \label{tab:uv2}}
\begin{tabular}{lcccccc}
\tableline\tableline
 $\log_{10}(z+1)$ & $\Gamma_{\HI}$ & $\Gamma_{\HeI}$  & $\Gamma_{\HeII}$ & $\dot{q}_{\HI}$ & $\dot{q}_{\HeI}$ & $\dot{q}_{\HeII}$\\
   & (s$^{-1}$) & s$^{-1}$) & (s$^{-1}$) & (erg s$^{-1}$) & (erg s$^{-1}$) & (erg s$^{-1}$)\\
\tableline
0.0000 & 5.700e-14 & 3.100e-14 & 1.122e-16 & 3.561e-25 & 4.486e-25 & 5.008e-27\\
0.0212 & 7.131e-14 & 3.942e-14 & 1.291e-16 & 4.466e-25 & 5.632e-25 & 5.729e-27\\
0.0414 & 8.817e-14 & 4.882e-14 & 1.564e-16 & 5.546e-25 & 6.944e-25 & 6.874e-27\\
0.0607 & 1.081e-13 & 6.037e-14 & 1.892e-16 & 6.806e-25 & 8.499e-25 & 8.215e-27\\
...\\
\tableline
\end{tabular}
\tablecomments{ 
This table is published in its entirety in a machine readable
format. A portion is shown here for guidance regarding 
its form and content.
Column 1: redshift (logarithm).
Column 2: $\HI$ photoionization rate.
Column 3: $\HeI$ photoionization rate.
Column 4: $\HeII$ photoionization rate.
Column 5: $\HI$ photoheating rate.
Column 6: $\HeI$ photoheating rate.
Column 7: $\HeII$ photoheating rate.}
\end{center}
\end{table}

\begin{table}
\begin{center}
\caption{Tabulated UV Background for Early \HI{} Reionization ($z_{\rm reion,\HI}=9.70$), $\Delta T_{\HI}=2\times10^{4}$ K;
He\_A Reionization, $\Delta T_{\HeII}=1.5\times10^{4}$ K. \label{tab:uv3}}
\begin{tabular}{lcccccc}
\tableline\tableline
 $\log_{10}(z+1)$ & $\Gamma_{\HI}$ & $\Gamma_{\HeI}$  & $\Gamma_{\HeII}$ & $\dot{q}_{\HI}$ & $\dot{q}_{\HeI}$ & $\dot{q}_{\HeII}$\\
   & (s$^{-1}$) & (s$^{-1}$) & (s$^{-1}$) & (erg s$^{-1}$) & (erg s$^{-1}$) & (erg s$^{-1}$)\\
\tableline
0.0000 & 5.700e-14 & 3.100e-14 & 1.122e-16 & 3.561e-25 & 4.486e-25 & 5.008e-27\\
0.0212 & 7.131e-14 & 3.942e-14 & 1.291e-16 & 4.466e-25 & 5.632e-25 & 5.729e-27\\
0.0414 & 8.817e-14 & 4.882e-14 & 1.564e-16 & 5.546e-25 & 6.944e-25 & 6.874e-27\\
0.0607 & 1.081e-13 & 6.037e-14 & 1.892e-16 & 6.806e-25 & 8.499e-25 & 8.215e-27\\
...\\
\tableline
\end{tabular}
\tablecomments{This table is published in its entirety in a machine readable
format. A portion is shown here for guidance regarding 
its form and content.
Column 1: redshift (logarithm).
Column 2: $\HI$ photoionization rate.
Column 3: $\HeI$ photoionization rate.
Column 4: $\HeII$ photoionization rate.
Column 5: $\HI$ photoheating rate.
Column 6: $\HeI$ photoheating rate.
Column 7: $\HeII$ photoheating rate.}
\end{center}
\end{table}

\bibliography{apj-jour,uvmodeling}    

\end{document}